**A nanoporous capacitive electrochemical ratchet for continuous ion separations**

Rylan Kautz[1†], Alon Herman[2†], Ethan J. Heffernan[1], Keren Shushan Alshochat[2], Eden Grossman[2], Rahul Saxena[2], Camila Muñetón[3,4], David Larson[5,6], Joel W. Ager III[5,6,7,8], Francesca M. Toma[5,6,9,10], Shane Ardo[1,3,11]* and Gideon Segev[2,5,6]**

[1]Department of Materials Science & Engineering, University of California Irvine, Irvine, CA 92697, USA

[2]School of Electrical Engineering, Tel Aviv University, Tel Aviv 6997801, Israel

[3]Department of Chemistry, University of California Irvine, Irvine, CA 92697, USA

[4]Department of Chemistry, University of Massachusetts Boston, Boston, MA 02125 USA

[5]Chemical Sciences Division, Lawrence Berkeley National Lab, Berkeley, CA 94720, USA

[6]Joint Center for Artificial Photosynthesis, Lawrence Berkeley National Lab, Berkeley, CA 94720, USA

[7]Materials Sciences Division, Lawrence Berkeley National Lab, Berkeley, CA 94720, USA

[8]Department of Materials Science and Engineering, University of California Berkeley, Berkeley, CA 94720, USA

[9]Institute of Functional Materials for Sustainability, Helmholtz Zentrum Hereon, Teltow, 14513, Germany

[10]Faculty of Mechanical and Civil Engineering, Helmut Schmidt University, Hamburg 22043, Germany

[11]Department of Chemical & Biomolecular Engineering, University of California Irvine, Irvine, CA 92697, USA

*Email: ardo@uci.edu

**Email: gideons1@tauex.tau.ac.il

† These authors contributed equally to this work

## Abstract

Directed ion transport in liquid electrolyte solutions underlies many phenomena in Nature and industry. While Nature has devised structures that drive continuous ion flow without Faradaic redox reactions, artificial analogs do not exist. Here we report the first demonstration of an ion pump that drives aqueous ions against a force using a capacitive ratchet mechanism that does not require redox reactions. Modulation of an electric potential between thin metallic layers on either face of a nanoporous alumina wafer immersed in solution resulted in persistent voltages and ionic currents. This occurs due to the non-linear capacitive nature of electric double layers, whose repeated charging and discharging sustains a continuous ion flux. Ratchet driven electrodialysis was demonstrated reaching a 50% decrease in the conductivity of the solution in a dilution cell. These ratchet-based ion pumps can enable continuous desalination and selective ion separation using an electrically powered device with no moving parts.

## Main

Ratchets are non-equilibrium devices that utilize temporally modulated input signals and spatial asymmetries to drive a steady state particle flux.[1–3] Ratchets have been studied theoretically and experimentally[1,4–8] for electronic signal rectification,[1,4,7,8] to drive net transport of uncharged species through induced-charge electrokinetics,[9] for microparticle and nanoparticle sorting,[10–16] and to drive net ionic current or water pumping by alternating redox reactions.[17–19] A recent theoretical study has shown that ratchet based ion pumps (RBIPs) can sort ions with unprecedented selectivity and desalinate water.[20,21]

An electric double layer forms at the interface between an electrode and a liquid electrolyte and consists of a compact layer of adsorbed solvent molecules and ions, and a diffuse layer of mobile ions. This results in a capacitance that varies with the magnitude of the electrode potential and in response to temporal changes in it.[22–24] These variations in capacitance affect the double layer charging and discharging time constants. Here we utilize this phenomenon to pump ions across a membrane. RBIP membranes are made of nanoporous anodized aluminum oxide (AAO) wafers with thin metal layers, serving as electrodes covering the two wafer surfaces without blocking the pores, thus forming nanoporous capacitor-like structures (Figure 1a-e, Figure S1). Details on the RBIP fabrication process can be found in the methods section.

We start by analyzing the performance of RBIPs in which both surfaces are coated with gold thin films. This configuration is termed Au-Au. Due to the rough and poly-crystalline nature of the gold films covering the AAO wafer (Figure 1b,c), the RBIP surface is inhomogeneous. Moreover, the energetics of $Cl^-$ adsorption processes are facet dependent.[25] Combined with the inherently non-linear nature of double layers,[22] these result in potential-dependent adsorption and charging phenomena on the RBIP surfaces leading to a dispersion of time constants and thus a non-linear capacitance.[23–26] Since no two poly-crystalline surfaces are identical, small differences in the RBIP surface properties result in an unintentional asymmetry, which is essential for ratchet-driven transport.

When an input signal, $V_{in}$, is applied between the two metal layers, the double layer at one surface is charged while the other is discharged with time constants that are determined by the properties of the input signal, the metal surface, and the aqueous electrolyte. The electrostatic potential difference between the bulk solution and the adjacent metal surface at each side of the RBIP ($V_L$ and $V_R$ in Figure 1a) fluctuates according to these time constants, and the electric potential difference between the bulk of the two electrolyte compartments is:

$$V_{out}(t) = V_L(t) + V_{in}(t) - V_R(t). \quad (1)$$

The RBIP performance was evaluated by placing an RBIP between two compartments of aqueous chloride-containing electrolyte and measuring the voltage between two Ag/AgCl wires, which are immersed in the solution on each side of the RBIP ($V_{out}$). Because of the extremely high reaction rate for reversible oxidation of $Cl^-$ at Ag to form insoluble AgCl, the voltage difference between the two Ag/AgCl wires is dominated by the electrochemical potential difference of $Cl^-$ at the wire surfaces. The measured resting voltage between the two Ag/AgCl wires, $V_{rest}$, when both compartments are filled with the same electrolyte (1 – 10 mM $Cl^-$ containing solution) is lower than 10 mV in magnitude and is typically about 3 mV. Application of a DC bias across the RBIP contacts ($V_{in}$ of +300 mV or –300 mV) resulted in the expected prompt observation of $V_{out} \approx V_{in}$ that decayed to $V_{out} \approx V_{rest}$ over ~1 sec due to capacitive charging.

Hence, under these conditions there is no steady state net driving force for ion transport through the RBIP.

To demonstrate the difference in charging and discharging time constants and their contribution to the ratchet output, the voltage signals $V_L(t)$ and $V_R(t)$ were measured while the RBIP was operating. The input signal was a rectangular wave in the form:

$$V_{in}(t) = \begin{matrix} V_a & 0 < t \leq d_c T \\ -V_a & d_c T < t \leq T \end{matrix}, \qquad (2)$$

where $T$ is the temporal period and $d_c$ ($\in [0,1]$) is the duty cycle. The charging and discharging time constants of every surface at every part of the period were found by fitting the averaged measured signals to a single exponential charging/discharging function:

$$V(t) = \begin{matrix} V_{f,1} + \left(V_{i,1} - V_{f,1}\right)\exp\left(-\frac{t}{\tau_1}\right) & 0 < t \leq d_c T \\ V_{f,2} + \left(V_{i,2} - V_{f,2}\right)\exp\left(-\frac{t-d_c T}{\tau_2}\right) & d_c T < t \leq T \end{matrix}, \qquad (3)$$

where $V_i$ is the initial voltage, and $\tau$ is the charging/discharging time constant. $V_f$ is the voltage at which the signal would saturate if the capacitances were linear. A subscript 1 notes the first part of the temporal period ($0 < t \leq d_c T$) and a subscript 2 the second part of the period ($d_c T < t \leq T$).

Because the RBIP is electrically floating (i.e., no part of the system is grounded), charging of one contact is accompanied by discharging of the other (Figure 1f-g). Thus, the time constants for charging the right and left surfaces are $\tau_{R,1}$ and $\tau_{L,2}$ respectively, and the time constants for discharging the right and left surfaces are $\tau_{R,2}$ and $\tau_{L,1}$ respectively. Figure 1h shows the extracted time constants for charging and discharging the surfaces as a function of the duty cycle. The other fitted parameters are shown in Figure S2. In linear systems the time constants are invariant. However, due to the non-linearity of the double layer capacitances, all time constants vary with the duty cycle. Particularly, there is a significant difference between the charging (and discharging) time constants of the two surfaces: $\tau_{L,1} \neq \tau_{R,2}$ and $\tau_{L,2} \neq \tau_{R,1}$. At a duty cycle of 0.5 the time constants are: $\tau_{L,1} = 6.59$ ms, $\tau_{R,2} = 8.21$ ms, and $\tau_{L,2} = 8.5$ ms, $\tau_{R,1} = 7.23$ ms. This asymmetry implies that a voltage rise due to the charging of a surface at the first part of the period is not fully negated by the charging of the opposite surface during the second part of the period. As a result, a non-zero time-averaged output voltage can be obtained even for time-symmetric input signals. For duty cycles above 0.4 the time constants for charging one surface and discharging the other are similar yet vary significantly between the first part and the second part of the period: $\tau_{L,1} \approx \tau_{R,1} \neq \tau_{L,2} \approx \tau_{R,2}$. A similar analysis for signals with different frequencies and $d_c = 0.5$ shows that the time constants decrease significantly with the signal frequency (Figure S4-5). Thus, the charging and discharging dynamics are mostly determined by the duration for charging and discharging. As a result, the difference in charging and discharging time constants is smallest for $d_c = 0.5$, and for this duty cycle the asymmetry is determined only by the non-linear nature of the capacitance with the potential and the difference in the microstructure of the two surfaces.

The net ratchet induced voltage, $\Delta\bar{V}_{out}$, was obtained by finding the mean value of $V_{out}$ while the ratchet is ON and reducing from it the mean value of $V_{out}$ while the ratchet was OFF (see the methods section and Figure S3 for more details). Figure 1i shows $\Delta\bar{V}_{out}$ as a function of the input signal duty cycle. Demonstrating the signature output of a flashing ratchet,[1,4] the RBIP output is very low when a constant bias is applied (duty cycle of 0 or 1). However, once the input signal alternates, a net voltage builds up between the two compartments providing a driving force for ion transport. Moreover, $\Delta\bar{V}_{out}$ reaches 29.1 mV for a moderate duty cycle of 0.375 where the time averaged input is -75 mV, and 23 mV for a duty cycle of 0.5 where the time averaged input voltage is 0 V. Since $\Delta\bar{V}_{out}$ is maximal when the time-averaged input voltage is near 0 V, and it is negligible for extreme duty cycles when the magnitude of the time-averaged input voltage is maximal, it can be concluded that the contribution of mechanisms such as conduction, ion electrophoresis and electro-osmosis to $\Delta\bar{V}_{out}$ is minimal. This was further verified by showing that mass transport of water through the membrane is negligible (Figures S17-S18). Furthermore, the amplitude of the input signal is significantly lower than the Gibbs free energy required for water splitting. Hence there are no electrochemical reactions taking place on the electrode surfaces. At such low voltages, even if the voltage is held constant, the applied potential bias supports only the formation of double layers, there is no direct current between the electrodes, the electric field outside the double layers is zero, and there is no electrophoretic ion transport.

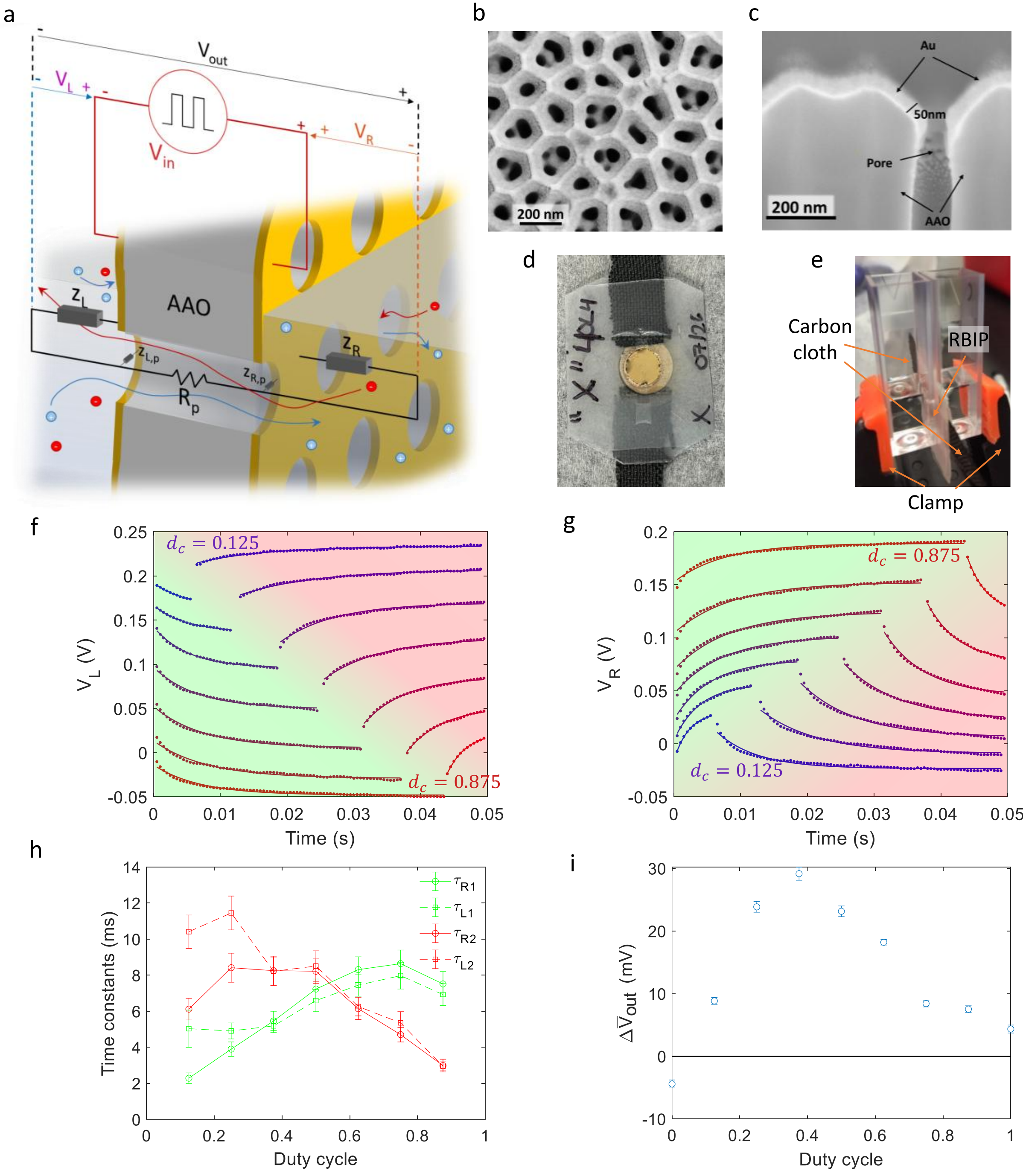

***Figure 1: The RBIP***|*(a) A schematic illustration of the RBIP structure and double layers impedances. (b-c) Plan and cross section view SEM images of fabricated RBIPs. respectively. (d) A photograph of a sample before its assembly. (e) A photograph of the RBIP performance characterization setup. (f-g) Experimentally measured signals* $V_L(t)$ *and* $V_R(t)$, *respectively, for rectangular-wave inputs with various duty cycles,* $d_c$, *over one time period,* $T = 50$ *ms. The green areas are for the first portion of the period* $0 < t \leq d_c T$ *and the red areas are for the second portion of the period* $d_c T < t \leq T$. *The dots are the measurements, and the solid lines mark the best exponential fit to the data. The color coding marks the duty cycle. The electrolyte is 1 mM KCl aqueous solution, and the input signal is alternating between 300 mV and -300 mV. (h) Time constants extracted from the fitted curves in (f) and (g). Subscripts 1 and 2 denote that the time constants were extracted for the first* ($0 < t \leq d_c T$ ) *and second* ($d_c T < t \leq T$ ) *parts of the input signal, respectively. The error bars indicate the fitting 95% confidence interval. (i) The time averaged output voltage* $\Delta \bar{V}_{out}$ *as a function of the input signal duty cycle.*

Next, we measured the ratchet performance for input signals with various frequencies and duty cycles. The charging and discharging time constants of $V_{out}(t)$ vary significantly with the input signal frequency and duty cycle (Figure S6 and Figure S8a-b). Figure 2a shows $\Delta\bar{V}_{out}$ as a function of the duty cycle and frequency. The input signal amplitude is $V_a$ = 0.3 V, the electrolyte is 1 mM HCl aqueous solution, the active area of the membrane is A = 0.32 $cm^2$. As in Figure 1i, here also $\Delta\bar{V}_{out}$ is practically zero for extreme duty cycles and low frequencies when the device is effectively at its steady state for significant parts of the temporal period. $\Delta\bar{V}_{out}$ increases with the input signal frequency reaching about -17 mV at a frequency of 250 Hz and a duty cycle of 0.6. Although the RBIP output increases with frequency within this range, it decreases for higher frequencies (Figure S14 and Figure S19).

$\Delta\bar{V}_{out}$ values for an aqueous 1 mM electrolyte were observed to be larger (in magnitude) than $\Delta\bar{V}_{out}$ values for an aqueous 10 mM electrolyte (Figure S14). Since the resistance for charge transport through the pore, $R_p$, effectively shunts $\bar{V}_{out}$ (Figure 1a), solutions with high ionic strength and membranes with a large pore diameter will tend to produce a lower RBIP output. The ionic strength of the solution also affects the double layer capacitances and the charging and discharging time constants at the RBIP surfaces leading to higher optimal frequencies (Figure S14). We observed that the diminished performance with higher concentration solutions can be partially mitigated using AAO membranes with smaller pores, but at the expense of higher overall ionic resistance per pore (Figure S19). Ratcheting was also demonstrated with $\Delta\bar{V}_{out}$ measured between two leak-free reference electrodes. Since reference electrodes measure the electric potential, this allows decoupling between the electric and chemical potentials that are induced by the ratchet (Figure S16). More details on the measurement of $\Delta\bar{V}_{out}$ and the output dependence on different input signals and solution parameters can be found in section 2 in the supporting information.

The buildup of $\bar{V}_{out}$ can result in the separation of cations and anions until columbic forces negate the ratchet action preventing further charge transport, or drive ambipolar transport.[21] To harness the ratchet induced voltage to drive a sustained net ionic current, charge neutrality must be maintained. This can be achieved by electrically shorting the two Ag/AgCl wires, which provides a low resistance path for removing chlorides from one compartment and generating them in the other. In this case, the current between the Ag/AgCl wires, $I_{out}$, is a result of oxidation of one silver wire and an aqueous chloride to form silver chloride, and reduction of silver chloride on the other wire to form silver and an aqueous chloride. The generation of chlorides in one compartment and their removal in the other balances deviations from electroneutrality due to ion pumping driven by charging and discharging of the RBIP contacts. Figure 2b shows the net ratchet induced current density output $\Delta\bar{I}_{out}/A$ as a function of the input signal duty cycle and frequency. As the voltage output, the current output shows a flashing rachet-like behavior with zero net output current at duty cycles of 0 and 1. Once an alternating input signal is injected to the device, a significant output current flows reaching as much as 3.15 μA/$cm^2$ at a frequency of 250 Hz and a duty cycle of 0.6. Figure S7 and Figures S8c-d show the average temporal response and the extracted time constants for all frequencies and duty cycles.

To demonstrate the RBIP ability to drive ions against a force, a current-voltage scan was taken between the two Ag/AgCl wires while the ratchet is OFF ($V_{in} = 0$ V) and when it was ON with a duty cycle of 0.5 and a frequency of 100 Hz (Figure 2c). When the ratchet is OFF, the curve shows an expected linear ohmic behavior. This is in contrast to membranes with conical nano-pores that were shown to have a rectifying transport behavior.[17,18] However, when the ratchet is ON, the curve shifts by approximately the same $\Delta\bar{V}_{out}$ that was measured under the same conditions (Figure 2a), thus entering a quadrant in which ions are driven against an electrostatic force. It is interesting to analyze this result through the redox potentials of the Ag/AgCl wires. When the ratchet is OFF, the CV curve of the two Ag/AgCl wires follows a simple resistive form in which $Cl^-$ is oxidized at the positively biased Ag/AgCl wire yielding an anodic current. However, once the ratchet is ON, the onset for this anodic reaction is shifted and $Cl^-$ is oxidized even at voltages below 0 V. Complementary, the onset of the cathodic reaction is also shifted such that a negative overpotential of about 15 mV is required to drive a cathodic current. Since the current voltage curve is linear, the energetic efficiency of the RBIP is approximately $\eta = 0.25 V_{oc} I_{sc} / \bar{P}_{in}$ where $V_{oc}$ and $I_{sc}$ are the open circuit voltage and the short circuit current as shown in Figure 2a,b respectively and $\bar{P}_{in}$ is the time averaged input power. At a frequency of 250 Hz and a duty cycle of 0.5, the input power is approximately 0.215 mW, and with the corresponding output from Figure 2a,b results in an efficiency of about $\eta = 1.96\cdot10^{-5}$.

The non-linear capacitance effects and the resulting ratcheting behavior were reproduced in simulation. The coupled Poisson-Nernst-Planck equations were solved for a one-dimensional system consisting of two electrolyte reservoirs separated by a membrane with a fixed pore charge. The two electrodes were modeled by defining the electric potential at the edges of the membrane. Figure S11a shows a schematic illustration of the simulation domain, and Figure S11b shows the spatial distribution of the electric potential and ion concentration in the system during a typical time-period. The input signal alternated between 0.8 V and 0V and its frequency was 5 kHz. The solution was 1 mM KCl. Figure 2d shows the simulated output voltage considering three different membrane pore charges. As demonstrated experimentally, the time constants for charging and discharging the surfaces vary significantly with the input signal duty cycle and potential (Figures S12-13). As a result, the net output voltage shows a ratcheting behavior: it approaches zero when a constant bias is applied (duty cycle of 0 or 1), but a significant time averaged voltage is obtained when an alternating input signal is introduced. The introduction of pore charge increases the output voltage and shifts the optimal duty cycle towards 1. Full details of the numerical model and the time constant extraction are given in the methods section and in the supporting information (section 6). Since the simulation describes the general physics of charging and discharging of the membrane surface, these results are not material specific and can be translated to material systems beyond those studied in this work.

Engineering devices for increased spatial asymmetry should result in increased asymmetric charging (and discharging) and thus enhanced performance. This can be achieved by varying the deposition process of the two metallic layers or using different materials. Asymmetric devices were fabricated by depositing gold on one surface of the AAO wafer and platinum on the other (a configuration that is termed Au-Pt). Figure 2e shows the extracted time constants for charging and discharging of the gold and platinum surfaces as a function of duty cycle. Although the input signal is the same as in Figure 1h, there is a larger difference between the charging (and discharging) time constants of the two electrodes. As a result, a dramatic increase in the output voltage (Figure 2f) and output current (Figure 2g) is observed. The measured signals and fitted curves are shown in Figure S9. The input voltage is not equally distributed between the two contacts, and the electric potential of the platinum surface alternates with an amplitude that is almost double that of the gold surface (Figures S9a,b). As discussed in the theoretical analysis (see section 13 of the supporting information), such an asymmetry can enhance device performance, yet it is not a prerequisite for device operation.

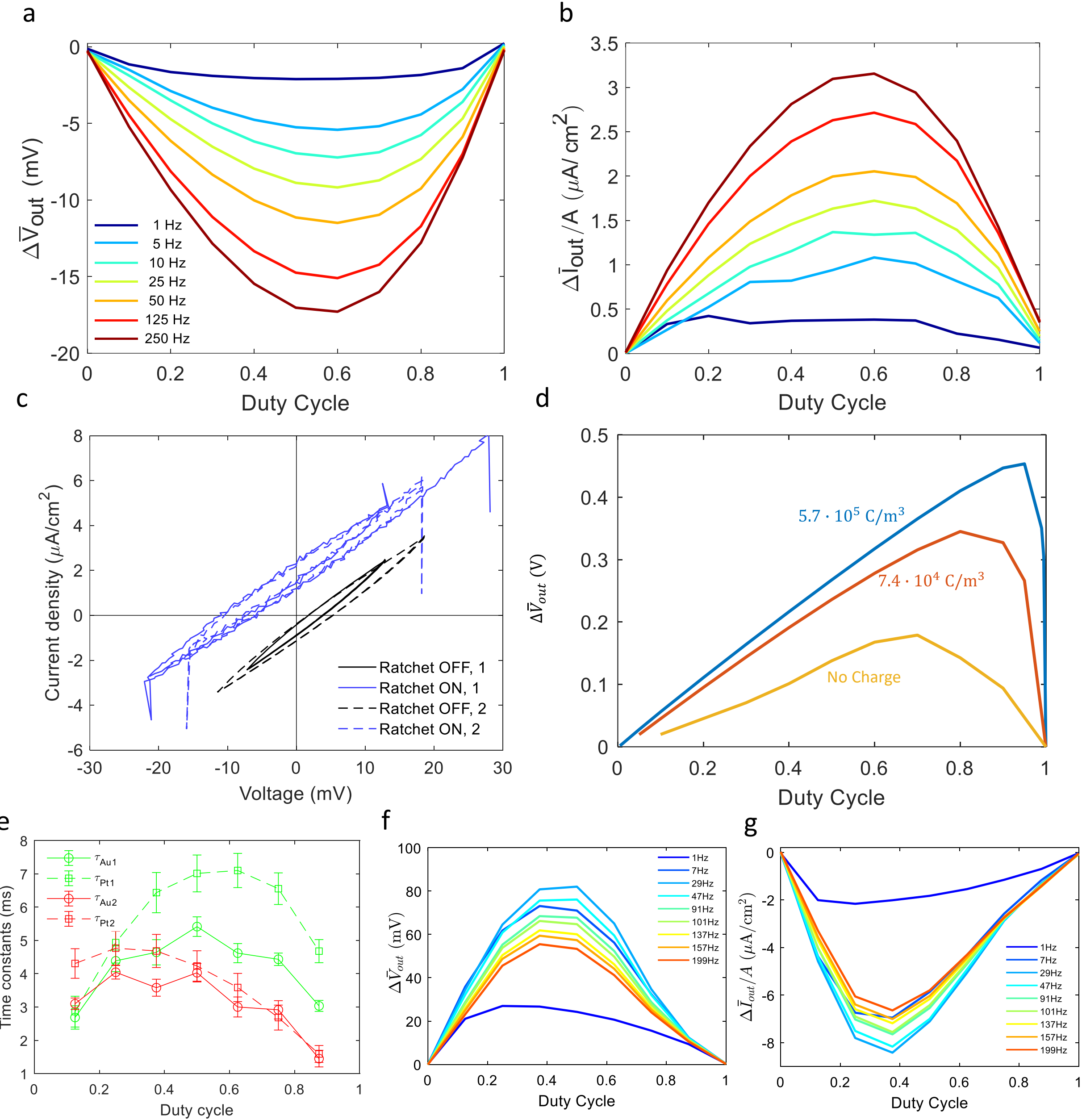

***Figure 2: RBIP performance examination****| Au-Au RBIP net output voltage (a), and net output current density (b) as a function of the input signal duty cycle and frequency. (c) Current–Voltage curves ($\bar{I}_{out}/A$ , $\bar{V}_{out}$ ) with Vin = 0 and with a square-wave input at a duty cycle of 0.5 and a frequency of 100 Hz. The input signal amplitude is $V_a$ = 0.3 V and the electrolyte is 1 mM HCl aqueous solution. (d) Numerical simulation: net output voltage as a function of the input signal duty cycle for different membrane pore charges. $V_{in}$ alternates between $0.8\ V$ to 0 V at a frequency of 5 kHz. (e) Au-Pt ratchet time constants as a function of the input signal duty cycle. Au-Pt ratchet net output voltage (f), and net output current density (g), as a function of the input signal duty cycle and several frequencies. The input signal amplitude is $V_a$ = 0.3 V and the electrolyte is 1 mM KCl aqueous solution.*

To demonstrate the RBIPs ion pumping functionality, we have demonstrated water deionization in a three-compartment electrodialysis system.[27] In this system, an RBIP is placed between two electrolyte reservoirs, with each reservoir connected to a central dilution cell by ion-selective membranes - a cation-exchange membrane (CEM) on one side and an anion-exchange membrane (AEM) on the other, thus forming a closed ionic circuit. Figure 3a shows a schematic illustration of the system. A photograph and a drawing of the setup are shown in Figures S20a-b. When an input signal is applied to the RBIP, it induces a non-zero time-averaged voltage across it, driving transport of predominantly cations through the CEM and predominantly anions through the AEM. Electroneutrality is maintained by the exchange of cations and anions through the RBIP itself. As a result, the initial 1 mM KCl concentration in all compartments decreases in the dilution cell, as indicated by a decrease in solution conductivity.

The conductance in the dilution cell was measured with two titanium wires, as described in the methods section and the supporting information (section 8), and the RBIP output voltage was monitored by measuring $V_{out}$ between two Ag/AgCl wires (Figure 3a). The system was placed inside an incubator, to reduce temperature related conductance changes, and temperature was also measured at the side wall of the dilution cell (externally). To further account for temperature effects, the conductance was also measured in a reference cell, which has the same volume and geometry as the dilution cell and was located next to it. The electrolyte in all compartments was 1mM KCl aqueous solution, and the ratchet was driven with a rectangular wave of $V_a = 1\ \mathrm{V}$, $f = 11\ \mathrm{Hz}$, $d_c = 0.5$, which maximizes its output (Figure S20c). Figure 3b shows the measured conductance in the dilution and reference cells normalized by their initial conductance. The absolute conductance and the measured temperature are shown in Figure S23. When the input signal is switched 'On', $V_{out}$ becomes negative, leading to a decrease in the solution conductivity (and the ion concentration) in the dilution cell. When the RBIP is switched 'Off', $V_{out}$ drops in magnitude, and the conductivity in the dilution cell begins to increase, indicating transport of cations and anions back into the dilution cell concomitant with transport of ions across the RBIP for charge neutrality. These 'On'/'Off' cycles were repeated, demonstrating a clear causal relationship between RBIP operation and conductivity changes in the dilution cell. At its lowest point, the conductivity in the dilution cell was reduced by 50% ($t = 37\ \mathrm{h}$), while the conductivity in the reference cell remained nearly constant. Notably, the concentration gradient between the dilution cell and the reservoirs resulted in a buildup of $V_{out}$ when the system was at rest (shaded areas in Figure 3b). A similar effect was used to probe the concentration difference in a second deionization experiment (see section 9 in the supporting information). The RBIP output started to diminish after approximately 30 hours, resulting in a lower deionization rate. After 44 hours a rapid decline in performance was observed. The system was then allowed to rest until $t = 94\ \mathrm{h}$ in which back diffusion almost brought the conductivity to its equilibrium value. The consistent decrease in the conductance in the dilution cell when an input signal was applied thus validates the RBIPs ion pumping functionality.

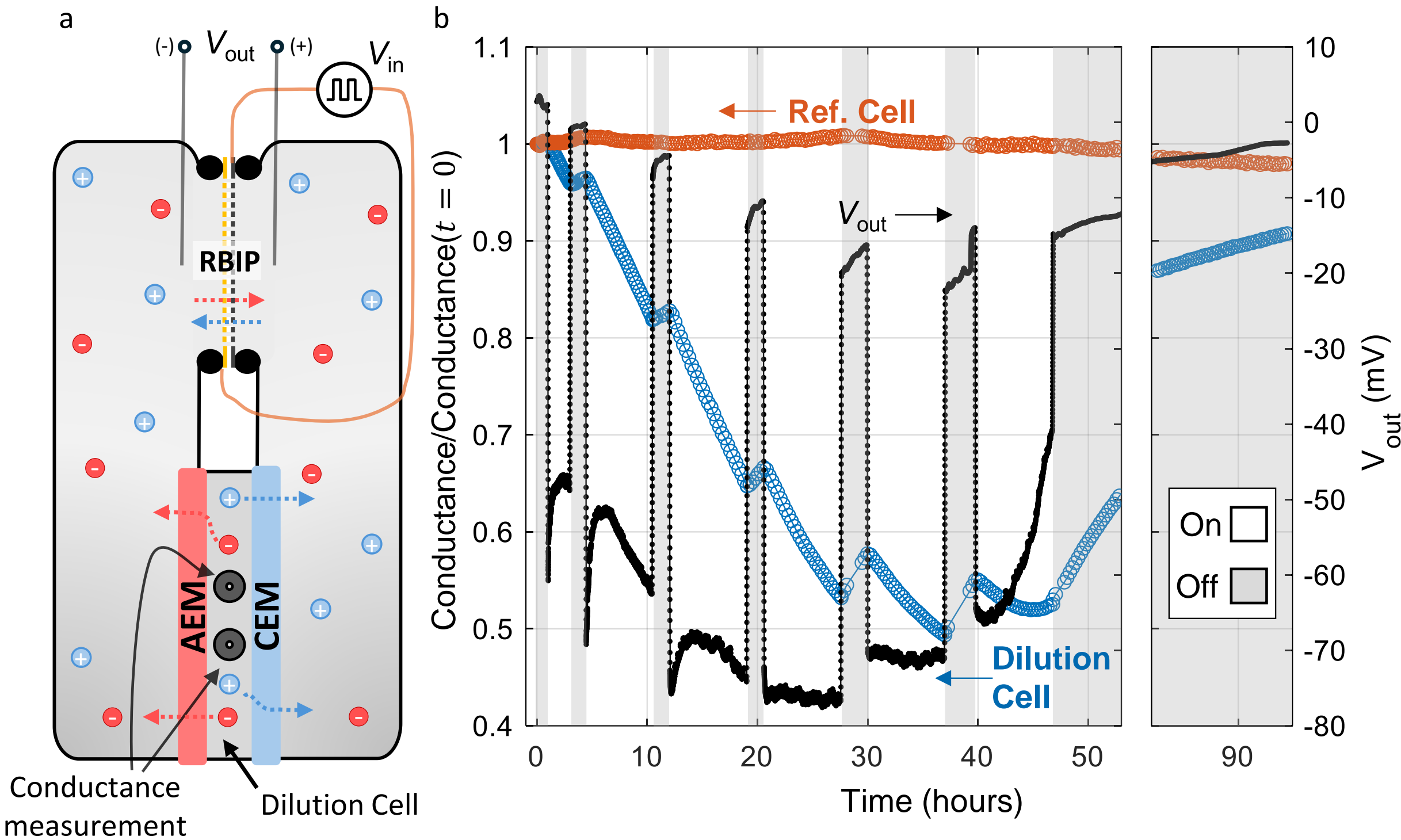

*Figure 3: **Electrodialysis (ED) deionization demonstration** |(a) Schematic illustration of the ED deionization system. $V_{out}$ is measured between two Ag/AgCl wires with the indicated polarity, and conductance is measured in the dilution cell using two titanium wires, as detailed in the methods section. (b) Normalized conductance (relative to conductance at $t = 0$) in the dilution cell and reference cell (left y-axis), and output voltage across the RBIP (right y-axis), as a function of time. Bright areas are time periods in which the RBIP is continuously operating, and in the shaded areas the input voltage is set to 0 V. The input signal during RBIP operation is a square wave with: $V_a = 1\ V,\ f = 11\ Hz,\ d_c = 0.5$, and the electrolyte was initially 1 mM KCl aqueous solution in all compartments. Ratchet induced ion pumping resulted in a 50% reduction in the conductivity of the solution in the dilution cell. The volume of the dilution and reference cells is $0.63\ cm^3$ and $0.66\ cm^3$, respectively.*

A simple analytical model predicts ratchet behavior similar to our experimental observations. The ratchet performance can be computed by assuming that the voltages $V_L$ and $V_R$ follow a charging and discharging behavior as in equation (3), and inserting them and the input signal into equation (1). The non-linear nature of the double layer impedance is approximated by assigning different time constants for the surfaces charging and discharging. Since these time constants are determined mostly by the duration of the charging and discharging period (Figure 1h, Figure S5a), it can be assumed that $\tau_{L,2} = \tau_{R,2}$ and $\tau_{L,1} = \tau_{R,1}$. $\Delta\bar{V}_{out}$ was calculated using this procedure for various input signals and time constants. More details on the analytical model can be found in the supporting information (section 13). Figure 4a shows the normalized output, $\Delta\bar{V}_{out}/V_a$, as a function of the duty cycle for several input signal frequencies. The charging and discharging time constants are $\tau_{L,2} = \tau_{R,2} = 8$ ms, $\tau_{L,1} = \tau_{R,1} = 2$ ms which are close to those extracted from experiment (Figure 1h). The output curves show a ratchet behavior with outputs near zero for extreme duty cycles and low frequencies, in which the temporal periods are significantly longer than the charging and discharging time constants. Like the experimental results, the output reaches largest magnitudes at moderate duty cycles and higher frequencies. Figure 4b shows similar curves for $\tau_{L,2} = \tau_{R,2} = 100$ ms and $\tau_{L,1} = \tau_{R,1} = 1$ ms. For such an extreme difference in time constants, the output is higher and the optimal duty cycle shifts toward 0. In both cases, the output saturates when increasing the frequency. Figure 4c shows the maximal normalized output (in terms of absolute values) as a function of $\tau_{R,1}$ and $\tau_{R,2}$, with $\tau_{L,1} = \tau_{R,1}$ and $\tau_{L,2} = \tau_{R,2}$, and Figure 4d shows corresponding duty cycles. The parameters used in Figure 4a,b are marked with a triangle and a diamond respectively. When the RBIP capacitance is linear, i.e. all the time constants are equal, the output is zero. However, the output increases with the difference in time constants reaching magnitudes that approach $2V_a$ at optimal duty cycles that shift toward 0 or 1. Hence, for the highest performance, the time constants for charging and discharging must be as far apart as possible. It should be noted that devices with a geometrical asymmetry but with a linear capacitance, i.e. $\tau_{R,1} = \tau_{R,2}$ and $\tau_{L,1} = \tau_{L,2}$, produce a zero output. Thus, optimizing electrodes for high frequency dispersion is essential for efficient RBIP operation. In devices where $\tau_{L,1} \neq \tau_{R,1}$ and $\tau_{L,2} \neq \tau_{R,2}$, the maximal obtainable output is determined by the surface with the larger difference in time constants, and the effective amplitude acting on each surface (the amplitude asymmetry coefficient, $a$, as discussed in section 13 in the supporting information). The larger the amplitude acting on one of the surfaces, the more dominant are the time constants of that surface in determining the total output (Figure S35). In the Au-Pt device (Figure 2e-g), the time constants of the Pt surface are further apart (Figure 2e) and the amplitude of the voltage that the Pt surface fluctuates is larger (Figure S9f) thus leading to a higher device performance.

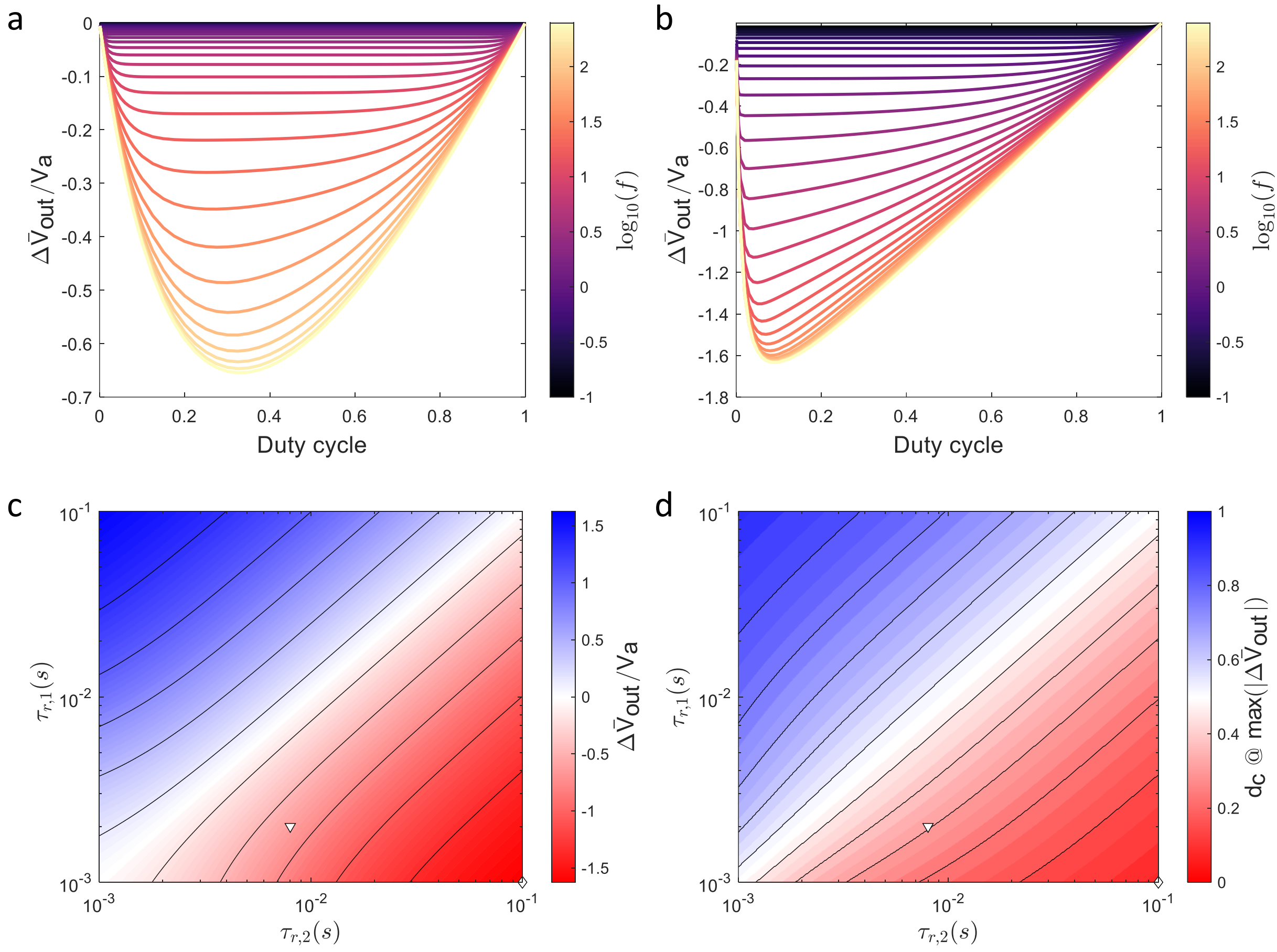

***Figure 4:*** *Analytical model | (a) The ratchet normalized output, $\Delta\bar{V}_{out}/V_a$, as a function of the input signal duty cycle for several input signal frequencies, $\tau_{L,2} = \tau_{R,2} = 8$ ms, $\tau_{L,1} = \tau_{R,1} = 2$ ms, and $V_a = 0.3\ V$. (b) $\Delta\bar{V}_{out}/V_a$ as a function of the input signal duty cycle for several input signal frequencies, $\tau_{L,2} = \tau_{R,2} = 100$ ms, $\tau_{L,1} = \tau_{R,1} = 1$ ms, and $V_a = 0.3\ V$. (c) The RBIP normalized output, $\Delta\bar{V}_{out}/V_a$, as a function of $\tau_{R,1}$ and $\tau_{R,2}$. For each combination of time constants the duty cycle is optimized to obtain*

*the highest output. The input signal frequency is 100 Hz. (d) The duty cycles for which the magnitude of $\bar{V}_{out}$ is maximal. The triangle and the diamond mark the parameters used in a and b respectively.*

The analytical model demonstrates that a non-linear capacitance, as observed in electric double layers, can result in a ratchet-like behavior. However, since the time constants in real devices vary significantly with the input signal properties, this model cannot predict the experimental performance of real devices and can only define optimal time constants for specific input signals. A detailed discussion of the model limitations can be found in the supporting information (section 13).

Scaling up RBIPs requires stable materials that can be processed on a large scale. Since the RBIP driving mechanism is not governed by the chemistry of specific materials, it can be expected that other membrane and electrode materials will also exhibit an RBIP functionality. Engineering surfaces for highly nonlinear capacitance while maintaining spatial asymmetry might be achievable without nano-scale precision thus allowing RBIP large area fabrication. Building on top of the suggested structures, more complex devices can be fabricated which support enhanced functionality. For example, layered structures can result in an ion pump driven with a flashing ratchet mechanism.[28] The energetic efficiency of flashing ratchet-based dialysis was compared to reverse osmosis and was shown to be higher under a wide range of conditions.[29] In addition, we showed that RBIPs can theoretically drive ions of the same charge, but different diffusion coefficients, in opposite directions as well as to transport both cations and anions in the same direction.[20,21] Such devices can pave the way towards ion selective pumps and single membrane dialysis systems. Further increasing the number of metal layers connected to independent voltage sources may allow tailoring the potential distribution within the pores which can further increase the device functionality and efficiency.[6]

In conclusion, we have realized a first-of-its-kind ion pump driven by a capacitive ratchet mechanism. This type of ion pump can drive a net ion flux in steady state with no redox reactions at the RBIP contacts. Ion pumping against an opposing force was demonstrated as well as electrolyte deionization. The driving mechanism stems from the non-linear capacitance of electrode double layers, which leads to deviation in time constants between charging and discharging. The demonstrated ion pump can pave the way for the development of high efficiency desalination and selective ion separation systems, requiring neither moving parts nor redox reactions.

## Methods

### Sample preparation

RBIP samples were fabricated by depositing metal contacts on both surfaces of annealed anodized aluminum oxide (AAO) wafers (InRedox Materials Innovation) with thickness 50 µm, and pore diameters of 20 nm, 40 nm, or 60 nm. The wafers were air annealed for ~10 h at a temperature of 650 °C. Several RBIP structures and recipes were tested all showing the RBIP functionality. The samples discussed in Figure 2a-b were made of AAO wafers with a pore diameter of 40 nm and their contacts were deposited with thermal evaporation of a chrome adhesion layer and then gold. The thickness of both layers was 40 nm (planar equivalent). The samples discussed in Figure 2e-g and Figure 3 had a pore diameter of 60 nm. The platinum contact was deposited with an electron beam evaporator and the gold contact was deposited with magnetron sputtering. The thickness of both contacts was 50 nm (planar equivalent). The sample discussed in Figure 1 had a pore diameter of 60 nm and the metal contacts were deposited with magnetron sputtering of gold on both sides of the wafer (50 nm planar equivalent) followed by an 8 nm thick $TiO_2$ coating deposited by ALD. The ALD recipe is as described by Vega et al.[30] where the exposure time to the precursors was set to 1 s and the purging time was set to 5 s. The edges of the sample were masked during the ALD process to allow contacting the sample. The metal contacts in the samples described in Figures S14, S19 were deposited with electron beam evaporation of a 10 nm thick titanium adhesion layer and then 40 nm of gold. The RBIP ion pumping properties were tested in an electrochemical cell in which the RBIP served as a membrane separating two aqueous electrolyte compartments, each containing an Ag/AgCl wire that was used to probe the voltage or current between the two compartments. In the experiments shown in Figures 2a-b, and in Figures S6-8,S15,S24-27,S30 the electrochemical cell was as shown in Figure 1e, where parafilm was used to secure the wafers and prevent leakage (Figure S1), and Carbon black cloth to contact the RBIP electrodes (no contact with the electrolyte). The RBIP active area was 0.32 $cm^2$. The electrodialysis cell discussed in Figure 3 was made of Formlabs Clear V4 resin. In all other experiments the electrochemical cell was constructed from PEEK, with Silicone rubber used to secure the wafers and prevent leakage, and ZOFLEX® CD45.1 conductive foam and copper tape to contact the RBIP electrodes (no contact with electrolyte). Here the RBIP active area was 0.16 $cm^2$.

## RBIP performance characterization

To ensure that all the observed performance is a result of a ratcheting mechanism, as opposed to artifacts induced by unwanted electronic feedbacks, all measurement instruments and power sources were electrically floating (i.e., no part of the system was grounded), and all input signals and measurements were performed in two electrode setups. In Figures 2a-b the input signal was produced with a BioLogic VSP-300 multichannel potentiostat. The voltage and current measurements were conducted with a separate channel from the same potentiostat. Voltage measurements were taken with a sampling time of 200 µs and current measurements were taken with a sampling time of 400 µs. All the potentiostat channels were in floating mode. The RBIP performance discussed in Figure 2a-b was characterized using the following procedures. First, $V_{in}$ was set to 0 V for 30 s. Next, the ratchet signal was introduced for 90 s. Finally, $V_{in}$ was set to 0 V for another 60 s. The ratchet induced voltage $\Delta\bar{V}_{out}$ was calculated according to $\bar{V}_{out} = \bar{V}_{out,ON} - (\bar{V}_{out,OFF_1} + \bar{V}_{out,OFF_2})/2$ and the ratchet induced current $\Delta\bar{I}_{out}$ was calculated according to $\bar{I}_{out} = \bar{I}_{out,ON} - (\bar{I}_{out,OFF_1} + \bar{I}_{out,OFF_2})/2$. $\bar{V}_{out,ON}$ and $\bar{I}_{out,ON}$ are respectively the voltage and current averaged over the last 30 s of the period in which the input signal was applied. $\bar{V}_{out,OFF_1}$ and $\bar{I}_{out,OFF_1}$ are respectively the voltage and current averaged over the last 15 s in the first (30 s long) period in which $V_{in}$ = 0 V. $\bar{V}_{out,OFF_2}$ and $\bar{I}_{out,OFF_2}$ are respectively the voltage and current averaged over the last 30 s of the second OFF period.

In Figure 1, Figure 2e-g, Figure 3b, and Figure S20c the input signal was supplied with a Keysight 33510B waveform generator and voltages were measured with Keysight 34465A multimeters. In Figure 1 and Figures 2f-g the response to every input signal was measured for 60 s after which the input was set to 0 V for 60 s. In Figures 1f-g the voltage $V_L$ was measured between the left gold contact and the adjacent Ag/AgCl wire, $V_{out}$ was measured between the two Ag/AgCl wires, and $V_R$ was calculated using equation (1). The mean temporal signals, $V_L(t)$ and $V_R(t)$, were found by obtaining the mean responses out of 300 periods. The sampling time was $5 \cdot 10^{-4}$ s. Figure S3 shows an example of the measured signal and the $\Delta\bar{V}_{out}$ calculation shown in Figure 1i. In Figures 2f-g the voltage and current measurements were conducted with a sampling time of $6 \cdot 10^{-4}$ s. The net output voltage and current were calculated according to the method described above for Figures 2a-b, where the averaged voltage and current were obtained by averaging measurements between 0.85 and 0.95 of each ON/OFF period. In Figure 3b the voltage measurement was conducted with an integration time of 2 s to reduce the output signal oscillations and obtain only the net averaged voltage. The sampling time was 4 s. In Figure S20c data was recorded with an integration time of $1 \cdot 10^{-4}$ s and sampling time of $2 \cdot 10^{-4}$ s.

In some samples, input of a constant bias higher than ±300 mV resulted in a constant, non-negligible $V_{out}$. This may suggest Faradaic reactivity, which was followed by degradation in the RBIP performance for some samples (section 11 in the supporting information). In other samples, non-zero $V_{out}$ under a constant bias was attributed to blocked pores (section 12 in the supporting information).

## Numerical Simulation

The electric potential and ion concentrations in response to a rectangular wave input signal was computed by solving the one-dimensional Poisson-Nernst-Planck (PNP) equations using COMSOL Multiphysics® v6.1. The model geometry is illustrated in Figure S11a. The simulation domain consists of two electrolyte reservoirs (thickness $W = 4.5\ \mu\text{m}$) which are separated by the RBIP membrane (thickness $L = 1\ \mu\text{m}$). The RBIP electrodes are modeled as point contacts (0D) and the boundary conditions at the edges of the simulation domain were of zero flux thus modeling closed reservoirs. Simulations were performed assuming a symmetric binary 1:1 electrolyte with equal diffusion coefficients ($D = 2 \cdot 10^{-9}\ \text{m}^2/\text{s}$) for both ion species, no faradaic reactions at the electrodes and no solvent convection. A full description of the governing equations, boundary conditions, input signal, meshing, simulation sequence and membrane surface charge calculation, is given in the supporting information (section 6).

## Conductance measurements within the electrodialysis experiment

Real-time solution conductance measurements were carried out by embedding two titanium electrodes into the dilution cell in the electrodialysis setup. Similar to two-electrode conductivity meters, the conductance was obtained by applying a sine wave voltage signal between the two electrodes and measuring the resulting sinusoidal current flowing through the electrolyte. A detailed description of the test methodology and measurement setup is given in the supporting information (section 8).

# Acknowledgments

RK acknowledges the U.S. National Science Foundation Graduate Research Fellowship Program (DGE-1321846). AH acknowledges the support of the Boris Mints Institute. EJH acknowledges a Graduate Assistance in Areas of National Need (GAANN) Fellowship. CM acknowledges the gracious support for summer research at UC Irvine from Research Corporation for Science Advancement through a Cottrell Scholars Collaborative (Award #27512) awarded to SA and the UCI Vice Provost for Teaching and Learning. JWA was supported by the Joint Center for Artificial Photosynthesis, a DOE Energy Innovation Hub, supported through the Office of Science of the U.S. Department of Energy under Award No. DE-SC0004993. SA acknowledges the support of the Gordon and Betty Moore Foundation under a Moore

Inventor Fellowship (GBMF grant #5641) and The Beall Family Foundation (UCI Beall Innovation Award). SA also acknowledges the Phase I Centers for Chemical Innovation (CCI) Program in the U.S. National Science Foundation Division of Chemistry under Grant CHE-2221599 for supporting collaborative student exchanges with the Sa Group at the University of Massachusetts Boston, as well as revisions to the initial manuscript submission.GS thanks the Azrieli Foundation for financial support within the Azrieli Fellows program. This work is partially funded by the European Union (ERC, ESIP-RM, 101039804). Views and opinions expressed are however those of the author(s) only and do not necessarily reflect those of the European Union or the European Research Council Executive Agency. Neither the European Union nor the granting authority can be held responsible for them. We would also like to acknowledge Prof. Reg Penner for allowing us to work with his group members and use their thermal evaporator for the fabrication of some of the RBIPs. We acknowledge the contribution of TAU Nano center for providing the sputtering and e-beam evaporation equipment, and the HRSEM. We thank G. Radovsky for taking the HRSEM images. We thank O. Nir and A. Chandra for guidance with the Electrodialysis setup.

## Conflict of Interest

GS, JWA, FMT, and SA filed patent applications US 16/907,076 and US 17/125,341 for ratchet-based ion pumping membrane systems. There are no other conflicts of interest to declare.

## Author contributions

GS, JWA and SA conceptualized this work, RK, AH, EJH, KSA, EG, RS, CM, DL and GS conducted the investigation, AH, SA and GS designed the methodology, RK, AH, EJH, KSA, RS, CM and GS conducted the formal analysis and acquired the data, visualizations were designed by EJH, AH, RK, SA and GS, the original draft was written by SA and GS, and all authors contributed to its reviewing and editing. GS and SA supervised the project, GS, SA and FMT administered it, and acquired the funding for it.

**Supporting information for A nanoporous capacitive electrochemical ratchet for continuous ion separations**

Rylan Kautz[1†], Alon Herman[2†], Ethan J. Heffernan[1], Keren Shushan Alshochat[2], Eden Grossman[2], Rahul Saxena[2], Camila Muñetón[3,4], David Larson[5,6], Joel W. Ager III[5,6,7,8], Francesca M. Toma[5,6,9,10], Shane Ardo[1,3,11]* and Gideon Segev[2,5,6]**

[1]Department of Materials Science & Engineering, University of California Irvine, Irvine, CA 92697, USA

[2]School of Electrical Engineering, Tel Aviv University, Tel Aviv 6997801, Israel

[3]Department of Chemistry, University of California Irvine, Irvine, CA 92697, USA

[4]Department of Chemistry, University of Massachusetts Boston, Boston, MA 02125 USA

[5]Chemical Sciences Division, Lawrence Berkeley National Lab, Berkeley, CA 94720, USA

[6]Joint Center for Artificial Photosynthesis, Lawrence Berkeley National Lab, Berkeley, CA 94720, USA

[7]Materials Sciences Division, Lawrence Berkeley National Lab, Berkeley, CA 94720, USA

[8]Department of Materials Science and Engineering, University of California Berkeley, Berkeley, CA 94720, USA

[9]Institute of Functional Materials for Sustainability, Helmholtz Zentrum Hereon, Teltow, 14513, Germany

[10]Faculty of Mechanical and Civil Engineering, Helmut Schmidt University, Hamburg 22043, Germany

[11]Department of Chemical & Biomolecular Engineering, University of California Irvine, Irvine, CA 92697, USA

*Email: ardo@uci.edu

**Email: gideons1@tauex.tau.ac.il

† These authors contributed equally to this work

## 1. Sample assembly

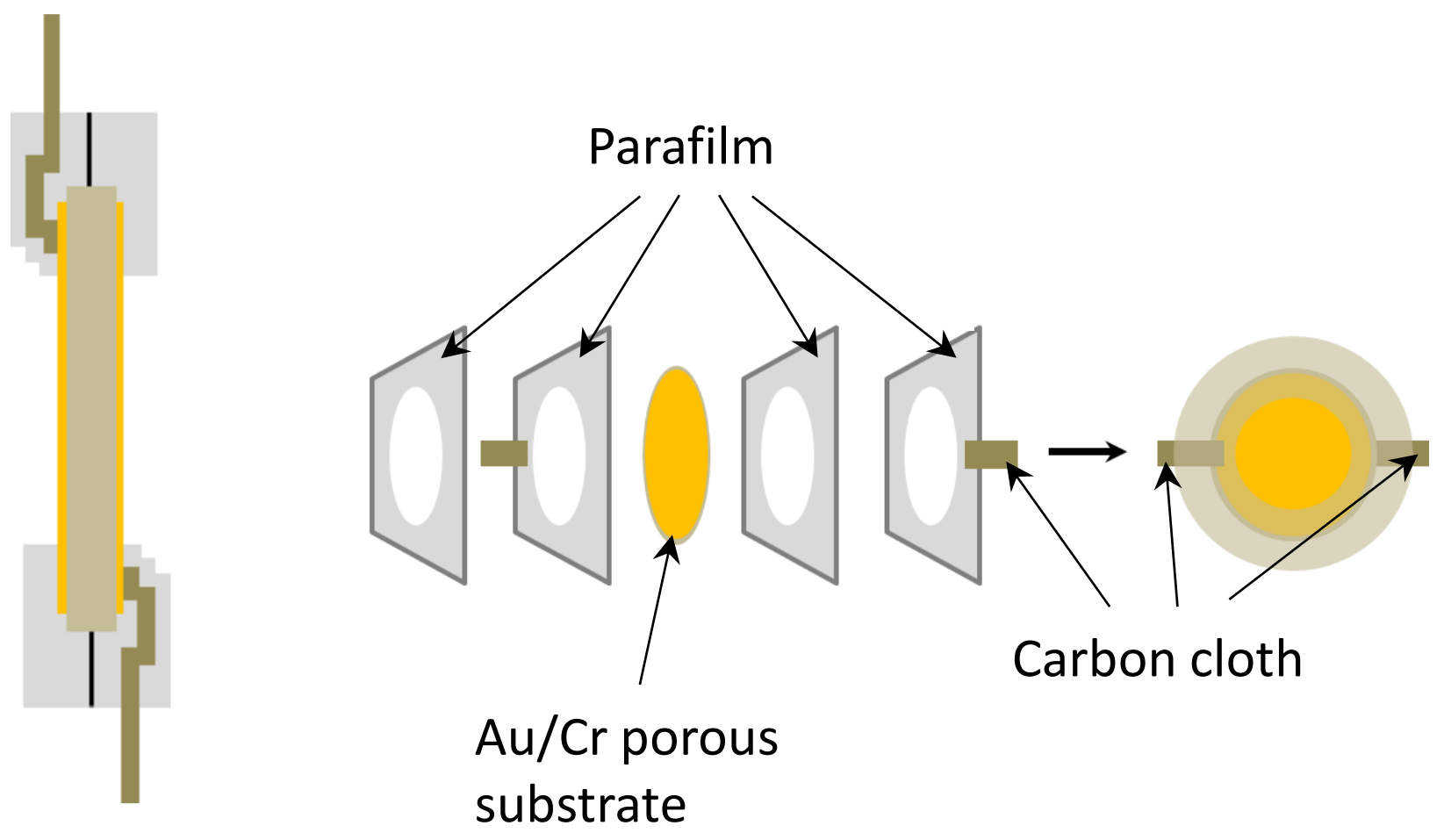

*Figure S1: Schematic illustration of an RBIP sample assembly in test setup.*

## 2. Duty cycle sweep additional data

Figure S2a-c show respectively the fitted effective amplitude, $V_i - V_f$, the final potential, $V_f$ , and the fit $R^2$ as a function of the input signal duty cycle obtained from the data presented in Figures 1f-g. Figure S2d shows the measured signals as well as the signals obtained by inserting the fitted values into equations 3 and 1. The charging/discharging response to a large voltage step is non-linear with the potential as it includes the fast relaxation of the electrochemical double layer, and the slower relaxation of other ionic processes.[1] As a result, the time constants describing the charging and discharging of the surfaces increase with the duration of the voltage step (Figure 1h). Furthermore, the fitted curves in Figure 1f-g do not account for the fast processes and consistently underestimate the temporal change near $t = 0$ and $t = d_c T$. Figure S2e compares the net output voltage, $\Delta\bar{V}_{out}$, obtained from the measurements (same as Figure 1i) and by integrating the fitted data from Figure S2d. The good agreement between the measured and the output obtained with the fitted data implies that the performance of the RBIP can be described with four time constants (charging and discharging each of the surfaces), and that the contribution to the net output of the faster processes at $t = 0$ and $t = d_c T$ is small.

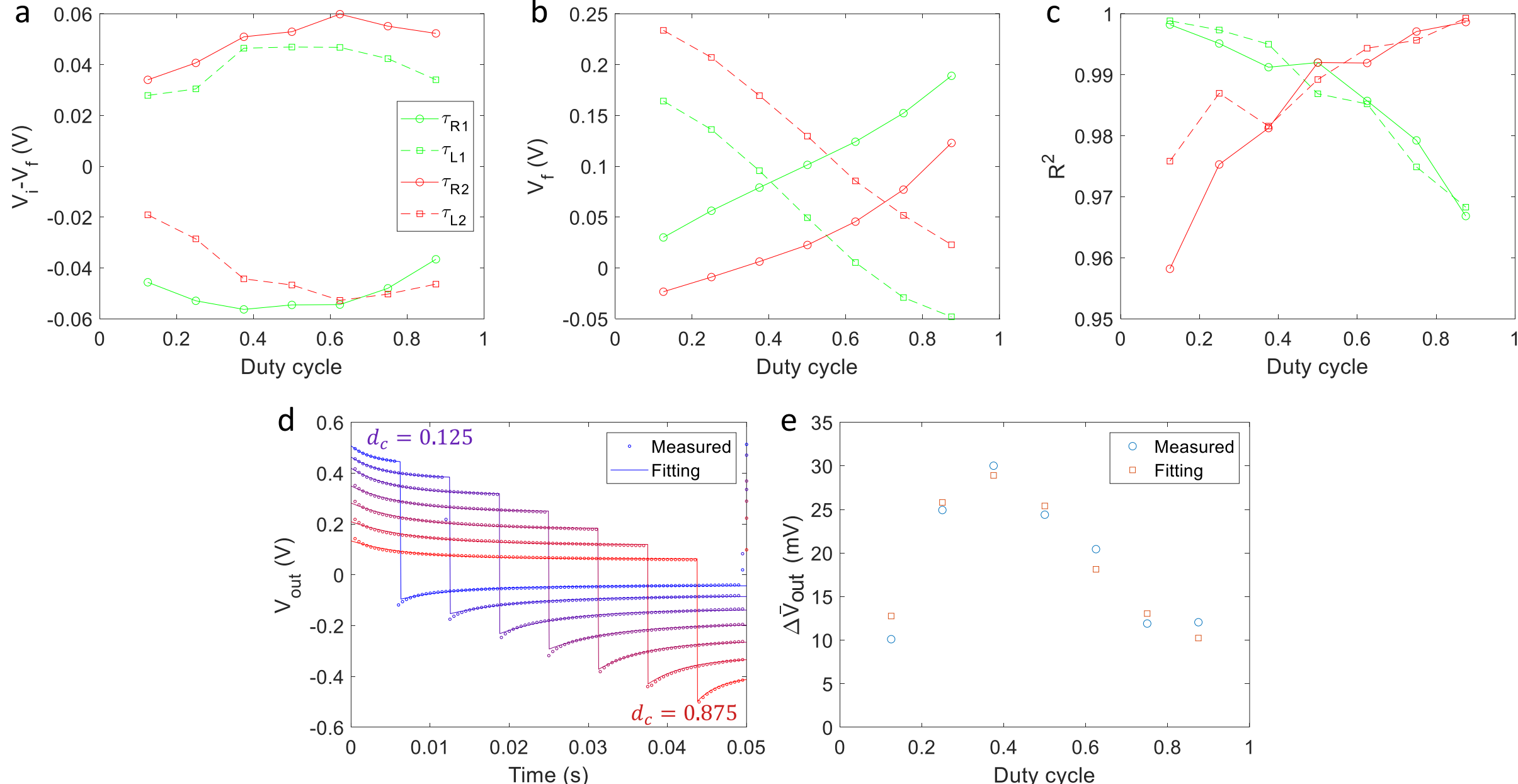

*Figure S2: the parameters obtained from the fitting process. (a) The fitted effective amplitude of the input signal $V_i - V_f$, (b) the final potential, $V_f$, and (c) the fit $R^2$ as a function of the input signal duty cycle. The color coding is as in (a). (d) the measured averaged $V_{out}$ and $V_{out}$ obtained from the fitted parameters and equations 1-3. (e) the measured ratchet voltage output and the output calculated with the fitted parameters.*

Figure S3a shows the raw data measured during the experiment described in figure 1 (the periodic signals cannot be resolved from Figure S3a because of the long duration of the experiment. This data is presented to clarify how the analysis is carried out). The grey shaded areas mark the times when the ratchet was OFF, and the white regions mark the times when the ratchet was ON. The input signal duty cycle is indicated near the top of every white region. The black curve is the measured voltage. $V_{out,ON}$ (light blue) is the voltage measured when the ratchet was ON and that was later used to calculate the time averaged voltage $\bar{V}_{out,ON}$ as described in the methods section. The time averaged voltage, $\bar{V}_{out,ON}$, for every duty cycle is indicated with markers. $\bar{V}_{out,OFF}$ (not shown in Figure S3a) is the temporal average of $V_{out}$ when the ratchet was OFF and was obtained by averaging the red areas in Figure S3a. Figure S3b shows the same measurement as in Figure S3a focusing on a short timespan with duty cycle of 0.5, allowing the recorded signal to be observed. Figure S3c shows the time averaged voltages $\bar{V}_{out,ON}$ and $\bar{V}_{out,OFF}$ and how the ratchet voltage output $\Delta\bar{V}_{out}$ is calculated: $\Delta\bar{V}_{out} = \bar{V}_{out,ON} - (\bar{V}_{out,OFF,1} + \bar{V}_{out,OFF,2})/2$.

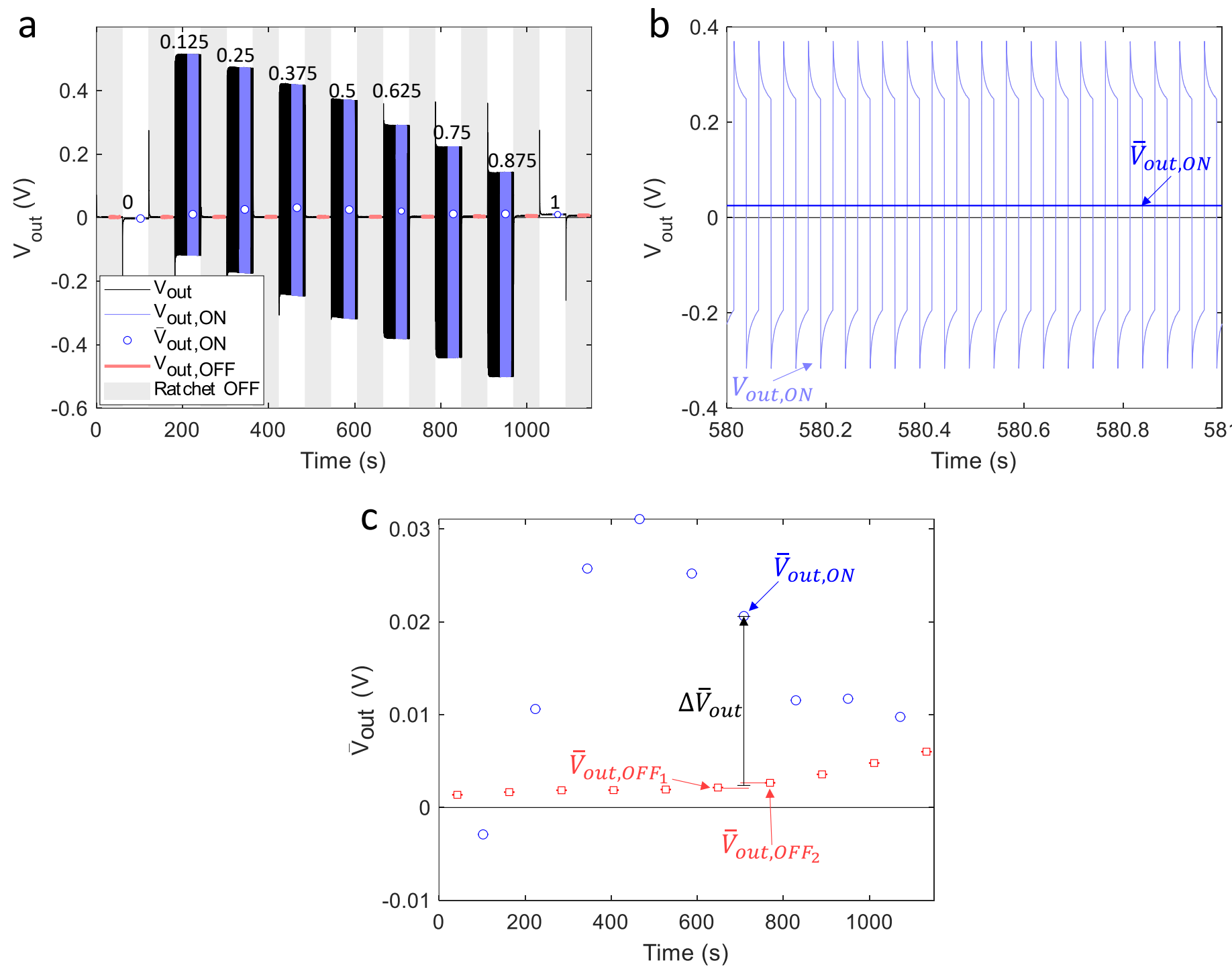

*Figure S3: $\Delta\bar{V}_{out}$ calculation, (a) The raw data recorded in the experiment described in Figure 1, and the ratchet output calculation. The grey shaded areas mark the time when the ratchet was OFF, and the white regions mark the time when the ratchet was ON. The black curve is the measured voltage. The numbers indicate the duty cycle of the input signal. (b) An example of $V_{out,ON}$ and $\bar{V}_{out,ON}$ within a short timespan with duty cycle of 0.5, which allows observing the recorded signal. (c) The time averaged voltages $\bar{V}_{out,ON}$ and $\bar{V}_{out,OFF}$ and the ratchet voltage output $\Delta\bar{V}_{out}$ calculation: $\Delta\bar{V}_{out} = \bar{V}_{out,ON} - (\bar{V}_{out,OFF,1} + \bar{V}_{out,OFF,2})/2$.*

## 3. Time constants extraction- frequency sweep

The sample analyzed in Figures 1f,g was also measured at various input signal frequencies. The duty cycle is 0.5, and the amplitude is $V_a$ = 0.7 V. The electrolyte is 0.2 mM NaCl aqueous solution. All other parameters are as described in Figure 1. Figure S4a-c show respectively the average temporal responses of $V_L$, $V_R$, and $V_{out}$. For clarity, the time for every signal was normalized by its period. The input signal frequency is indicated by the color bar. For the lowest frequency (0.1 Hz) three temporal periods were averaged to obtain the average response, for a frequency of 0.2 Hz, 8 periods were averaged, and for higher frequencies at least 15 periods were averaged. The solid lines show the best exponential fit to the measured data. Figure S5a shows the extracted time constants of the measured signals for the two parts of the input signal ($0<t<d_cT$, and $d_cT<t<T$). The time constants for all signals drop significantly with the input signal frequency demonstrating the frequency dispersion. Differences between the time constants for the two parts of the input signals are also observed. Figure S5b-c shows respectively the extracted values of $V_i$-$V_f$ and $V_f$ for each of the signals. Because the effective time constants increase as the frequency decreases, $V_f$ reaches the applied voltage only for very low frequencies. As a result, the effective amplitude of the potential modulation at the ratchet contacts, $V_i - V_f$, also decreases with frequency. This implies that at high frequencies the voltage $V_L$ and $V_R$ alternate according to an effective input signal with an amplitude that is smaller than $V_a$, potentially reducing the output of the device.

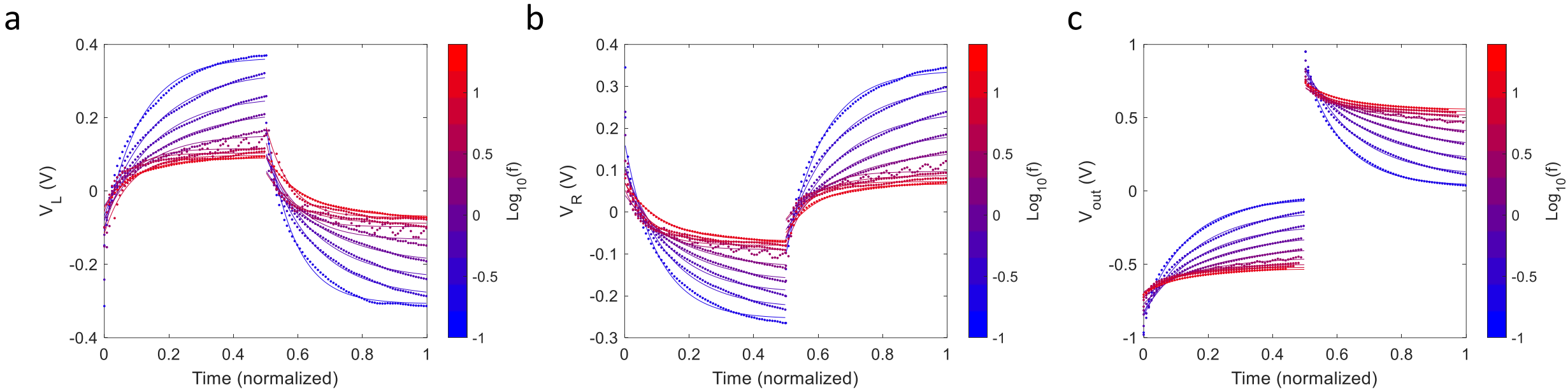

*Figure S4: (a-c) The averaged responses of $V_L(t)$, $V_R(t)$ and $V_{out}(t)$ respectively for various input signal frequencies. The sample is as in Figure 1. The duty cycle is 0.5, and the amplitude is $V_a$ = 0.7 V. The electrolyte is 0.2 mM NaCl aqueous solution. For each signal, the time was normalized by the input signal temporal period. The color bar indicates the input signal frequency. The symbols mark the measured signals, and the solid lines are the exponential fit. The plotted measured signals were under-sampled for visualization purposes.*

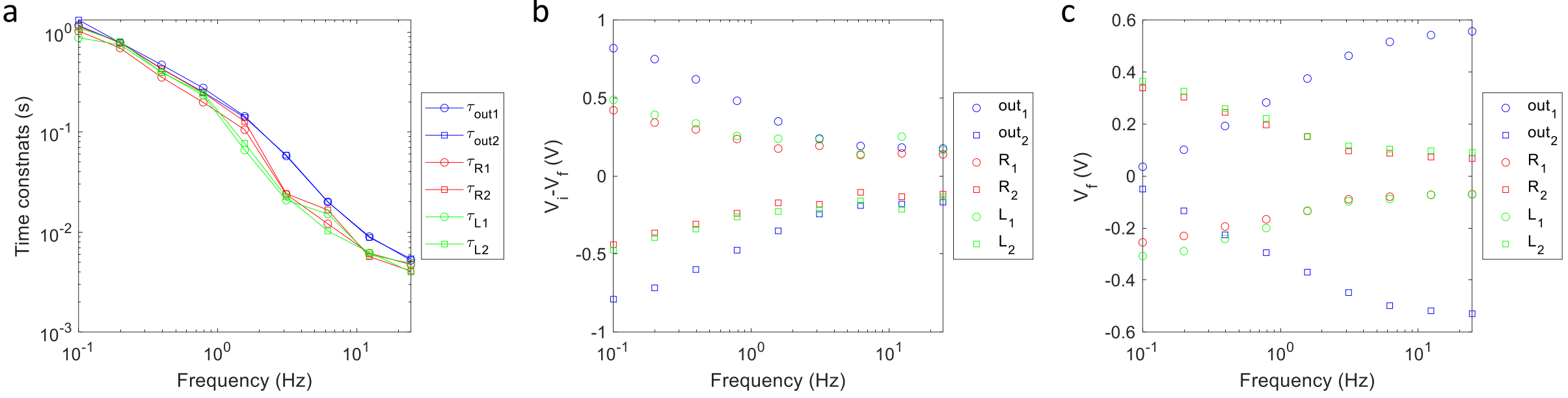

*Figure S5: (a) Extracted time constants, (b) the fitted effective amplitude $V_i - V_f$, and (c) $V_f$ for the signals shown in Figure S4.*

## 4. Time constants extraction for the Au-Au samples - additional data

Figure S6a-e and Figure S7a-e shows the measured average response of $V_{out}(t)$ and $I_{out}/A(t)$, respectively, for various duty cycles and frequencies of 1, 5, 10, 25, 50 Hz in (a-e), respectively. Each output is the averaged response of at least 30 temporal periods. The sample is as in Figure 2a-b, the electrolyte is 1 mM HCl aqueous solution, the RBIP pore diameter is 40 nm, the active area of the membrane is A = 0.32 cm$^2$, and the input signal amplitude, $V_a$, is 0.3 V. For frequencies of 125 and 250 Hz there were not enough measured data points per period to carry out the fitting process properly.

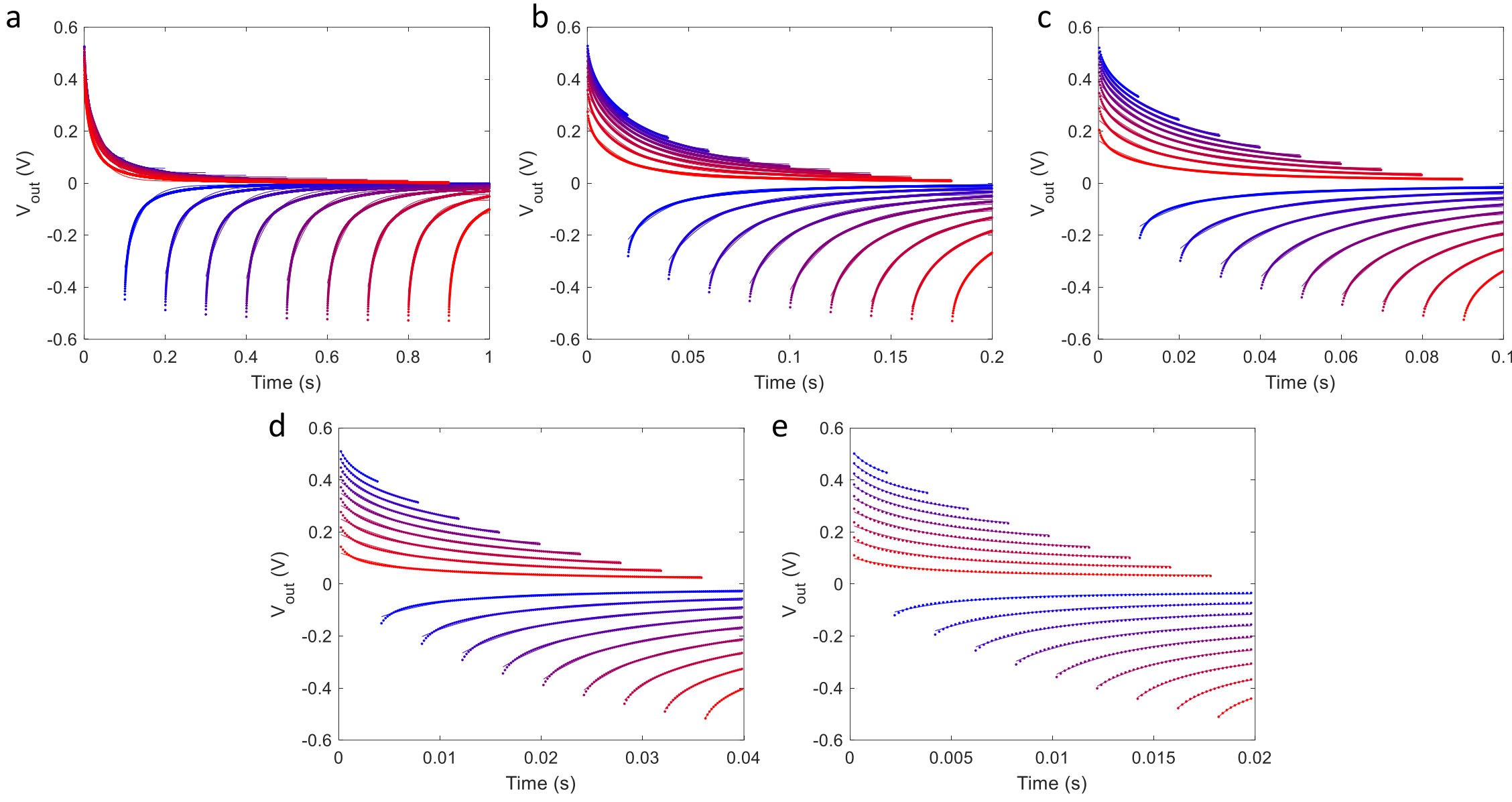

*Figure S6: Measured $V_{out}(t)$ for various duty cycles (dots) and the best fit to single exponential functions. The color coding matches the input signal duty cycle. The input signal frequencies are 1, 5, 10, 25, 50 Hz in (a-e) respectively. Each output is the averaged response of at least 30 temporal periods. The electrolyte is 1 mM HCl aqueous solution, the RBIP pore diameter is 40 nm, the active area of the membrane is A = 0.32 cm$^2$, and the input signal amplitude, $V_a$, is 0.3 V.*

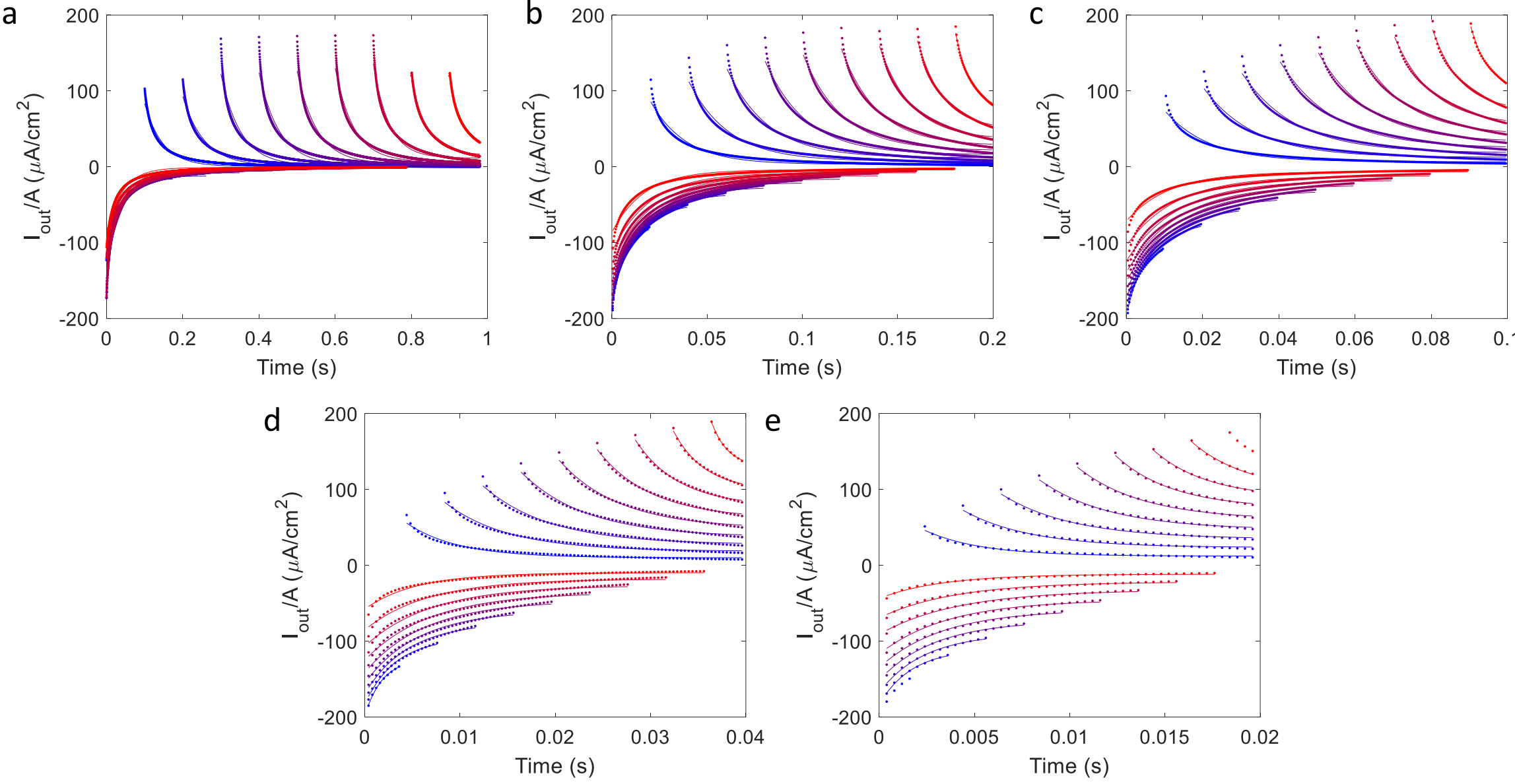

*Figure S7: Measured $I_{out}(t)$ for various duty cycles (dots) and the best fit to single exponential functions. The color coding matches the input signal duty cycle. The input signal frequencies are 1, 5, 10, 25, 50 Hz in (a-e) respectively. Each output is the averaged response of at least 30 temporal periods.*

Figure S8a,c show the mean temporal signal response of $V_{out}$ and $I_{out}/A$, respectively. The RBIP and input signal parameters are as in Figures 2a-b. The measured current is normalized by the RBIP active area ($A$ = 0.32 cm$^2$), and the time is normalized by the temporal period of every signal. Figure S8b,d shows the time

constants obtained from $V_{out}$ (Figure S6) and $I_{out}/A$ (Figure S7), respectively. The time constants vary significantly with frequency and duty cycle, and the time constants of the first part of every period are close to those of the second part of the period flipped with respect to a duty cycle of 0.5. The time constants for frequencies of 125 and 250 Hz were not extracted because there were not enough measured data points per period to carry out the fitting process properly.

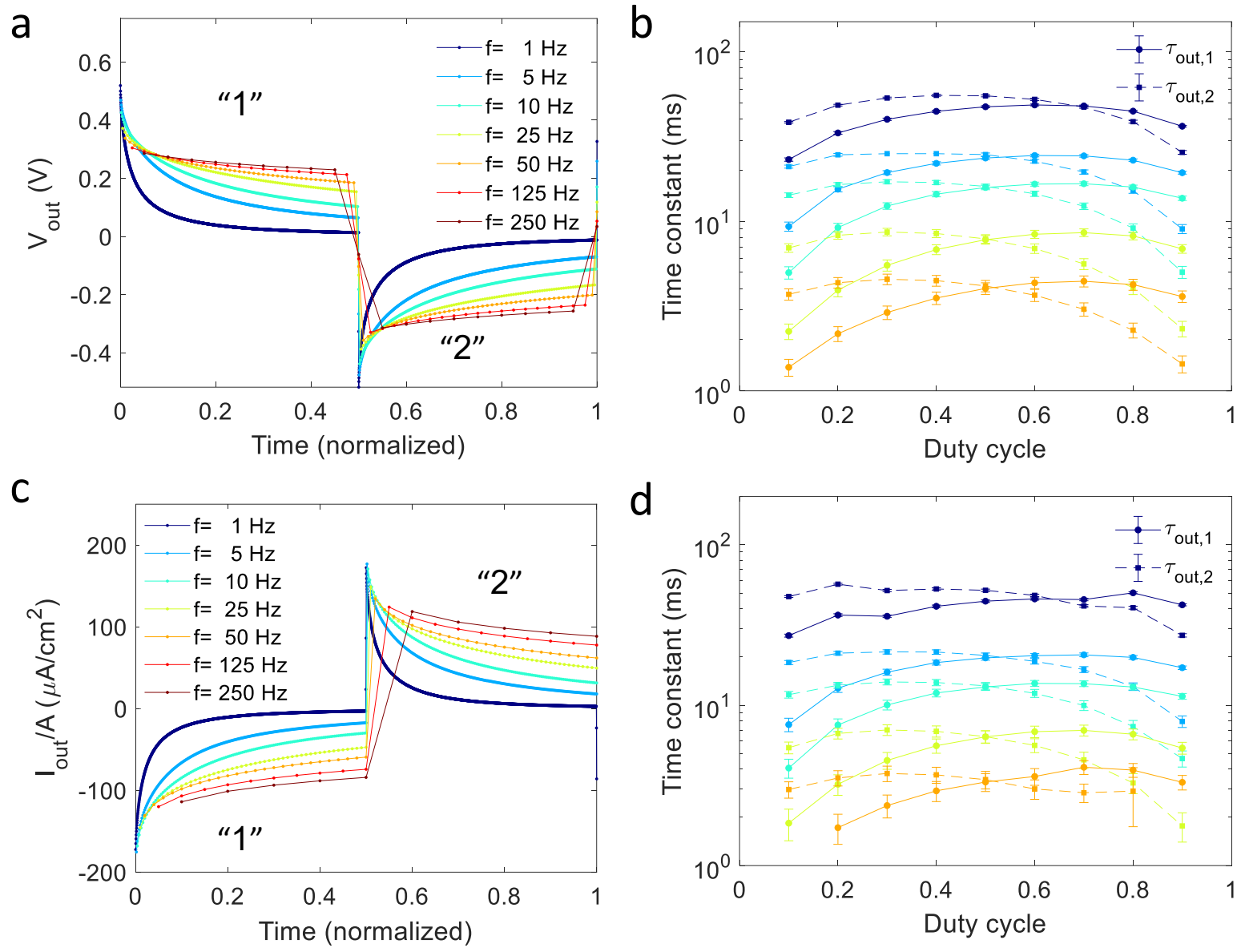

*Figure S8: (a) $V_{out}$ as a function of time (normalized by the period, T) for several input signal frequencies and a duty cycle of 0.5. (b) The time constants of $V_{out}$ for several input signal frequencies and duty cycles. (c) $I_{out}/A$ as a function of time (normalized by the period, T) for several input signal frequencies and a duty cycle of 0.5. (d) The time constants of $I_{out}/A$ for several input signal frequencies and duty cycles. The sample is as in Figure 2a-b, the RBIP pore diameter is 40 nm, the active area of the membrane is A = 0.32 cm$^2$, the electrolyte is 1 mM HCl aqueous solution, and the input signal amplitude, $V_a$, is 0.3 V.*

## 5. Time constants extraction for the Au-Pt sample

Figures S9a-b show the voltage signals measured between the surfaces of the Au-Pt sample and the Ag/AgCl wires next to each surface for various duty cycles. Figure S9a shows $V_{Au}(t)$, the voltage measured between the gold surface and the adjacent Ag/AgCl wire (dots) and the best fits to equation (3) (solid line). Figure S9b shows $V_{Pt}(t)$, the voltage measured between the platinum surface and the adjacent Ag/AgCl wire. Each presented signal is the average response of 300 temporal periods. The standard deviation between the 300 periods is smaller than the thickness of the markers. Figures S9c-e shows, respectively, the fitted $V_f$, the effective amplitude $V_i - V_f$, and the fitting $R^2$. To obtain the best fit for all duty cycles, the first three data points of each signal were excluded from the fitting process. When excluding the first part of the signal, the fitting process accounts for slow processes and neglects the fast charging-discharging processes. As shown in Figures S2d-e, S12c, S13c, and the corresponding discussion, the fast processes hardly contribute to the device output. Hence, this exclusion has little effect on the main parameters defining the RBIP performance. Figures S10a-c show respectively the fitted time constants without excluding any datapoints, when excluding the first 3 data points of every signal, and when excluding the first 5 data points of every signal. When data points are not excluded, the fast processes are accounted for in the fitting process leading to shorter extracted time constants. Conversely, when more data points are excluded, longer time constants are extracted. Nevertheless, all cases show a significant difference between the time constants of the gold surface, and the platinum surface as well as a difference

between the time constants for charging and discharging the surfaces. As shown with the analytical model (Figure 4, and section 13 in the supporting information), the output voltage increases with the difference in time constants. Thus, the difference between the time constants of the Au and Pt surfaces leads to a higher device output. For a duty cycle of 0.125, there was not enough information to fit the signals to the first part of the period with sufficient certainty when 5 points were excluded. Figure S9f shows the amplitude asymmetry coefficient, $a$, calculated using the fitted parameters and equation (S10). Figures S9g-h show the measured output voltage and current density used to create Figures 2e-g, respectively. To focus on the outputs and average out the signal oscillations, a moving average of 10,000 points was applied. The sampling time was $5 \cdot 10^{-4}$ s. These measurements, (prior to applying the moving average) were used to create Figures 2e-g. As shown in Figure S9g-h, during each 60-second On/Off cycle, the output voltage and current stabilize at different rates that vary with the duty cycle and frequency. An analysis of these longer-timescale dynamics is left for future work. Nevertheless, extended measurements (Figure 3b and Figure S24b) have shown that an output voltage can be sustained for many hours.

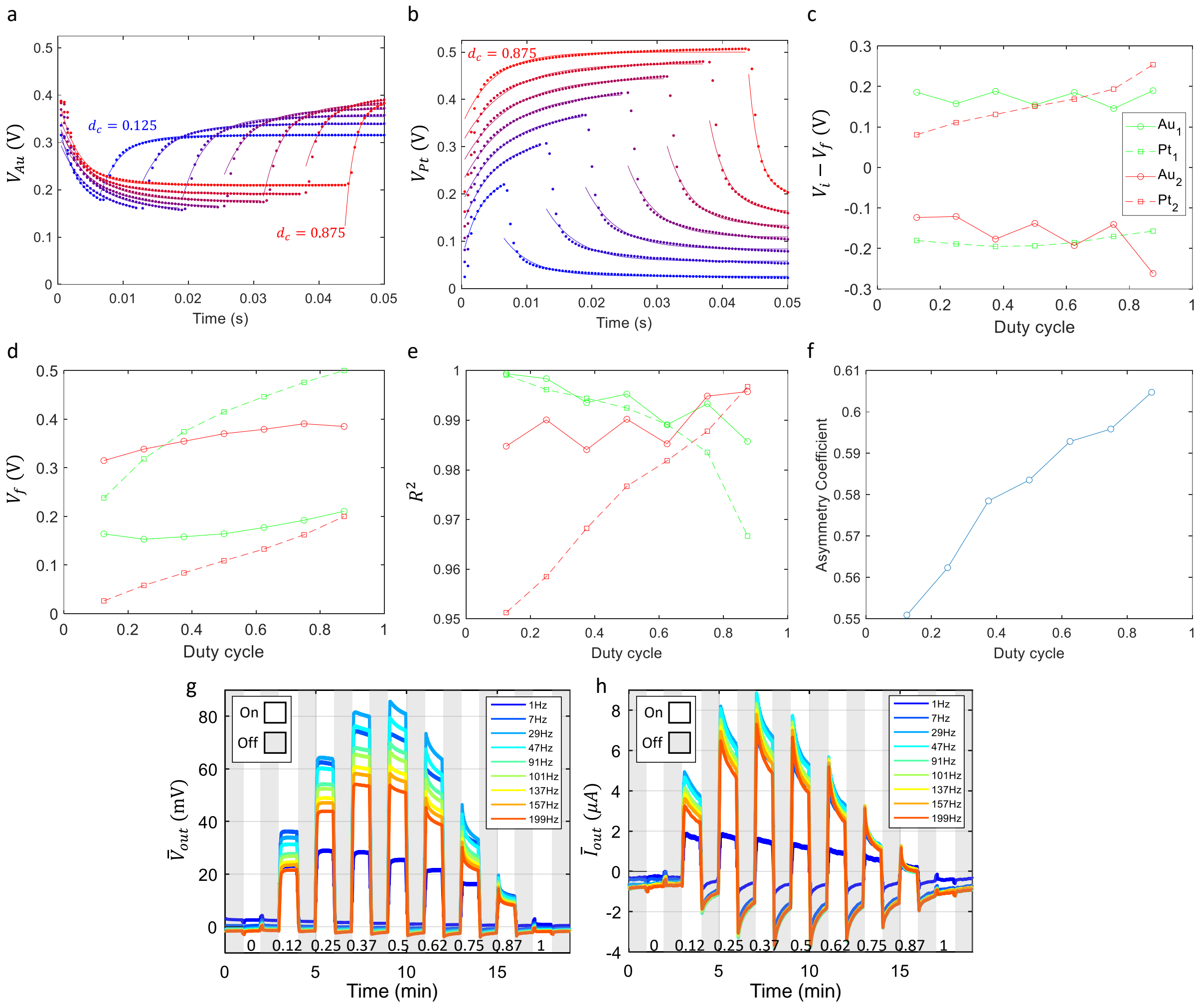

*Figure S9: (a,b) The Experimentally measured signals $V_{Au}(t)$ and $V_{Pt}(t)$, respectively, for rectangular-wave inputs with various duty cycles. The temporal period is $T = 50\ ms$ ($f = 20\ Hz$). The dots are the average response of 300 measured temporal periods, and the solid lines mark the best exponential fit of the data. The color coding marks the duty cycle. The time constants of the fitted curves were used to create Figure 2d. (c-e) The fitted $V_f$, the effective amplitude, $V_i - V_f$, and the fitting $R^2$ respectively. The color coding for (d) and (e) is as in (c). (f) The amplitude asymmetry coefficient, a, calculated using the fitted parameters and equation*

*(S10). (g-h) Moving average (N=10,000) of the measured output voltage and output current density, respectively, for several input signal duty cycles and frequencies. The grey shaded areas mark the time when the ratchet was OFF, and the white regions mark the time when the ratchet was ON. The duty cycle for each ratchet ON period is indicated below the curves. The electrolyte is 1 mM KCl aqueous solution, and the input signal is alternating between 0.3 V and -0.3 V.*

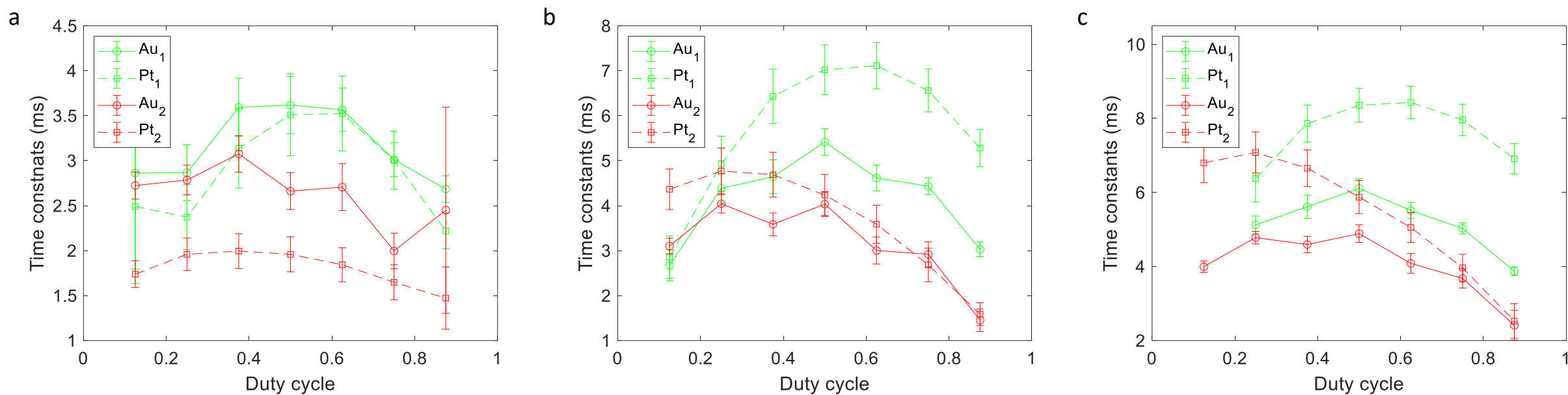

*Figure S10: The time constants extracted from the signals $V_{Au}(t)$ and $V_{Pt}(t)$ (Figure S9 a, b) without excluding any datapoints (a), when excluding the 3 first datapoints of every signal (b), and when excluding the 5 first datapoints of every signal (c).*

## 6. Numerical simulation – Methods and additional data

The cation, anion, and electric potentials distributions in response to an input signal were found by solving the one-dimensional Poisson-Nernst-Planck (PNP) equations using COMSOL Multiphysics® v6.1. Assuming a symmetric binary z : z electrolyte, no solvent convection, and no bulk reactions, the governing equations are as follows:

$$-\varepsilon\frac{\partial^2\phi}{\partial x^2} = \rho = Fz(C_+ - C_- + \rho_m), \tag{S1}$$

$$\frac{\partial C_\pm}{\partial t} = -\frac{\partial J_\pm}{\partial x} = -\frac{\partial}{\partial x}\left(-D_\pm\frac{\partial C_\pm}{\partial x} \mp \frac{zD_\pm}{V_{th}}C_\pm\frac{\partial\phi}{\partial x}\right). \tag{S2}$$

where $\phi$ is the electric potential in the solution, and $\rho$ is the total space charge density that includes the contributions of the cation and anion concentration ($C_+$ and $C_-$, respectively), and the membrane fixed charge, $\rho_m$. $z$ is the ion valence, $V_{th} = RT_r/F$ is the thermal voltage. $J_\pm$ and $D_\pm$ are the cation and anion flux and diffusion coefficient, respectively. $\varepsilon, F, R$, and $T_r$ are the dielectric permittivity of the solvent, the Faraday constant, universal gas constant, and temperature, respectively. The membrane is modeled through a fixed space charge density, which has a value of $\rho_m$ (units in $\mathrm{C/m^3}$) at the membrane domain ($-L/2 < x < L/2$) and 0 elsewhere.

The input signal $V_{in}(t)$ is a periodic square wave signal that is defined for an integer temporal period $n$ as follows:

$$V_{in}(t) = \begin{matrix} V_a, & nT < t < (n+\delta)T \\ \alpha V_a, & (n+\delta)T < t < (n+1)T \end{matrix}. \tag{S3}$$

Each temporal period is described by a duty cycle $\delta = \tau_1/T$, which is the ratio between the duration of the first step $\tau_1$, where the potential is at its maximum value $V_a$, to the total duration of the period, $T(= 1/f)$. The temporal period is completed with a second step, in which the potential is multiplied by a potential symmetry factor in the range $-1 \le \alpha \le 0$. Figure S12 and Figure S13 show the results for $\alpha = 0$ and $\alpha = -1$, respectively. In all simulations the amplitude is $V_a = 0.8$ V and frequency $f = 5$ kHz.

Figure S11a shows an illustration of the simulation domain. The input signal, $V_{in}(t)$, is applied to the right electrode ($x = L/2$). The left electrode is defined as a reference point for the potential, i.e. $\phi(x = -L/2) = 0$. The signal is electrically floating in the sense that it does not impose a strict potential difference between the contacts and the bulk solution. To simulate a closed system (the total amount of ions in the system is constant), the ionic fluxes are set to zero at the reservoir's outer edges ($x = \pm(W + L/2)$). Hence, the boundary conditions can be formulated as:

$$\phi(x = -L/2, t) = 0, \tag{S4a}$$
$$\phi(x = L/2, t) = V_{in}(t), \tag{S4b}$$
$$J_{\pm}(x = -L/2 - W, t) = J_{\pm}(x = L/2 + W, t) = 0. \tag{S4c}$$

The initial conditions for the transient simulation are calculated by solving a preliminary stationary problem ($\partial C_{\pm}/\partial t = 0$), while applying $C_{\pm}(x = -L/2 - W, t) = C_{\pm}(x = L/2 + W, t) = C_0$, and $V_{in}(t) = 0\ V$. The initial guess for the stationary problem is: $\phi(x, 0) = 0$, and $C_{\pm}(x, t = 0) = C_0$, where $C_0 = 0.5\ \mathrm{mM}$ in all simulations. The transient solution is run until the per-period average output voltage, $V_{out}$, at the reservoir outer edges reaches a steady state.

The mesh spatial discretization used quadratic Lagrange elements. The one-dimensional mesh was defined by an exponential growth in element size away from each of the electrodes, with at least 20 elements within 1 nm from each side of the electrodes. The time-steps were automatically generated by the solver, using COMSOL's *Events* interface to define the abrupt and periodic changes in the electric potential input at the electrodes. The convergence criterion for the estimated error in each time step used a relative tolerance of 0.005 and an absolute tolerance of 0.0005.

The AAO membrane space charge density was derived from both theoretical and experimental sources. Using the MUSIC model,[2,3] theoretical calculations yielded a surface charge density of 5.7 mC/m² for 0.5 mM electrolyte concentration. In contrast, experimental measurements on untreated AAO membrane showed a lower value of 1.08 mC/m².[4] Given that the Debye length at 0.5 mM (13.6 nm) is on the same order of the pore radius, we assumed uniform charge distribution within the pores. The surface charge density in a cylindrical nano-channel can be converted to an effective space charge density using $\rho_m = 2\sigma/r$, where $\sigma$ is the surface charge density and $r$ is the pore radius. To establish an upper bound for the space charge, the theoretical surface charge was used with $r$ = 20 nm yielding $\rho_m =$ 5.7×10$^5$ C/m³. A space charge density of 7.4×10$^4$ C/m³ matches approximately the experimentally measured surface charge[4] with $r \approx$ = 30 nm. Unless stated otherwise, the former value (5.7×10$^5$ C/m³) was used in all the simulations.

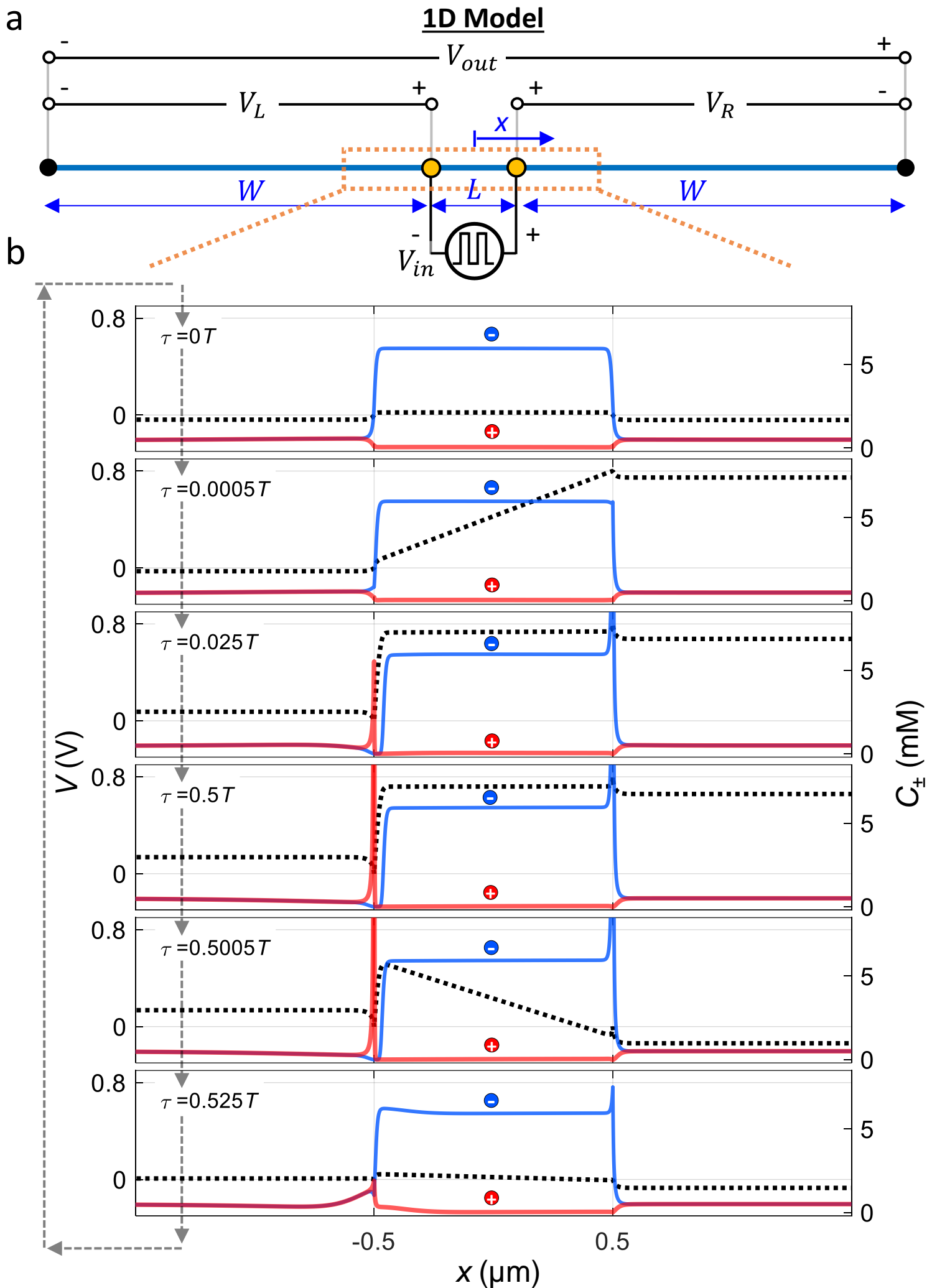

*Figure S11: (a) Numerical simulation: one-dimensional model of the system. (b) Spatial distribution of cations (solid red line), anions (solid blue line), and electric potential (dot-dot black line), at a few key times during a typical period ($d_c = 0.5$). Slow charging of cations/anions can be observed at the left/right electrode during the 1st half period, while a much faster discharging takes place during the 2nd half period. The simulation parameters are: $L = 1\ \mu m$, $W = 4.5L$, $D_{\pm} = 2 \cdot 10^{-9}\ m^2/s$, $\rho_m$ = 5.7×10⁵ C/m³, $C_0 = 0.5\ mM$, $V_a = 0.8\ V$, $\alpha = 0$, $d_c = 0.5$, and $f = 5\ kHz$.*

Figure S12 shows the numerical simulation results for, $\mathrm{f} = 5\ \mathrm{kHz}$, $\alpha = 0$, and duty cycles between 0.1 and 0.9. Figures S12a,b show respectively the mean calculated signals $V_L(t)$ and $V_R(t)$ and the best fits to equation (3). Figures S13a,b show the same results for $\alpha = -1$. As in the experimental results, the fitted curves in Figures S12a,b and Figures S13a,b do not account for the fast processes near $t = 0$ and $t = d_c T$. To verify that the contribution to the net output of the faster processes at $t = 0$ and $t = d_c T$ is small, the net output voltage, $\Delta\bar{V}_{out}$, obtained from the simulation was compared to the net output voltage obtained from the fitted data. Figure S12c and Figure S13c show $\Delta\bar{V}_{out}$ calculated directly from the simulation results and $\Delta\bar{V}_{out}$ obtained from the fitted data. The excellent agreement between the two calculations shows that $\Delta\bar{V}_{out}$ is almost entirely determined by the slow processes. Figure S12d and Figure S13d show the fitted time constants as a function of the input signal duty cycle for $\alpha = 0$ and $\alpha = 1$, respectively. Figure S12e and Figure S13e show the corresponding $R^2$. As in the experimental results, the time constants vary significantly with the duty cycle. The time constants for discharging the surfaces towards a potential of 0, which is at the point of zero charge ($\tau_{R2}, \tau_{L2}$ in Figure S12d), are significantly shorter compared to charging and discharging towards other potentials ($\tau_{R1}, \tau_{L1}$ in Figure S12d and all the time constants in Figure S13d). This may be because when discharging to the point of zero charge, less ions must diffuse towards the surface from the bulk solution which is the rate limiting step in the charging/discharging process. For $\alpha = 0$, $\tau_{R2}$ and $\tau_{L2}$ hardly change with the duty cycle because there is sufficient time to fully discharge the surfaces at this frequency and duty cycle range. For $\alpha = -1$, the time constants change with duty cycle is mirrored about a duty cycle of 0.5. This response is similar to an "odd ratchet effect" induced in systems with antisymmetric potential distributions driven with time symmetric signals.[5] Figure S12f shows the output voltage for a voltage step response for three values of pore charge. As expected, the RBIP output drops to zero for a constant applied potential bias. These outputs were used for the data point at a duty cycle of 1 in Figure 2d in the main text. Figure S13f shows $\Delta\bar{V}_{out}$ as a function of the duty cycle for $\alpha = -1$ and several values of pore charge. All outputs are antisymmetric with respect to a duty cycle of 0.5. Similar to the results for $\alpha = 0$, $\Delta\bar{V}_{out}$ increases and the optimal duty cycle shifts towards 0 and 1 as the pore charge increases.

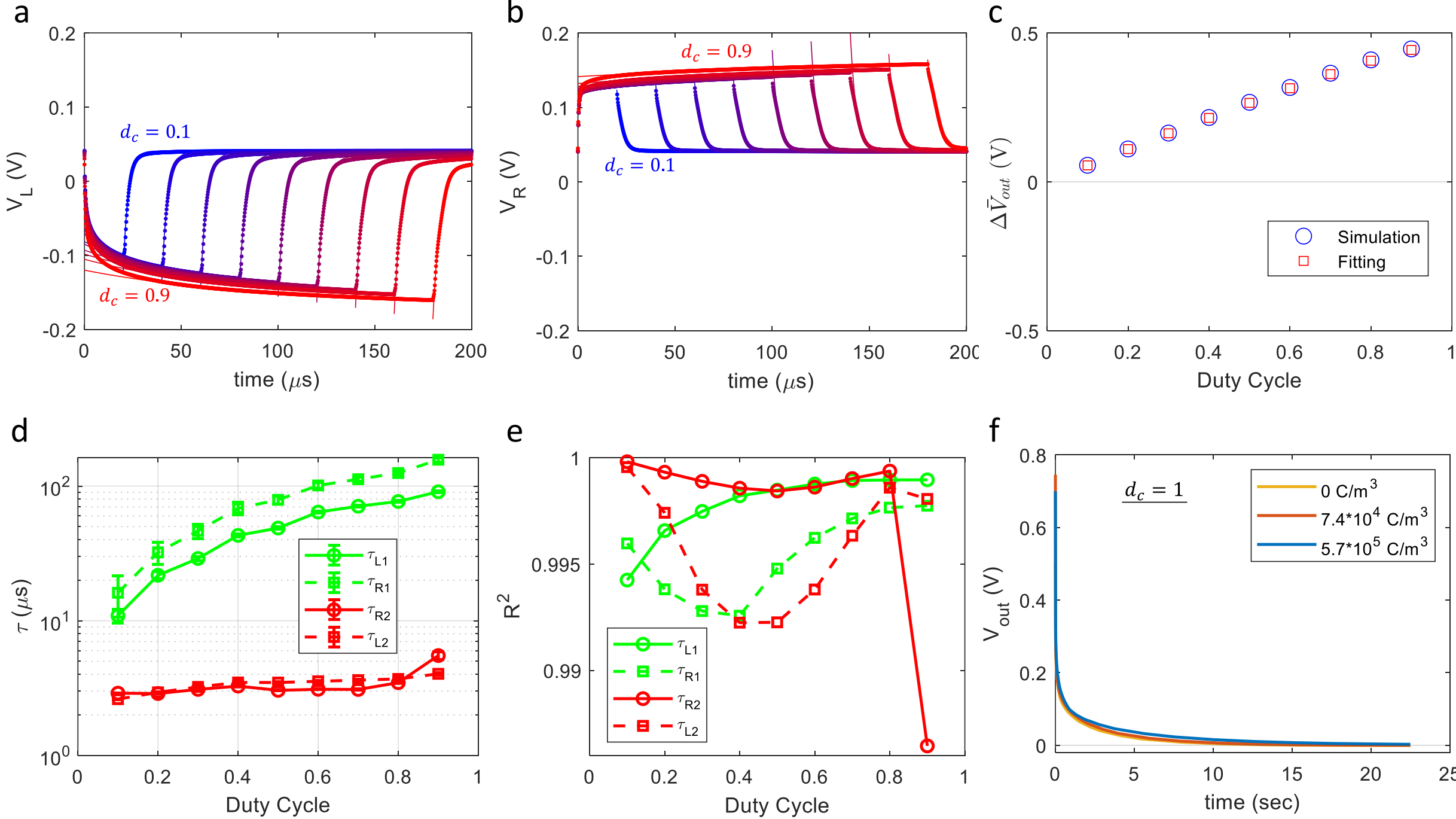

*Figure S12. Simulation results for $\alpha = 0$. (a,b) The calculated and fitted $V_L(t)$, and $V_R(t)$ ,respectively, for various duty cycles. The dots are the calculated data point from the simulation, and the solid lines mark the best fit to eq. (3) in the main text. The color coding marks the duty cycle. (c) Comparison of $\Delta\bar{V}_{out}$ from simulation (circle) and calculated with the fitted curves (square). (d) Time constants extracted from the fitted curves in a and b. Subscripts 1 and 2 denote the first ($0 < t \leq d_c T$ ) and second ($d_c T < t \leq T$ ) parts of the input signal, respectively. The error bars indicate the fitting 95% confidence interval. (e) $R^2$ values of the fitted curves. (f) $V_{out}$ as a function of time for a step input of $V_{in} = V_a$ (duty cycle of 1). In a-e the input signal period is $T = 200\ \mu s$ ($f = 5\ kHz$), and in a-f the amplitude is $V_a = 0.8\ V$.*

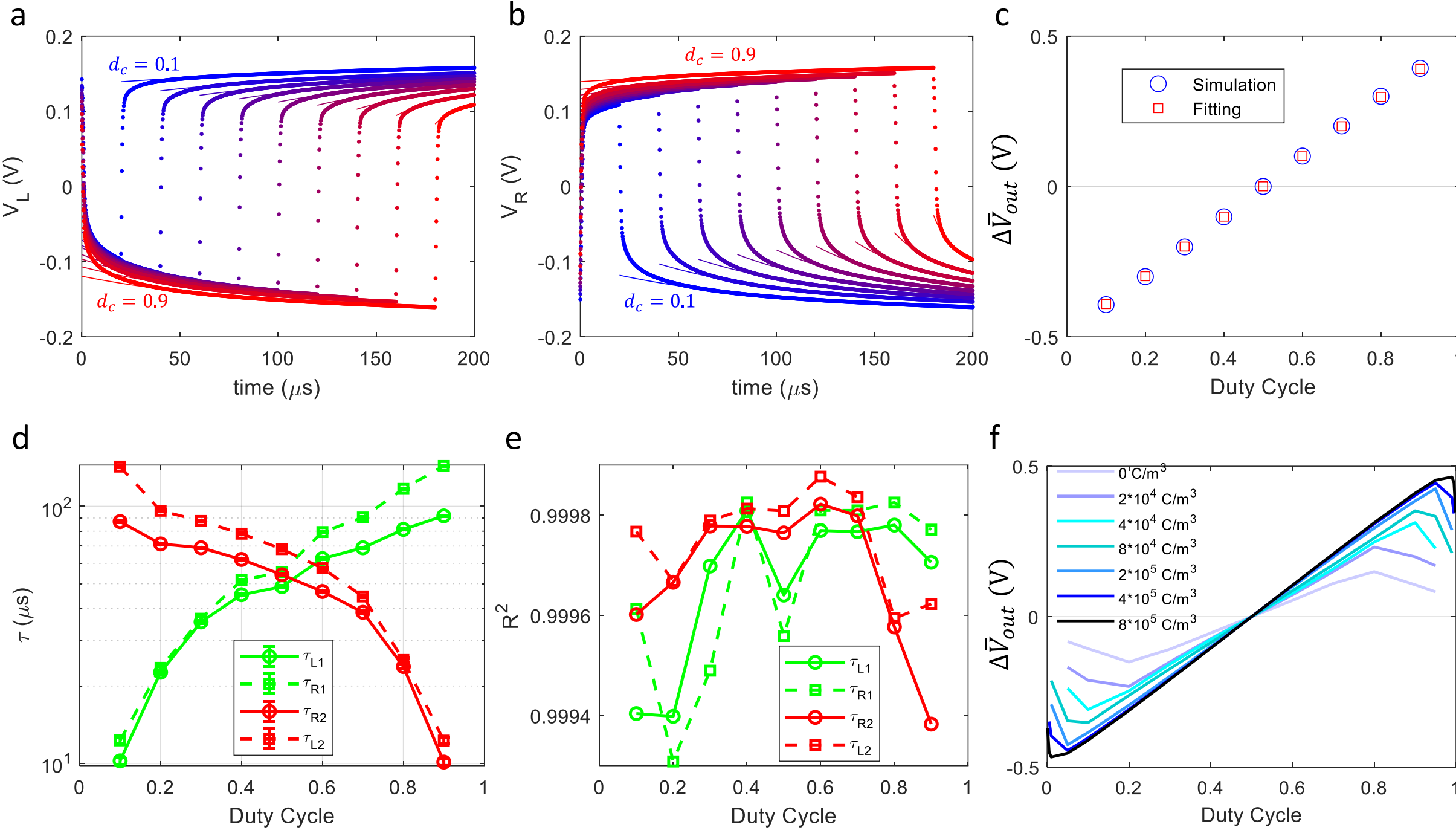

*Figure S13. Simulation results for $\alpha = -1$. (a,b) The calculated and fitted $V_L(t)$, and $V_R(t)$, respectively, for various duty cycles. The dots are the calculated data point from the simulation, and the solid lines mark the best fit to eq. (3) in the main text. The color coding marks the duty cycle. (c) Comparison of $\Delta\bar{V}_{out}$ from simulation (circle) and calculated with the fitted curves (square). (d) Time constants extracted from the fitted curves in a and b. Subscripts 1 and 2 denote the first ($0 < t \leq d_c T$ ) and second ($d_c T < t \leq T$ ) parts of the input signal, respectively. The error bars indicate the fitting 95% confidence interval. (e) $R^2$ values of the fitted curves. (f) The calculated $\Delta\bar{V}_{out}$ as a function of the input signal duty cycle and several levels of pore charge. In a-f the input signal period is $T = 200\ \mu s$ ($f = 5\ kHz$) and the amplitude is $V_a = 0.8\ V$.*

To decrease computational efforts, the length scale of the modeled membrane and reservoirs is significantly smaller than in our experimental system. Consequently, the time constants are much shorter, and the operation frequency is correspondingly larger.[6] While the pore charge values were taken from literature, our experimental results typically show more moderate optimal duty cycles (Figure 2). This discrepancy may arise from the one-dimensional nature of our model, which neglects potential screening effects toward the pore center. Moreover, while the performance of experimental system is determined also by the point of zero charge of the contact materials, the simplified model does not directly account for these variations. To provide upper and lower bounds to this aspect, we examined two limiting cases of $\alpha = 0$ (Figure 2d, Figure S11, Figure S12), and $\alpha = -1$ (Figure S13). In the first case, $V_{in}$ alternates between $V_a$ and zero potential while in the second case, $V_{in}$ alternates between $V_a$ and $-V_a$. These cases represent the boundaries within which experimental results typically fall. A more exhaustive study of the numerical model, discussing ion pumping, and the effects of the pore charge on the transport mechanism, is left for future work.

## 7. Additional results and methods

### *RBIP with a pore diameter of 40 nm in KCl aqueous electrolyte*

Here we have tested the performance of an RBIP fabricated on AAO wafers with 40 nm diameter pores. The wafers were air annealed for 11 h at 650 °C. The contacts were deposited with electron beam evaporation of 10 nm of titanium and 40 nm of gold (planar equivalent). For the results presented in this section, unless stated otherwise, the input electric signal $V_{\text{in}}(t)$ is a rectangular wave at a frequency of 100 Hz, and the amplitude, $V_a$, is 0.2 V. The ratchet input signal, $V_{in}$, was applied with an HP 3245A universal source and the voltage between the Ag/AgCl wires was measured with an Agilent 34401A multimeter, where both instruments shared the same ground. The voltage measurement was conducted with an integration time of 1.67 s to reduce the output signal oscillations and obtain only the net time averaged voltage, $\bar{V}_{out}$. The response to every input signal was measured for 5 min after which the input was set to 0 V for 5 min. $\bar{V}_{out,ON}$, $\bar{V}_{out,OFF_1}$ and $\bar{V}_{out,OFF_2}$ were averaged over the last 150 s of every period. Figure S14a shows the recorded voltage for duty cycles between 5% and 100% (the duty cycle is the portion of the time in every period where the voltage is at its high value. The input signal duty cycle is marked next to the output curve in Figure S14a).

Once a ratchet signal commences, $V_{out}$ quickly builds up to a level determined by the duty cycle. The ratchet-induced voltage reaches its largest values for duty cycles close to 50%, i.e., a temporally averaged input voltage of 0 V. For a duty cycle of 100%, which is the response to a voltage step from 0 V to 0.2 V, the voltage signal shows the well expected, fast capacitive charging behavior corresponding to polarization of the metal contacts. However, unlike the response to a duty cycle of 100%, for a duty cycle of 95% (a temporal average input voltage of 0.19 V), a ratchet action is observed and $V_{out}(t)$ has a negative sign with a much slower decay time. This provides a simple distinction between steady-state ratchet-driven transport and capacitive charging–discharging behavior.

To estimate the ratchet output voltage, the measured voltage was averaged over the last 2.5 min of every cycle and the difference between the ratchet ON and ratchet OFF average voltages was calculated. Figure S14b shows the ratchet time averaged output voltage, $\Delta\bar{V}_{out}$, as a function of the input signal amplitude, $V_a$, for 1 mM and 10 mM KCl aqueous solutions. More details on how $\Delta\bar{V}_{out}$ is calculated can be found in the Methods section. The input signal was an unbiased square wave with a frequency of 100 Hz, and a duty cycle of 50%. For the 1 mM KCl solution, a noticeable ratchet output is visible for ratchet signals with an amplitude as small as 0.05 V. This extremely low voltage threshold provides a clear distinction between ratchet-induced ion transport, where the voltage threshold is about *2kT* (*k* is the Boltzmann constant ($1.381 \times 10^{-23}$ J $K^{-1}$) and *T* is the temperature (K)), and ion transport induced by chemical reactions where the threshold bias is determined by the Gibbs free energy and the reaction overpotentials. The ratchet output is significantly smaller with the 10 mM KCl solution. This is an expected result, because at a higher ion concentration potential screening is more significant, and as a result the center of the pore is less affected by the input signal and can serve as a shunt for back-diffusing ions. Figure S14c-d shows the ratchet output voltage as a function of the frequency and duty cycle in 1 mM and 10 mM KCl, respectively. Since ratchet systems have no long-term output when a DC voltage is applied, the RBIP output voltage is close to 0 V for duty cycles near 0% and 100%. Similarly, at low frequencies, the RBIP fully charges and discharges the double layers, which is similar to operation under DC bias. Thus, the output is near 0 V at low frequencies as well. As a result, the RBIP shows a significant output only when operated with duty cycles near 50% and at input signal periods that are close to the characteristic charging–discharging time constants of the RBIP. When the input signal period is significantly shorter than the RBIP charging and discharging time constants the output again goes to 0 V.

The output of the ratchet was measured when switching the leads connected to the RBIP contacts, and the leads of the Ag/AgCl wires (used to measure $V_{out}$). Figure S15a shows the measurement setup and Figure S15b shows $\bar{V}_{out}$ as a function of the duty cycle for every measurement configuration. According to equation (2) in the main text, $V_{in}(t,d_c) = -V_{in}(t,1-d_c)$. Thus, switching the positive and negative leads for $V_{in}$ ('Ratchet Switch' in Figure S15a-b) resulted in $\bar{V}_{out}$ values that were reflected around 50% duty cycle such that $\bar{V}_{out,1}(d_c) = \bar{V}_{out,2}(1 - d_c)$; switching the positive and negative leads for $\bar{V}_{out}$ ('Ag/AgCl Switch' in Figure S15a-b) resulted in oppositely-signed values for $\bar{V}_{out}$, such that $\bar{V}_{out,1}(d_c) = -\bar{V}_{out,2}\,(d_c)$.

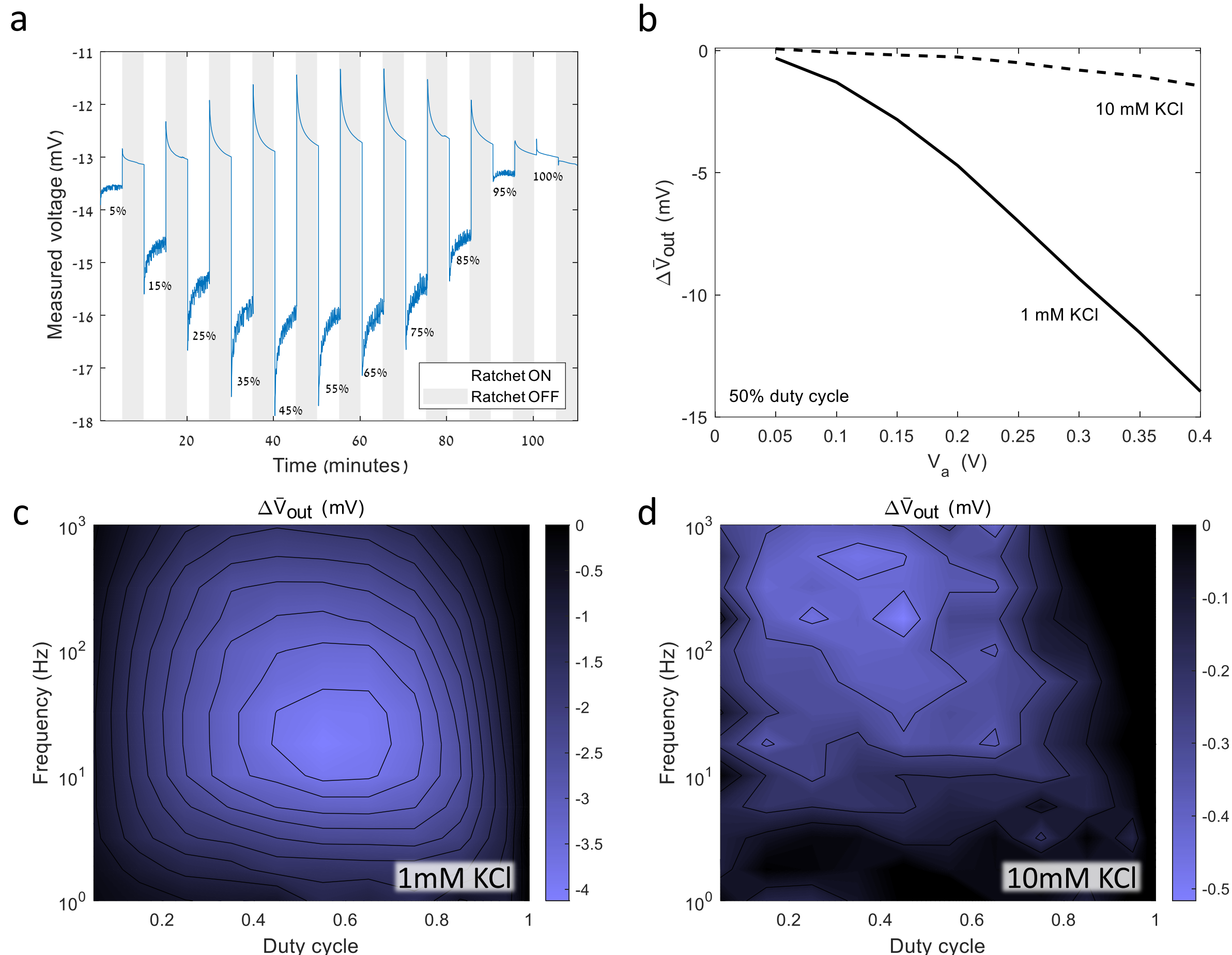

*Figure S14: The output of a RBIP sample with 40 nm pores. (a) Measured output voltage, $V_{out}$, for an input signal with various duty cycles. The input signal is a rectangular wave with $V_a$ of 0.2 V, a frequency of 100 Hz, and the electrolyte is 1 mM KCl aqueous solution. (b) The ratchet output voltage, $\bar{V}_{out}$, as a function of the input signal amplitude, the frequency is 100 Hz, and the duty cycle is 50%. (c,d) the ratchet output voltage as a function of the input signal frequency and duty cycle. The electrolyte is 1 mM KCl aqueous solution (c), and 10 mM KCl aqueous solution (d).*

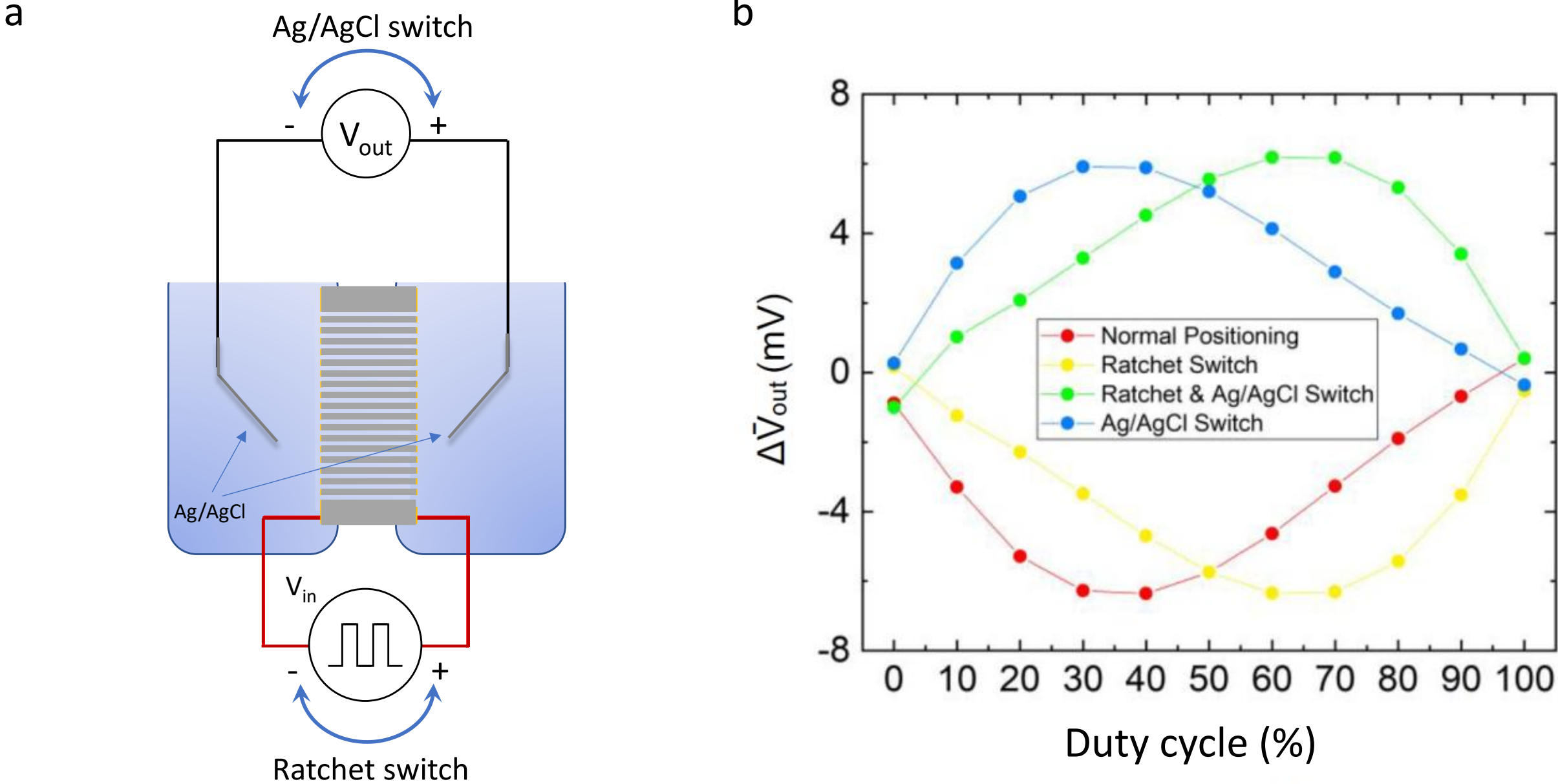

*Figure S15: (a) Illustration of the lead-switching experiment, (b) The voltage output of an RBIP, $\Delta\bar{V}_{out}$, when switching the input signal leads, and the leads for the Ag/AgCl voltage-sensing leads. The input signal frequency is 50 Hz and the electrolyte is 1 mM HCl aqueous solution. All other RBIP and input signal parameters are in as in Figure2.*

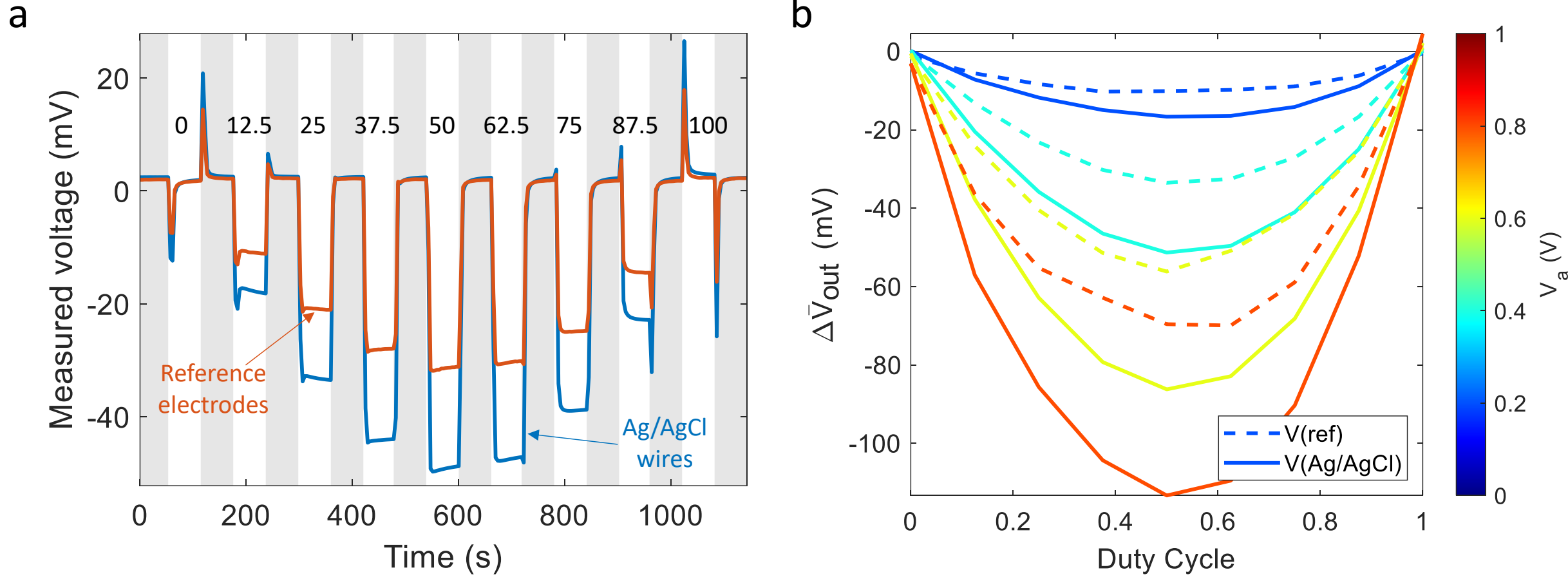

*Figure S16: (a) Voltage output, $\bar{V}_{out}(t)$, measured between two Ag/AgCl wires and two leak-free reference electrodes (LF-1-45, Innovative Instruments, Inc.). The pore diameter is 40 nm, the electrolyte is 0.2 mM KCl aqueous solution, the signal frequency is 100 Hz, and the amplitude is $V_a$ =0.4 V. The time periods in which the ratchet operated are the bright areas and the shaded areas are time periods in which the input voltage was set to 0 V. The duty cycle (in percent) used is noted next to the curves. (b) The average ratchet output voltage, $\bar{V}_{out}$, as a function of the duty cycle and several input signal amplitudes measured between two leak-free reference electrodes (dashed line) and Ag/AgCl wires (solid line).*

Electro-osmosis was shown to be negligible by running the ratchet continuously while observing the level of the solution within each compartment. The experimental setup is as the one described in Figure S24a. The ratchet was driven continuously for 16 minutes and 40 seconds with the auxiliary electrodes shunted. Then, the ratchet was turned OFF and $V_{out}$ was measured for 90 seconds. This process was repeated 9 times for a total of nearly 3 hours. The input signal was a square wave with a frequency of 125 Hz, an amplitude of 300mV and a 50% duty cycle. Each compartment was filled with 1.75 mL of 1mM HCl aqueous solution. The RBIP parameters are as in Figure S24. Although the system was operated for 3 hours, no visible change in water level was observed. Figure S17 shows optical photographs of the setup before and after operation. To further demonstrate that the RBIP prevents any water mass transport between the two compartments, this experiment was repeated with a solution height difference introduced between the two compartments. The solution, RBIP and input signal parameters are identical to those in Figure S17 except that one compartment was filled with 2.25 mL of solution and the other with 1.75 mL. The RBIP was operated for 3 hours as described in Figure S17 after which it was operated continuously for another 45 hours. Figure S18(a-c) shows optical images of the setup at the beginning of the experiment, after 24 and 48 hours respectively. No water transport through the membrane was observed during the experiment.

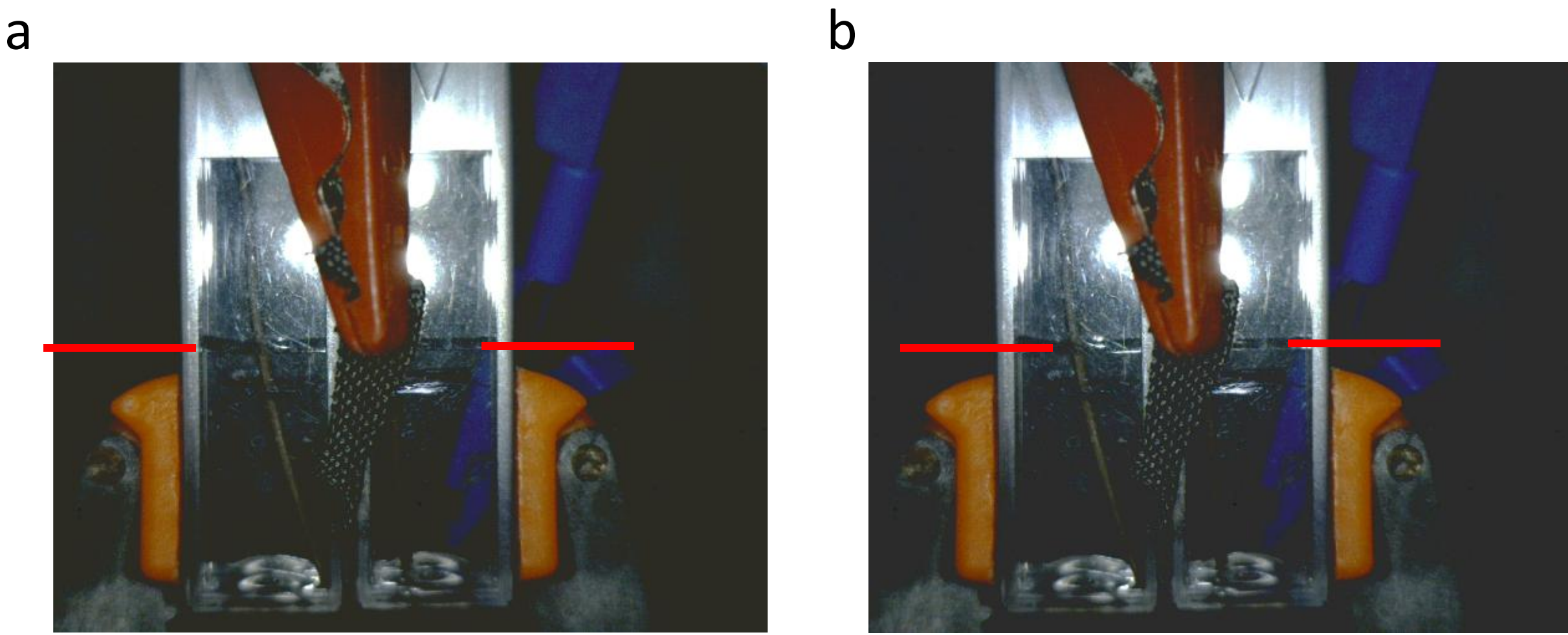

*Figure S17: water pumping experiment. Initial water level (a) and after 3 hours of ratchet operation (b). The red lines are guides for the eye indicating the solution level in each compartment. The experimental setup is as the one described in Figure S24a. The input signal was a square wave with a frequency of 125 Hz, an amplitude of 300mV and a 50% duty cycle. Each compartment was filled with 1.75 mL of 1mM HCl aqueous solution.*

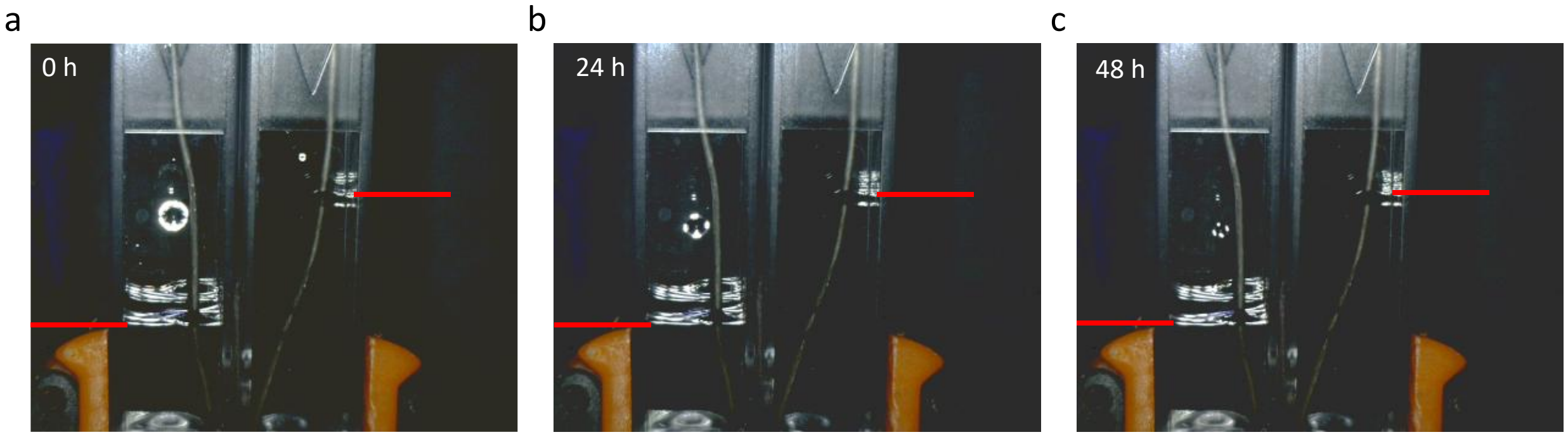

*Figure S18: Optical images of the electrochemical cell while an intentional solution level difference is introduced. (a) the initial water level, (b) after 24 hours, (c) after 48 hours. The red lines are guides for the eye indicating the solution level in each compartment. No water transport through the membrane was observed during this time. The solution, RBIP and input signal parameters are identical to those in Figure S17 except that one compartment was filled with 2.25 mL of solution and the other with 1.75 mL.*

*RBIP with a pore diameter of 20 nm operation in KCl aqueous electrolyte*

The same set of measurements as performed in the previous section (using RBIPs with a pore diameter of 40 nm) was repeated using a RBIP constructed on an AAO wafer with 20 nm diameter pores. The sample was air annealed for 11 h at 650 °C and the contacts were deposited by electron beam evaporation of 30 nm of titanium (adhesion layer) and 20 nm of gold (planar equivalent). Figure S19a shows typical data for the measured ratchet signal for an input signal frequency of 100 Hz and duty cycles between 5% and 95%. The electrolyte is 1 mM KCl aqueous solution. The input signal parameters are as in Figure S14a. The input signal was generated with a HP 3245A universal source. The voltage between the Ag/AgCl wires was measured with an Agilent 34401A multimeter where both instruments shared the same ground. The voltage measurement was conducted with an integration time of 1.67 s to reduce the output signal oscillations and obtain only the net averaged voltage. The response to every input signal was measured for 300 s after which the input was set to 0 V for 300 s. $\bar{V}_{out,ON}$, $\bar{V}_{out,OFF_1}$ and $\bar{V}_{out,OFF_2}$ were averaged over the last 150 s of every period. Unlike the 40 nm pore sample, the response to the ratchet signal does not go to 0 V for duty cycles near 0 and 1. This can be attributed to partial blocking of the pores, caused by their smaller diameter. Further discussion of the effect of blocked pores and other failure modes in RBIP devices can be found in the failure modes section of the supporting information.

Figure S19b shows the 20 nm pore RBIP output, $\Delta\bar{V}_{out}$, as a function of the input signal amplitude, $V_a$, in 1 mM KCl and 10 mM KCl aqueous solutions. The ratchet output is defined as in Figure S14. The frequency is 100 Hz and the duty cycle is 50%. A noticeable output was recorded for amplitudes as low as 50 mV demonstrating an extremely low voltage threshold. When compared to the 40 nm pore RBIP, the performance of the 20 nm pores RBIP is significantly higher at the higher ionic strength solution. Since in these samples the pore diameter is closer to the Debye length, the electric potential modulation affects a larger portion of the pore. As a result, the decrease in performance with the ionic strength is less pronounced in these RBIPs. Figure S19c-d shows $\Delta\bar{V}_{out}$ as a function of the input signal duty cycle and frequency for 1 mM KCl and 10 mM KCl aqueous solutions, respectively. For 1 mM KCl aqueous solution, the RBIP optimal input signal frequency and duty cycle are 56 Hz and 0.25, respectively, which yield an output of -7.5 mV. For 10 mM KCl solution, the optimal input signal frequency and duty cycle are 316 Hz and 0.35, respectively, and the output is -1.9 mV. The optimal operating frequency increases with the ionic strength and the 20 nm pores RBIP outperforms the 40 nm pores RBIP in higher concentration solutions.

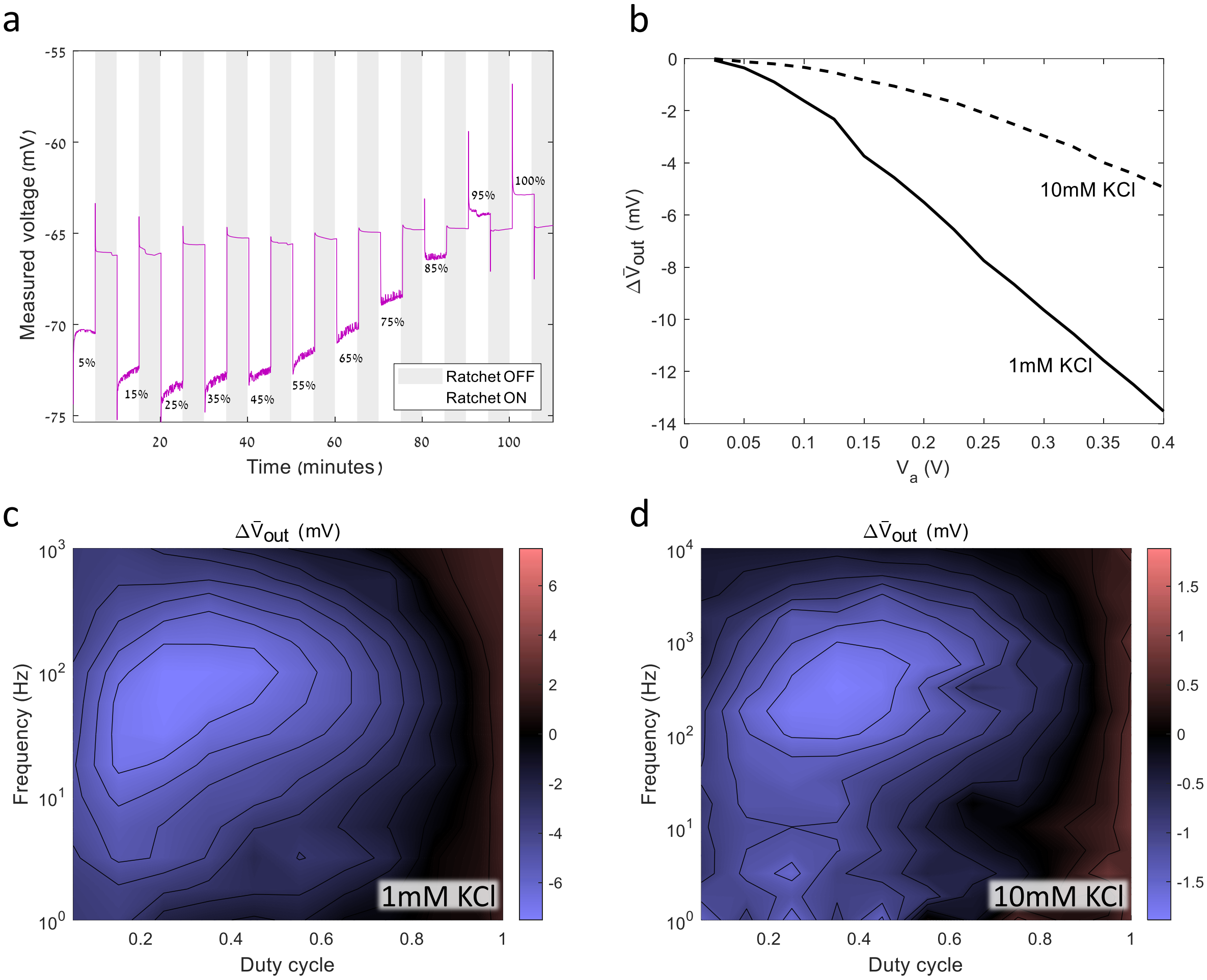

*Figure S19: The output of RBIPs with 20 nm pores. (a) Measured voltage for input signals with various duty cycles. The input signal is a square wave with $V_a$ of 0.2 V, the frequency is 100 Hz, and the electrolyte is 1 mM KCl aqueous solution. (b) Ratchet output voltage as a function of the input signal amplitude, $V_a$, the frequency is 100 Hz and the duty cycle is 50%. (c,d) Ratchet output voltage as a function of the input signal frequency and duty cycle. The input signal is a square wave with $V_a$ of 0.2 V. The electrolyte is 1mM KCl aqueous solution (c) and 10mM KCl aqueous solution (d).*

## 8. Additional data for the electrodialysis demonstration

### *Conductance measurement and the electrodialysis deionization test setup*

A photograph of the electrodialysis (ED) deionization test setup is shown in Figure S20a, and the system is shown to scale in Figure S20b. All parts were 3D printed using a Formlabs Form3 SLA printer with Clear Resin V4 material that was post-cured in UV. The length of the U-shaped reservoirs is 3.5 cm (as shown in Figure S20b), and the effective cross section is ~1 $cm^2$. The electrolyte volume in the dilution cell is 0.63 $cm^3$, and in each of the two reservoirs the volume is 4.1 $cm^3$. The submerged length of the conductivity measurement electrodes in the dilution cell is 18.5 mm, and the inter-electrode distance is 3.4 mm. The reference cell has a volume of 0.66 $cm^3$, the submerged electrodes length is 16 mm, and the inter-electrode distance is 6 mm. The CEM and AEM that were used are fumasep® FKS-PET-75 and FAS-PET-75, respectively. Laser cut Silicone gaskets were used to seal interfaces between the cell parts, IEMs and RBIP. Viton O-rings were used to seal the titanium electrodes. The RBIP active area was 0.16 $cm^2$. The dilution cell temperature was measured with a Type-K thermocouple that was inserted into the dilution cell wall and was not in direct contact with the solution. The temperature was recorded using Keysight 34465A multimeter with 4 s intervals between data points.

Figure S21a-b show respectively the Nyquist and Bode plots of an Electrochemical Impedance Spectroscopy (EIS) measurement of the titanium electrode pair inside the dilution cell. Within a frequency range of approximately 2 to 200 kHz, the imaginary component of the impedance was found to be negligible. In this range, the total impedance is $Z \approx \mathrm{Re}(Z) = R_{Ohmic} + R_{ct}$ , where $R_{Ohmic}$ represents the electrolyte's ohmic resistance and $R_{ct}$, often referred to as the charge-transfer resistance, corresponds to the effective resistance associated with Faradaic reactions occurring at the electrode surface.[7,8] Assuming a strong electrolyte that is fully dissociated in the solvent and applying a small perturbation amplitude, both $R_{Ohmic}$ and $R_{ct}$ are proportional to $1/C$, where $C$ is the electrolyte concentration. Within the identified frequency range, the conductance, $G$, is $G = 1/Z = |I|/|V|$, where $|I|$ is the current amplitude measured through the Ti electrodes, and $|V|$ is the voltage amplitude applied to the Ti electrodes. Since conductance is directly proportional to electrolyte concentration, variations in conductance can be used to track changes in concentration during the deionization process. The electrodes were cut from a titanium wire (annealed, 99.6+% purity, Supplier: Merck) with a diameter of 1 mm. EIS was measured with a BioLogic VSP-300 Potentiostat using an amplitude of 10mV between 200 Hz to 3MHz. The electrolyte that was used during EIS measurement was 1 mM KCl aqueous solution, which had a conductivity of $162\ \mu\mathrm{S/cm}$ (@25°C).

For conductance measurements the input signal was set as a sine wave with an amplitude of 20mV and a frequency of 5013 Hz, supplied with a Keysight 33510B waveform generator. Current was measured with Keysight 34465A multimeter, using an integration time of 20 μs and a sampling rate of 3.8 data points per cycle. Since the measured data is under-sampled, to find the current amplitude the data was then fitted to a sine signal, as shown in Figure S23a-b. Fitting was done by applying a discrete Fourier transform, zero padding in the frequency domain, and then reconstructing the signal in the time domain with an inverse Fourier transform.[9] For each conductance measurement, the input signal was applied for 2 seconds, with current data collected from the final 0.8 seconds to eliminate transient effects. The current amplitude was calculated as the average of the amplitudes from 8 segments, each 0.1 seconds in duration (approximately 500 signal periods), using the method outlined above. The standard deviation of segment amplitudes was determined for each measurement and was consistently below 0.5% across all data points.

To validate the conductance measurement method, a calibration curve was generated by measuring the conductance of pre-prepared electrolyte samples filled into the dilution cell and correlating these values with their known conductivities. As shown in Figure S22, there is an excellent linear fitting ($R^2 = 0.99$) within the chosen conductance range. Conductance was measured once per minute over a 10-minute period, and the reported values represent the average across all measurements during that time. Conductivity of the pre-prepared solutions was measured before being filled into the dilution cell, using Mettler Toledo SevenExcellence S470 conductivity meter and InLab 731 probe (a 4 graphite poles conductivity cell with an integrated temperature probe). The conductivity probe was calibrated at the beginning of the same day in a 2-points calibration using conductivity standard solutions (Merck Potassium chloride solution147 $\mu$S/cm & 1410 $\mu$S/cm).

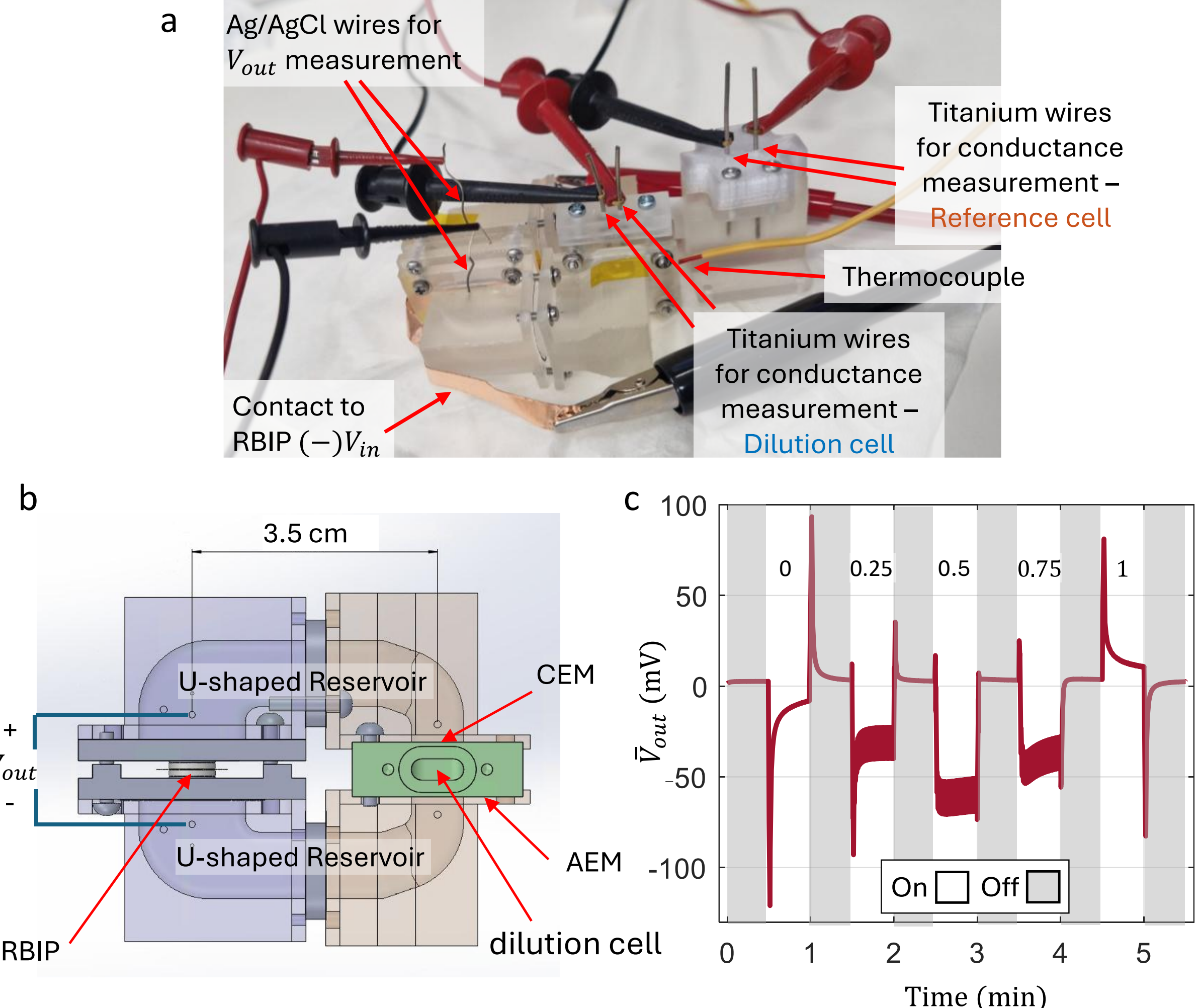

*Figure S20: (a) A photograph of the electrodialysis setup, showing the external connections and probes (the system was placed inside an incubator for control of ambient temperature). (b) CAD model of electrodialysis setup. The length of the U-shaped reservoirs is 3.5 cm, and the effective cross section area is ~1 cm². The electrolyte volume in the dilution cell is 0.63 cm³. (c) Voltage output, $\bar{V}_{out}(t)$, measured between two Ag/AgCl wires. The RBIP is a Au-Pt sample with physical properties as described in the methods section. The input signal is a square wave with $V_a$ =1 V, a frequency of 11 Hz, and the electrolyte is 1 mM KCl aqueous solution. The time periods in which the ratchet operated are the bright areas and the shaded areas are time periods in which the*

*input voltage was set to 0 V. The duty cycle is noted in each 'On' period. Data was recorded with sampling time of $2 \cdot 10^{-4}$ s and averaged over 6000 data points (moving average).*

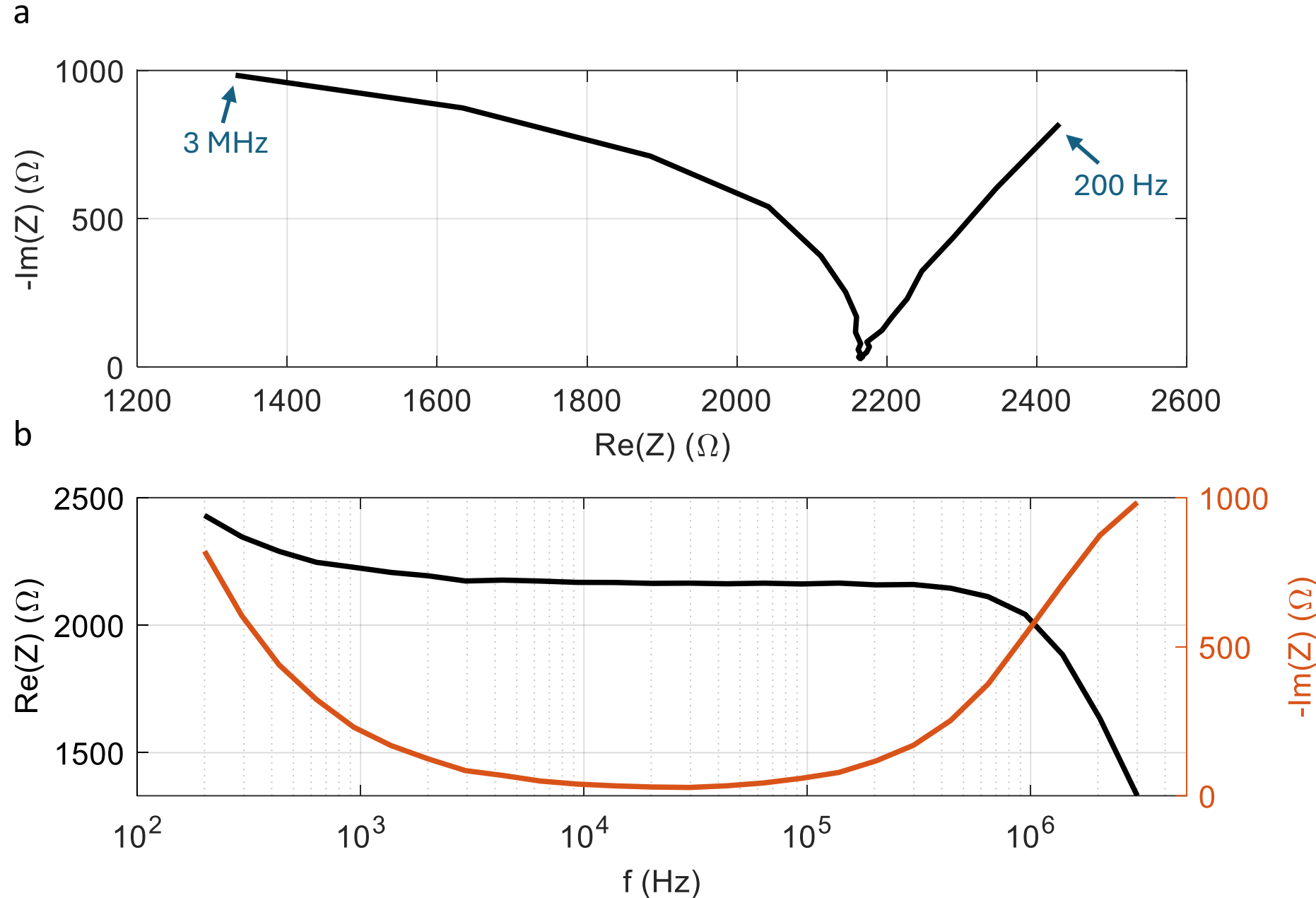

*Figure S21: Electrochemical Impedance Spectroscopy (EIS) results, measured in a 2-electrodes setup between titanium wires in the dilution cell with an amplitude of 10 mV. (a) Nyquist plot showing the negative imaginary impedance, $-Im(Z)$, versus the real part of the impedance, $Re(Z)$. (b) Bode plot showing $-Im(Z)$ and $Re(Z)$ as a function of frequency. The electrolyte is 1 mM KCl aqueous solution, which had a conductivity of $162\ \mu S/cm$ (@25℃).*

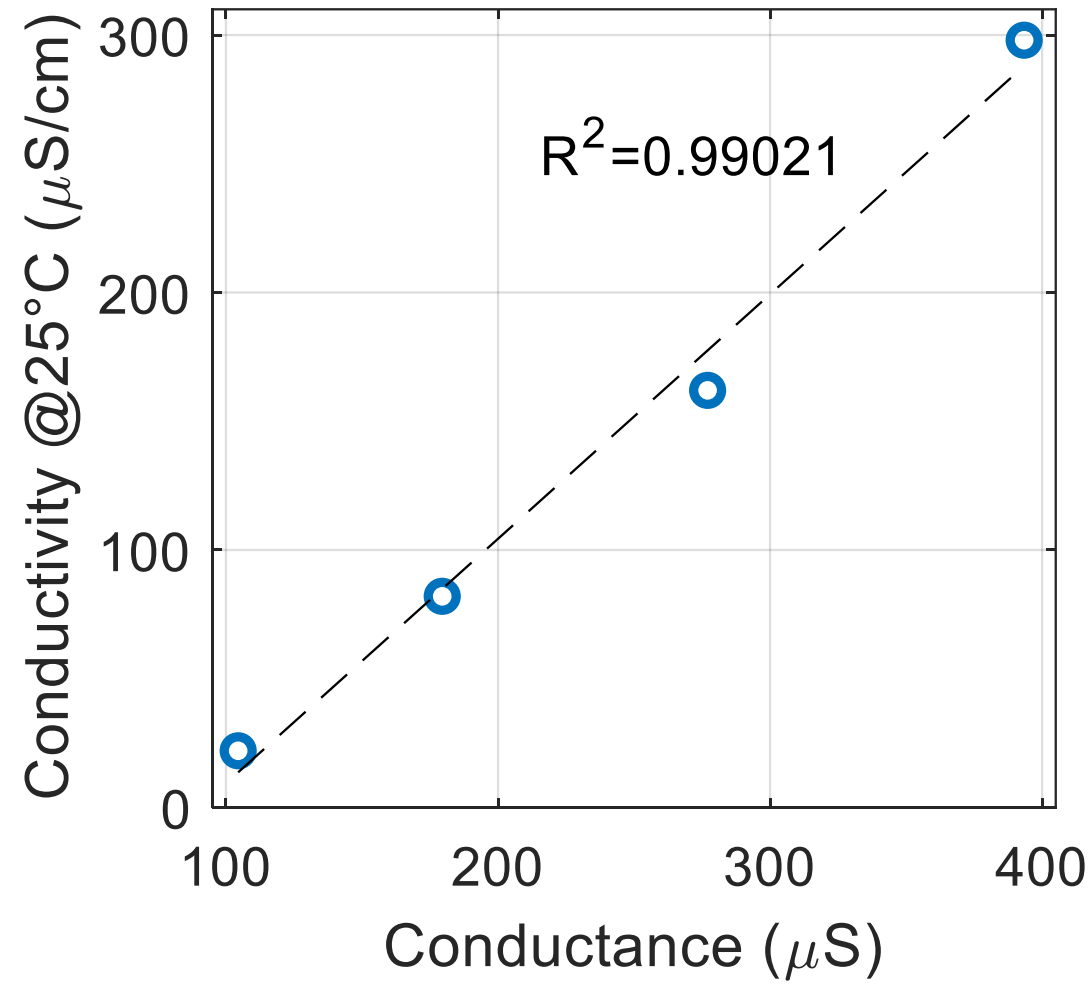

*Figure S22: Conductance calibration curve. Conductivity (@25℃) of pre-prepared electrolyte (KCl aqueous solution) as a function of the conductance measured within the dilution cell. The conductance was measured once per minute over a 10-minute period, and the reported values represent the average across all measurements during that time.*

*Raw data examples of electrodialysis experiment*

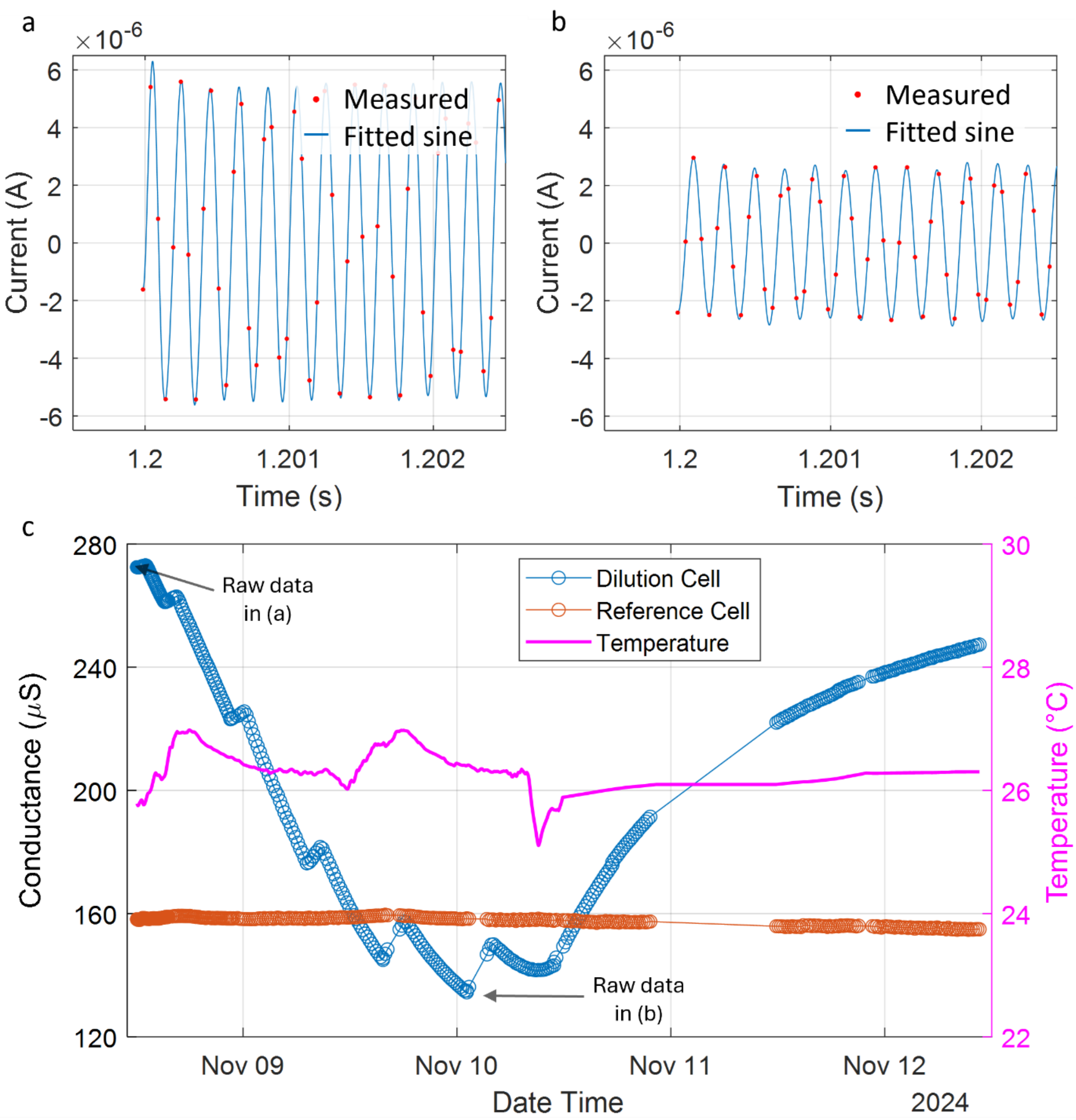

*Figure S23: Raw data of the electrodialysis deionization experiment. (a)-(b) Measured current between titanium wires in the dilution cell at $t = 0$ (a), and at the minimum conductance point (b). Red dots represent the measured data, and the solid blue curve is the sine wave fitting. (c) The conductance measured in the dilution and reference cells (left y-axis), and temperature (right y-axis), as a function of time. The temperature curve is a moving average of N=1000 data points. The arrows indicate the times where the measurements in a and b were taken.*

## 9. Two-compartment ratchet driven deionization

### *Methods and results*

For a significant concentration gradient to develop, charge neutrality must be maintained within the two compartments. This can be achieved by driving the ratchet while two Ag/AgCl auxiliary electrodes are shunted across the RBIP to provide a low-impedance pathway for electronic current to flow and an ample capacity of solid AgCl. Figure S24a shows a schematic illustration of such a system. By removing $Cl^-$ from the solution through oxidation at one auxiliary electrode and generating $Cl^-$ on the other, the shorted auxiliary electrodes assure that charge neutrality is maintained regardless of differences in ratchet-induced chloride and cation currents. Thus, when cations are pumped by the RBIP into one compartment, chlorides are generated there to compensate for any additional positive charge, leading to an increase in electrolyte concentration. At the same time, since cations leave the other compartment and chlorides are removed through the auxiliary electrode, the electrolyte concentration in this compartment is reduced. As the cation concentration builds in one compartment and decreases in the other, the ability to net pump cations decreases due to back diffusion of cations. After a sufficiently long time, the force exerted by the ratchet will be equal to the force induced by the opposing concentration gradient and cations will no longer be transported through the RBIP. At this stage $\bar{I}_{aux}$ will be at a magnitude that supports the ratchet induced current and back diffusion without violating electroneutrality in either compartment.

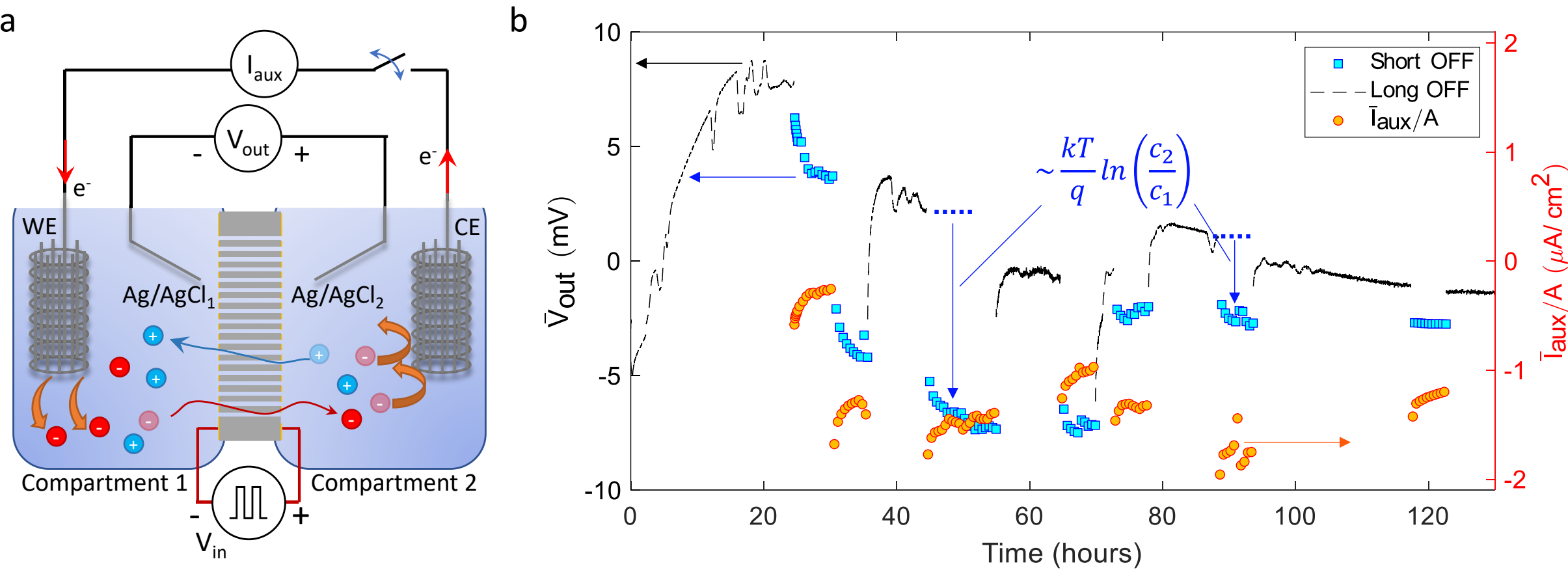

*Figure S24: Two-compartment deionization demonstration. (a) Schematic illustration of an electrochemical cell used to demonstrate electrolyte deionization. (b) The Ag/AgCl wires time-averaged voltage, $\bar{V}_{out}$. Blue markers are measurements taken while the ratchet was OFF for short intervals between periods of ratchet operation, and the black dashed lines are measurements taken while the system is continuously at rest. The orange markers are the average current density measured between the Ag/AgCl auxiliary electrodes during every ratchet ON period. The input signal is a rectangular wave with a duty cycle of 0.5, frequency of 125 Hz and an amplitude, $V_a$, of 0.3 V. The solution is 1 mM HCl aqueous solution, the RBIP pore diameter is 40 nm, the active area of the membrane is A = 0.32 $cm^2$.*

RBIP-driven electrolyte deionization was demonstrated with a setup as illustrated in Figure S24a. Two Ag/AgCl wire meshes were used as auxiliary electrodes, and a set of Ag/AgCl wires was used to measure the chloride electrochemical potential difference between the two compartments. The RBIP sample had the same parameters as the one discussed in Figure 2a-b and the input signal was a square wave with duty cycle of 0.5, frequency of 125 Hz and an amplitude, $V_a$, of 0.3 V. Output characterization measurements for this sample are shown in Figure S25. For the first 24 hours the system was at rest, allowing it to equilibrate. Then, to demonstrate deionization, the RBIP was operated continuously with the auxiliary electrodes shunted, thus concentrating the solution in one compartment, and diluting the solution in the other compartment as described above. Next, the ratchet was turned OFF (i.e., $V_{in}$ = 0 V) for short intervals of 30 or 90 seconds and $V_{out}$ was measured. The auxiliary electrodes current, $I_{aux}$, was not measured while the ratchet was OFF, thus the auxiliary electrodes were not shunted during these intervals (illustrated by the opening of the switch in Figure S24a). After several cycles of long operation and short ratchet OFF intervals, the ratchet was turned OFF for several hours and the system was allowed to equilibrate. This entire process was repeated multiple times as described below. The duration of the first set of ON cycles was 5 min after which the ratchet was turned OFF ($V_{in}$ = 0 V) for 30 s. In later sets, the ratchet ON duration was 30 min and the duration of the OFF interval was 90 s. To reduce noise, in the long duration ratchet OFF period (black curve in Figure S24b) every 20 data points were averaged together. $V_{out}$ was measured during every ratchet OFF interval and the auxiliary electrodes were not shunted during this time. Figure S24b shows the measured voltage for six periods of operation over the course of 5 days. Figure S26 shows in more detail the first three operation periods. The blue markers in Figure S24b mark the average voltage within each short OFF interval. The dashed black curves are the voltages measured during the long OFF durations. The orange markers are the time-averaged current density, $\bar{I}_{aux}/A$, measured while the ratchet is ON (normalized by the RBIP active area). The slow drift in voltage observed while the system is at rest (dashed black curves) can be a result of slow changes in the Ag/AgCl wires or changes in the AAO wafer pore charge. When the cell was allowed to equilibrate with $V_{\mathrm{in}}$ = 0 V over the course of at least one day, it attained an exceptionally stable and low voltage near 0 V. However, following ratchet operation, a non-negligible and consistent offset of $\bar{V}_{out}$ is observed demonstrating electrolyte deionization. The short time interval in which the ratchet was OFF and $\mathrm{V}_{out}$ was measured is sufficient for discharging the double layers, but too short to allow for any meaningful back diffusion of ions. Thus, the voltage offset is a result of the chemical potential difference of chlorides between the two compartments and the electrostatic potential difference induced by the permselectivity of the membrane.[10] Examples for this offset, which is approximately $kT/qln(c_2/c_1)$, are marked with the vertical arrows in Figure S24b ($k$, $T$, and $q$ being the Boltzmann constant, the temperature, and the elementary charge, respectively, and $c_1$ and $c_2$ are respectively the $Cl^-$ concentration in compartments 1 and 2). The cathodic current induced by the ratchet results in reduction of AgCl(s) to generate aqueous $Cl^-$ in compartment 1 where the working electrode is located (WE in Figure S24a), and a decrease in the aqueous $Cl^-$ concentration in compartment 2 where $Cl^-$ is consumed by the counter electrode (CE in Figure S24a) via oxidation of Ag(s) to form AgCl(s). As a result, the developed chemical potential difference, $kT/qln(c_2/c_1)$, is negative as shown in Figure S24b. The correlation between the concentration ratio and the measured voltage was calibrated by measuring the resting voltage between the two compartments with various concentration ratios (Figure S27). The maximal voltage offset induced by the ratchet (relative to adjacent resting periods) was -7 mV. Using the calibration curve, this translates to a maximal concentration ratio of about 0.78 between the two compartments. Over the course of the week, the output of the ratchet was reduced due to sample degradation. Further discussion of device degradation and other failure modes can be found in sections 11-12 in the supporting information.

Figure S25a,b show $\bar{V}_{out}$ and $\bar{I}_{out}$, respectively, as a function of duty cycle for the sample used for the deionization experiment (see Figure S24).

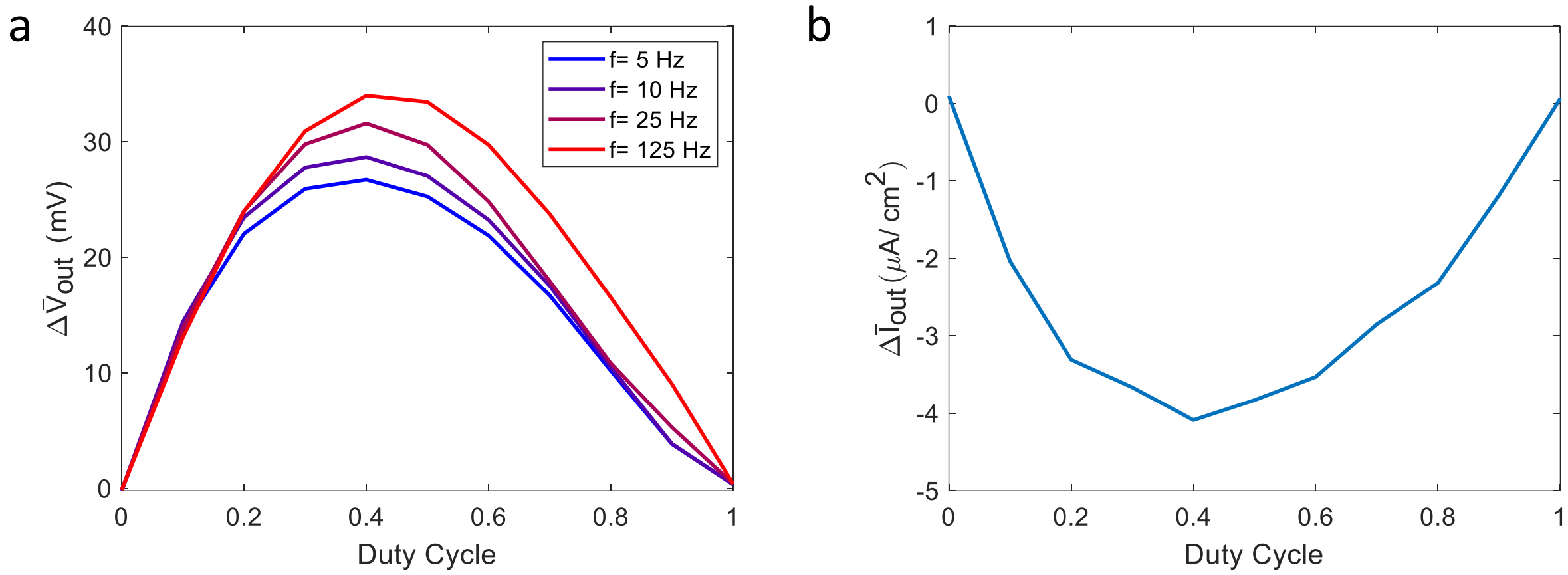

*Figure S25: $\Delta\bar{V}_{out}$ (a) and $\Delta\bar{I}_{out}$ (b) as a function of duty cycle for the sample used for the deionization experiment discussed in Figure S24. The input signal amplitude is $V_a$ =0.3 V, and for (b), the input signal frequency is 125 Hz. The electrolyte is 1 mM HCl aqueous solution.*

Figure S26a-c shows the same measurements as in Figure S24b focusing on the first three periods in which deionization was demonstrated. The grey shaded areas are times when the ratchet was turned OFF, and the voltage $V_{out}$ was measured. The dashed black curves (long OFF) are the voltages measured while the system was at rest for long durations, and the colored lines (short OFF) are measurements taken in the brief durations between ratchet ON periods. The markers ($\bar{V}_{out}$) are the temporal average of $V_{out}$ in the short OFF intervals. The white areas are the times when the ratchet was ON, and the current between the auxiliary electrodes was measured (red curves). Figure S26d-f shows the voltages measured during the short ratchet OFF periods in (a-c) respectively. Each curve in Figure S26d-f is an expanded view of the voltage measurement shown in Figure S26a-c with the same color. The input signal is a rectangular wave with a duty cycle of 0.5, frequency of 125 Hz, and an amplitude, $V_a$, of 0.3 V. The electrolyte is 1 mM HCl aqueous solution, the RBIP pore diameter is 40 nm, the active area of the membrane is A = 0.32 cm$^2$. The red curves in Figure S26a-c are the current densities measured between the auxiliary electrodes (normalized by the ratchet active area). Every data point is obtained by averaging the current over 1 s. The current response can be described in terms of two transient components. The first component is a fast current drop which decays after 5 min. This is the response for pumping ions that back-diffused while the ratchet was turned OFF during voltage measurements. After this initial response, the current magnitude decreases gradually. This slow decrease in current magnitude may be a result of the build-up of a cation concentration gradient as described in the main text. This assumption is supported by measured voltage which saturates in a similar manner.

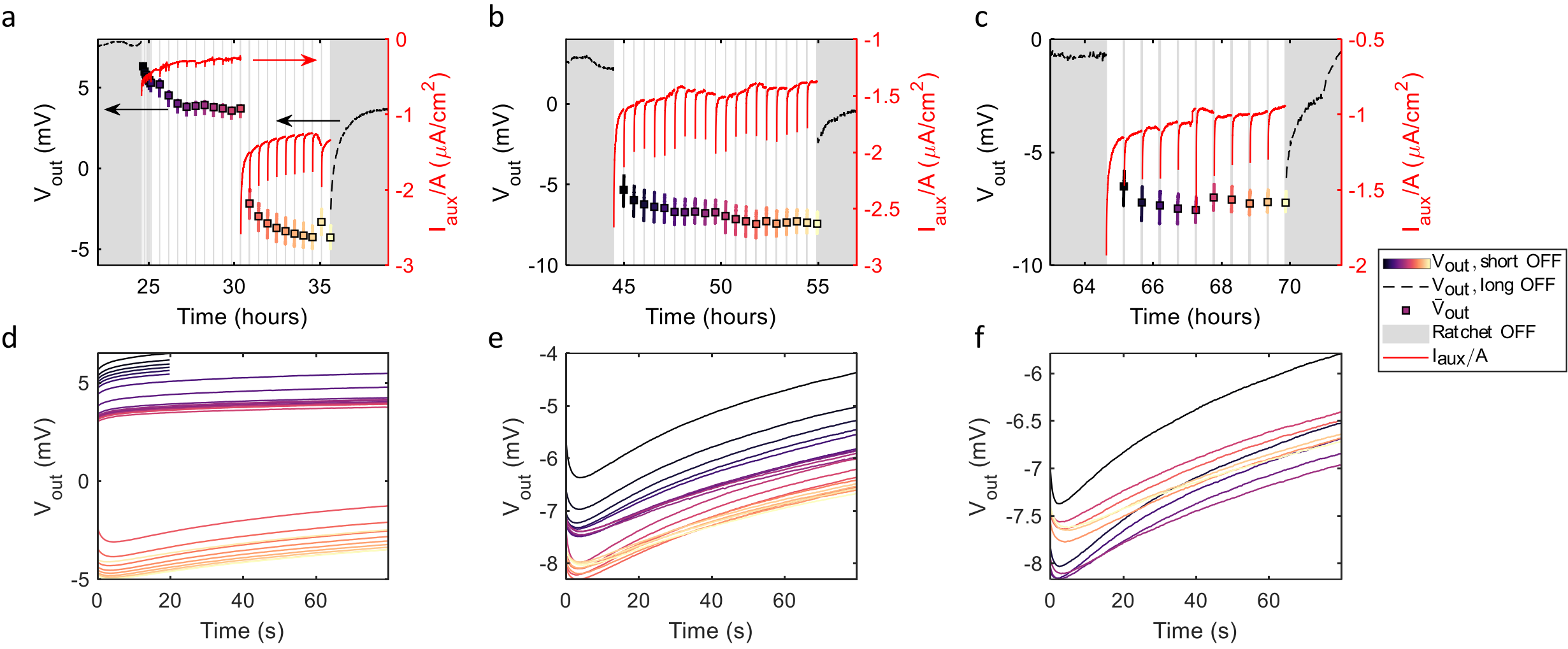

*Figure S26: (a-c) Same measurements as in Figure S24b focusing on specific times. The grey shaded areas are times when the ratchet was turned OFF, and the voltage $V_{out}$ was measured. The black curves ($V_{out}$, long off) are the voltages measured during the prolonged ratchet OFF periods, and the colored lines ($V_{out}$, short off) are voltages that were measured during the short OFF intervals. The markers are $\bar{V}_{out}$, the temporal average of $V_{out}$, short off. The red curves are the currents measured between the shorted Ag/AgCl auxiliary electrodes normalized by the RBIP active area. (d-f) Voltages measured during the short ratchet OFF intervals in (a-c) respectively. Each measured curve in (d-f) is an expanded view of the voltage measurement in (a-c) with the same color. The input signal is a rectangular wave with a duty cycle of 0.5, frequency of 125 Hz, and an amplitude, $V_a$, of 0.3 V. The electrolyte is 1 mM HCl aqueous solution, the RBIP pore diameter is 40 nm, the active area of the membrane is A = 0.32 cm$^2$.*

### Calibration of $V_{out}$ response from aqueous HCl concentration ratios across an RBIP

The correlation between the aqueous HCl concentration differences and the Ag/AgCl voltage was obtained by measuring the voltage between two Ag/AgCl wires placed in a cell filled with two solutions with predefined concentrations. An annealed AAO wafer with a pore diameter of 40 nm, and without deposited contacts, was used to separate the two cell compartments. Figure S27 shows the measured voltage as a function of the concentration ratio between the cells. However, it should be noted that variability between samples, the effect of the contacts, and changes in the pore surface charge due to the long exposure to acidic conditions, may lead to an inaccuracy in the concentration difference estimation.

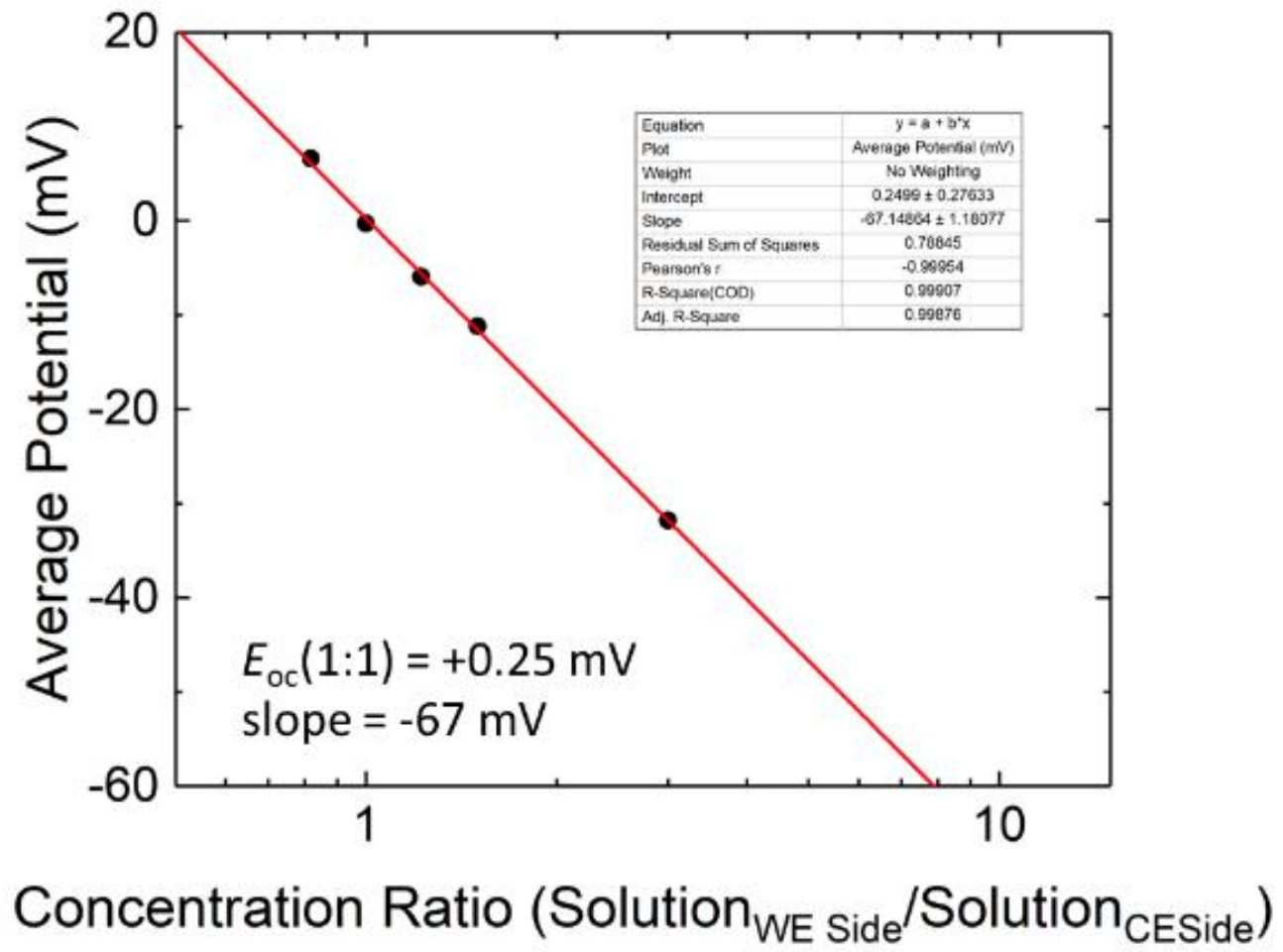

*Figure S27: Measured voltage between two Ag/AgCl wires across an annealed AAO wafer with no contacts deposited, when a concentration gradient was intentionally introduced between the two compartments. The reference concentration is aqueous 1 mM HCl.*

## 10. Comparison with electronic flashing ratchets

The RBIP described here is similar in function to flashing ratchets in the sense that particle transport is driven by internal potential fluctuations[11,12] and not by a voltage that is applied to external electrodes (as in rectifying ionic diodes[13,14] and rocking ratchets[11,12]). Thus, it is interesting to compare the obtained output to those reported in previous demonstrations of flashing ratchets. The ratio of the ratchet input signal amplitude, $V_a$, to the magnitude of the average output voltage, $\bar{V}_{out}$, or current, $\bar{I}_{out}$, are key parameters used to quantify the efficiency of a ratcheting process. In many of the previous demonstrations of flashing ratchets (which are closest in function to the RBIP reported here, but drove electrons or holes in semiconductors),[15–17] $V_a$ is applied to electronic conductors that are insulated from the charge transport layer to avoid shunts. Since there is a significant electric potential drop across the insulation layers, a large input voltage amplitude is required. With the RBIP architecture, the input signal, $V_a$, is applied to the metal electrodes that are deposited on top of the AAO membrane and charge transport across the metal|solution interface is possible only by inducing redox reactions. Thus, charge transport across this interface is negligible, if no exogenous species are added and/or the amplitude of $V_a$ is kept small. As a result, there is no need to insulate the contacts between the ratchet and the media in which the output charge transport takes place, and thus lower input signals can be used. In the case of the RBIP discussed in Figure 1, a $TiO_2$ ALD layer is deposited on top of the contacts to protect it from degradation and to assure that no redox reactions take place at the contacts. Nevertheless, as demonstrated with the other tested samples, this insulation layer is not strictly necessary, since the RBIP can drive ion transport with no associated redox reactions even without this layer (Figures 2-3, and section 7 in the supporting information).

## 11. Sample degradation

As prepared samples have shown a pumping performance that is significantly higher than presented in Figure S14. However, this higher performance degraded in the first several hours of operation. Figure S28a shows the recorded output of the RBIP discussed in Figure S14 during its first 20 h of its operation. In this measurement, the as-prepared RBIP was operated continuously with an input signal as described in Figure

S14a, with an amplitude of 0.4 V, until its performance stabilized. Only once the performance has stabilized the measurements presented in Figure S14a-d where conducted. In many of the tested samples post-mortem analysis revealed no noticeable changes in morphology. In such cases, failure may be a result of changes in the surface charge distribution within the pores leading to a change in the electric potential distribution within the RBIP. In other samples structural changes were visible. Figure S28b shows an EDS map overlaid on top of an SEM image of an as-prepared RBIP in which the contacts were deposited with electron beam evaporation. As shown in Figure S28b, the as-prepared samples have a uniform coverage of gold which allows effective biasing of the pores. Samples prepared with other methods had a similar structure. Figure S28c-d shows respectively an SEM image and an EDS map overlaid on top of the SEM image of an RBIP that was operated for about 24 h. As can be seen in Figure S28c-d, clustering of gold forms islands that block some of the pores but also prevents effective biasing of the pores that are not blocked thus leading to a significant reduction in the pumping performance. In other samples, failure was a result of contact delamination from the AAO substrate. Another degradation mechanism is dissolution.[18] To test for sample dissolution, a sample was placed in the electrochemical cell and was operated for 12 hours and then kept in the solution for another 25 hours. ICP-MS measurements of the solution show the presence of aluminum and gold at a concentration of about 35 ng/mL in both compartments. Annealing the wafers at higher temperatures for longer periods may improve the wafer crystallinity and thus reduce its dissolution. An optimization of the annealing process for improved device stability is left for future work.

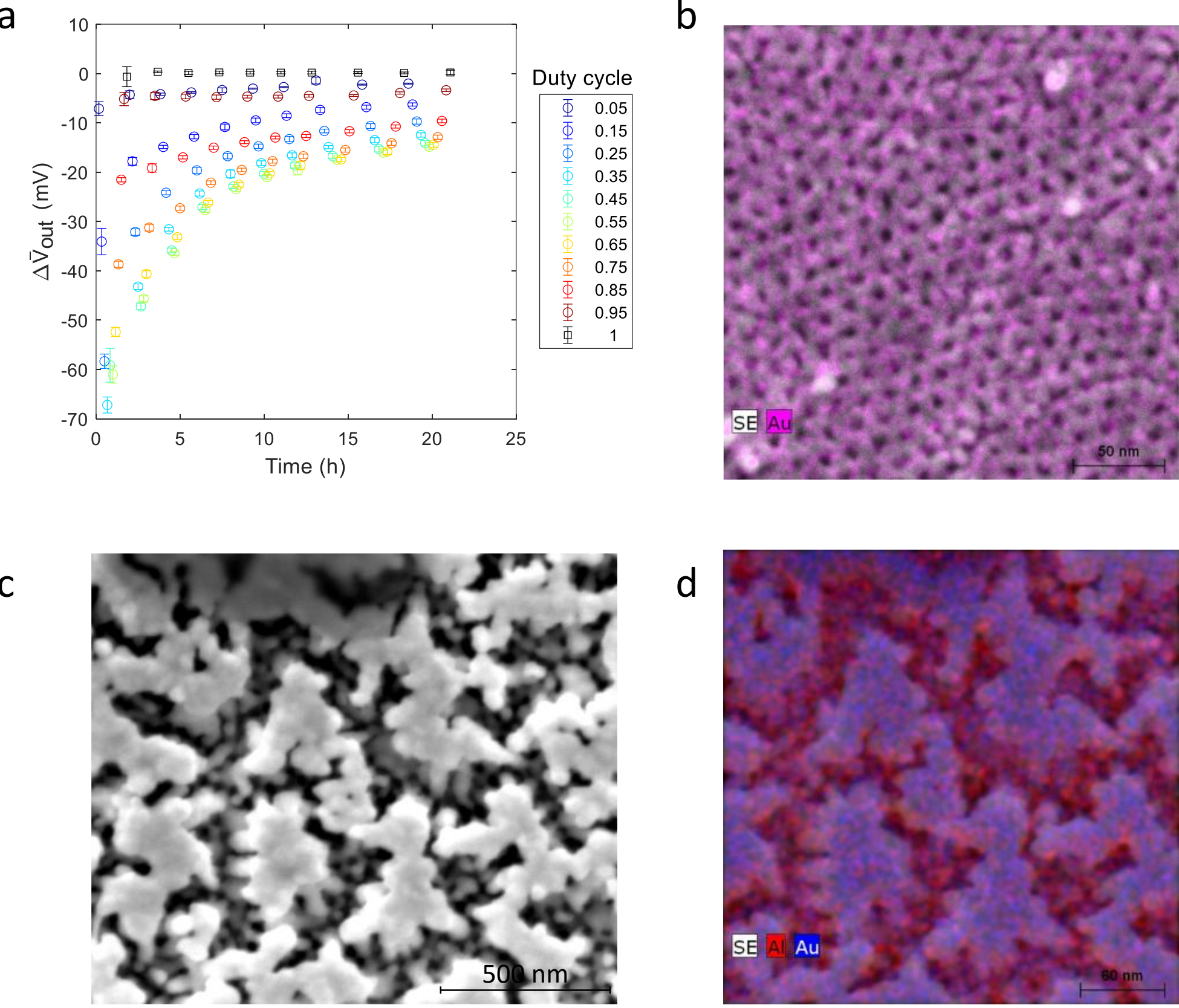

*Figure S28: Stability analysis of an RBIP. (a) Ratchet output of an as-deposited sample cycled between different duty cycles for 21 h. (b) SEM and EDS overlay of an as-deposited sample with 20 nm pores. (c-d) SEM image and EDS of a sample after 25 h of operation.*

## 12. Failure modes

The trivial failure modes that can affect the RBIP performance are the shunting and blocking of pores. Understanding the effect of these failure modes can help shed light on the contributions to the performance of different devices. When entire pores, or parts of the pores, are shunted, the electric potential distribution within the shunted regions is unaffected by the ratchet signal. As a result, the shunted regions provide a path for ions to diffuse back toward their equilibrium distribution, and thus limiting the RBIP pumping performance. When the salinity level is too high, such that the Debye length is significantly smaller than the pore radius, the center of the pore will not be perturbed by the ratchet signal thus forming a shunt. Another cause for pore shunting is improper biasing of the pore, for example when the deposited metal layer does not cover the entire surface of the AAO wafer.

When pores are blocked, or when the resistance to transport through them is too high, the device can be effectively described as two separate compartments where the voltage difference between the two compartments is determined by the ratchet signal. In such case, the time averaged voltage output, $\Delta\bar{V}_{out}$ is simply the average voltage applied to the RBIP, and thus it is linear with the duty cycle. Furthermore, if the pores are almost entirely blocked, the output is less affected by changes in the solution conductivity. Figure S29a-b shows an example of the output of an RBIP with blocked pores. The RBIP pore diameter is 20 nm and the deposited metal layers are 40 nm of titanium (adhesion layer) and 40 nm of gold (planar equivalent). Figure S29a shows the measured $V_{out}$ for an input signal with $V_a$ of 0.2 V, a frequency of 100 Hz, and duty cycles between 5% and 95%. All the measurement parameters are as in Figure S14a. The aqueous solution is 1 mM KCl and 10 mM KCl. Figure S29b shows the extracted $\Delta\bar{V}_{out}$ as a function of the input signal duty cycle. The linear relation between the output and the duty cycle, and the small change of output with the salinity indicates that transport through the pores is too resistive or that the pores are blocked. Similarly, when an alumina membrane was used that did not intentionally have pores in it, $\Delta\bar{V}_{out}$ values were largest for 0% and 100% duty cycle, and zero for 50% duty cycle, further supporting a ratchet-based mechanism for our RBIPs.

Clearly, some devices may have regions that operate well, regions that are fully or partially shunted and other regions that are fully or partially blocked. In such case, the measured signal will be a convolution of the ratchet output, which is zero at duty cycles of 0% and 100% but is nonzero elsewhere; the contribution of shunted regions, which is zero for all duty cycles; and the contribution of highly resistive or blocked regions, which is linear with the duty cycle. Thus, partially blocked areas, or high resistance for ion transport within the pore, may be the reason for the non-zero $\Delta\bar{V}_{out}$ obtained in RBIPs with pore diameter of 20 nm. Figure S30 shows another example for $\Delta\bar{V}_{out}$ as a function of the duty cycle for an RBIP where the pores are partially blocked. The sample and input signal parameters are as in Figures 2a-b and the aqueous solution is 1mM HCl. The output curve is superimposed on a linear trend which is indicative of partial blockage of the pores.

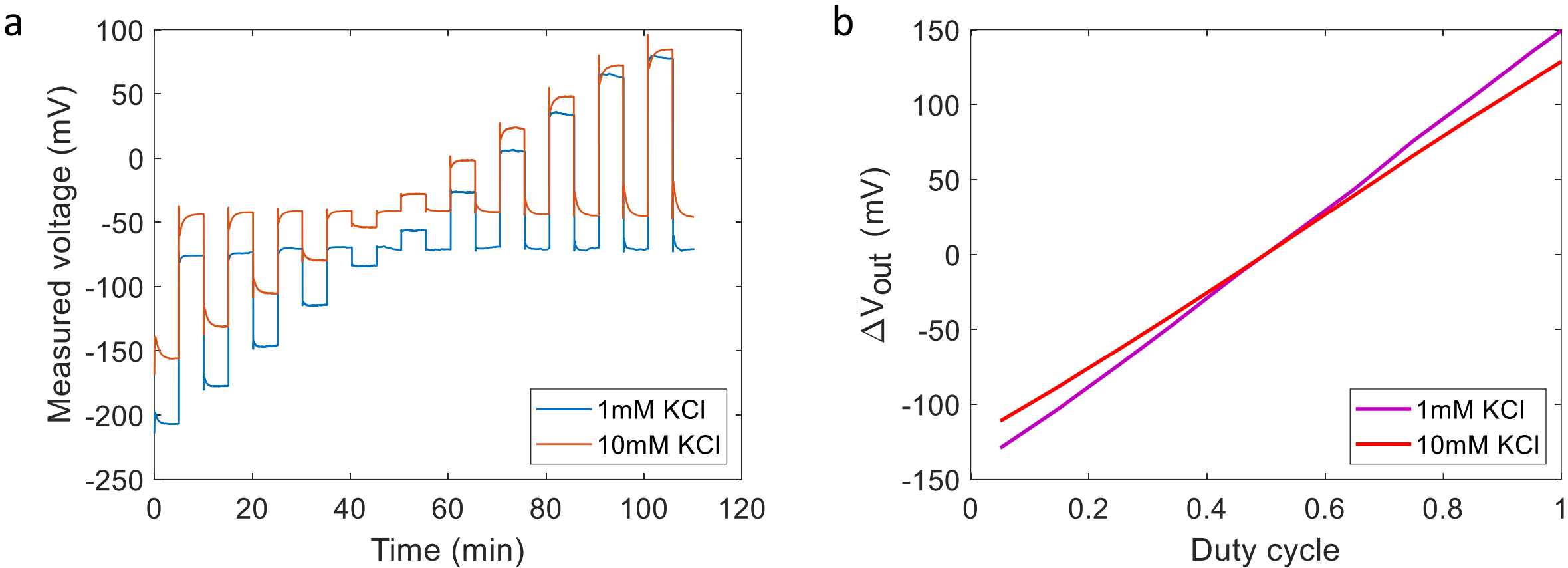

*Figure S29: Ratchet output for an RBIP with partially blocked 20 nm pores. (a) Measured output voltage for an input signal with various duty cycles. The input signal is a square wave with $V_a$ of 0.2 V and the frequency is 100 Hz. (b) Time averaged output voltage, $\Delta\bar{V}_{out}$, as a function of the input signal duty cycle.*

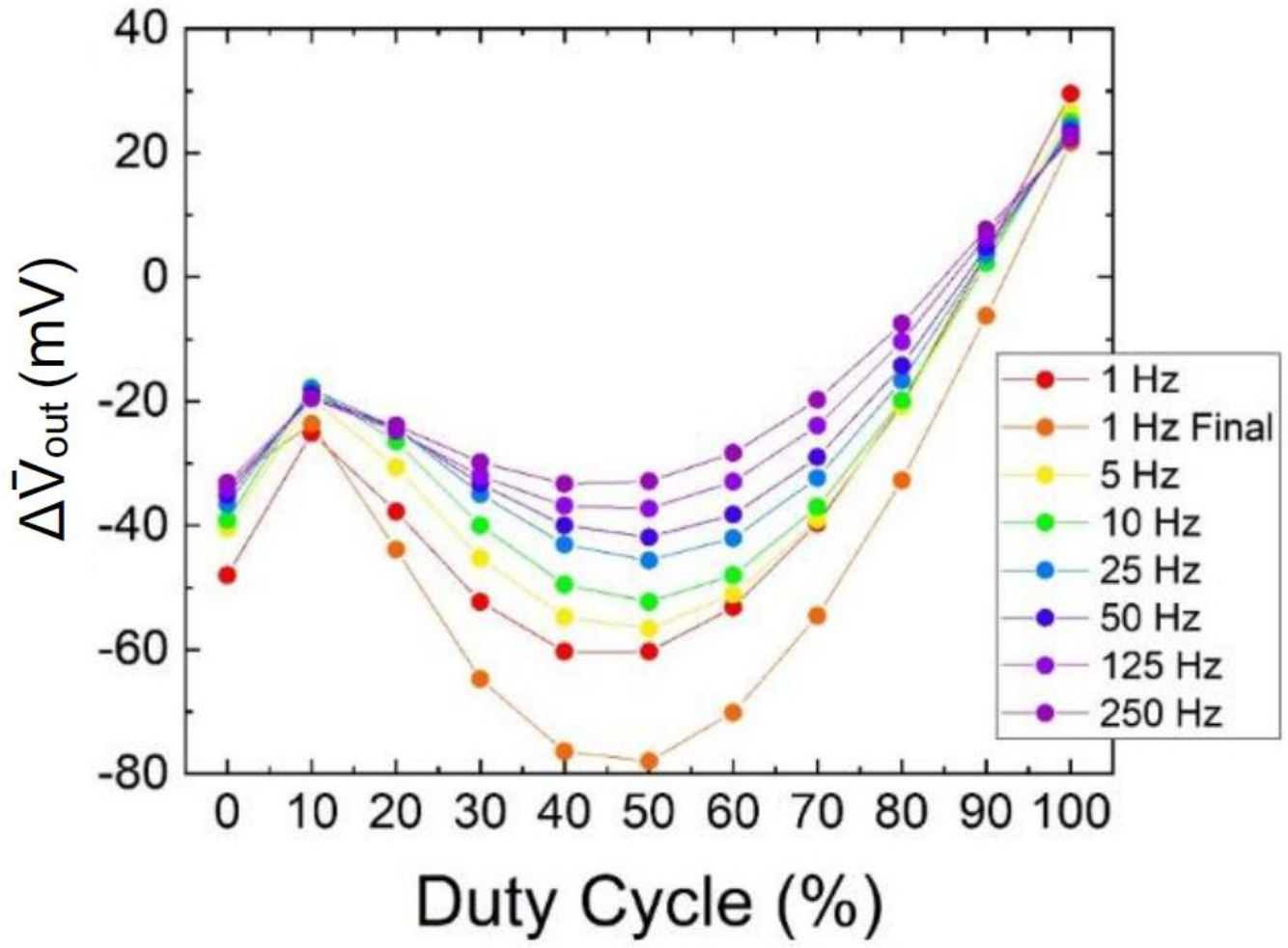

*Figure S30: Output voltage as a function of the duty cycle for an RBIP with partially blocked pores. The sample and input signal parameters are as in Figures 2a-b and the electrolyte is 1 mM HCl aqueous solution.*

## 13. RBIP Analytical Model

To estimate the ratchet-induced voltage between the two Ag/AgCl wires we assume that the voltages between the RBIP contacts and the Ag/AgCl wires next to them ( $V_L$ and $V_R$) can be modeled as charging and discharging capacitors. Figure S31a shows an illustration of the system and an equivalent circuit describing its operation.

The resting potential difference between each of the ratchet contacts and the wire next to it are $V_{R,eq}$ and $V_{L,eq}$, where subscript L,R denote left and right in Figure 1a and Figure S31a. In equilibrium $V_L = V_{L,eq}$ and $V_R = V_{R,eq}$. Since the ratchet is floating with respect to the Ag/AgCl wires, when a bias $V_a$ is applied, one metal contact is charged, and the other is discharged. However, the bias is not necessarily shared equally between the two contacts. We assign the parameter $a$ to describe this source of asymmetry . If the input signal has a very long temporal period compared to the time constants for charging and discharging, the voltages in the system will reach their steady state values. For the first part of the period, the steady state voltages $V_{Lf}$ and $V_{Rf}$ are:

$$V_L = V_{Lf,1} = V_{L,eq} - aV_a; \quad V_R = V_{Rf,1} = V_{R,eq} + (1-a)V_a. \tag{S5}$$

We use subscript 1 to denote the first part of the temporal period ($t < d_c T$) where $d_c$ is the duty cycle and $T$ is the signal period. The second part of the period ($t > d_c T$), is noted with subscript 2. In this case the steady state voltages follow:

$$V_L = V_{Lf,2} = V_{L,eq} + aV_a; \quad V_R = V_{Rf,2} = V_{R,eq} - (1-a)V_a. \tag{S6}$$

According to the definitions in Figure 1a, the output voltage, $V_{out}$, follows:

$$V_{out} = V_L + V_{in} - V_R. \tag{S7}$$

In steady state conditions for a positive voltage applied on the ratchet:

$$V_{out} = V_{L,eq} - aV_a + V_a - V_{R,eq} - (1-a)V_a = V_{L,eq} - V_{R,eq} = \Delta V_{eq}. \tag{S8}$$

In steady state conditions for a negative voltage applied on the ratchet:

$$V_{out} = V_{L,eq} + aV_a - V_a - V_{R,eq} + (1-a)V_a = V_{L,eq} - V_{R,eq} = \Delta V_{eq}. \tag{S9}$$

Thus, in steady state conditions, the application of a constant bias does not drive the device away from its equilibrium voltage. Furthermore, in a perfectly symmetric system, or if the equilibrium voltages of the two surfaces are the same, the steady state output voltage will be 0. The asymmetry coefficient can be obtained from equations (S5) and (S6):

$$a = \frac{1}{2} + \frac{V_{Lf,1} + V_{Rf,1} - V_{Lf,2} - V_{Rf,2}}{4V_a} \tag{S10}$$

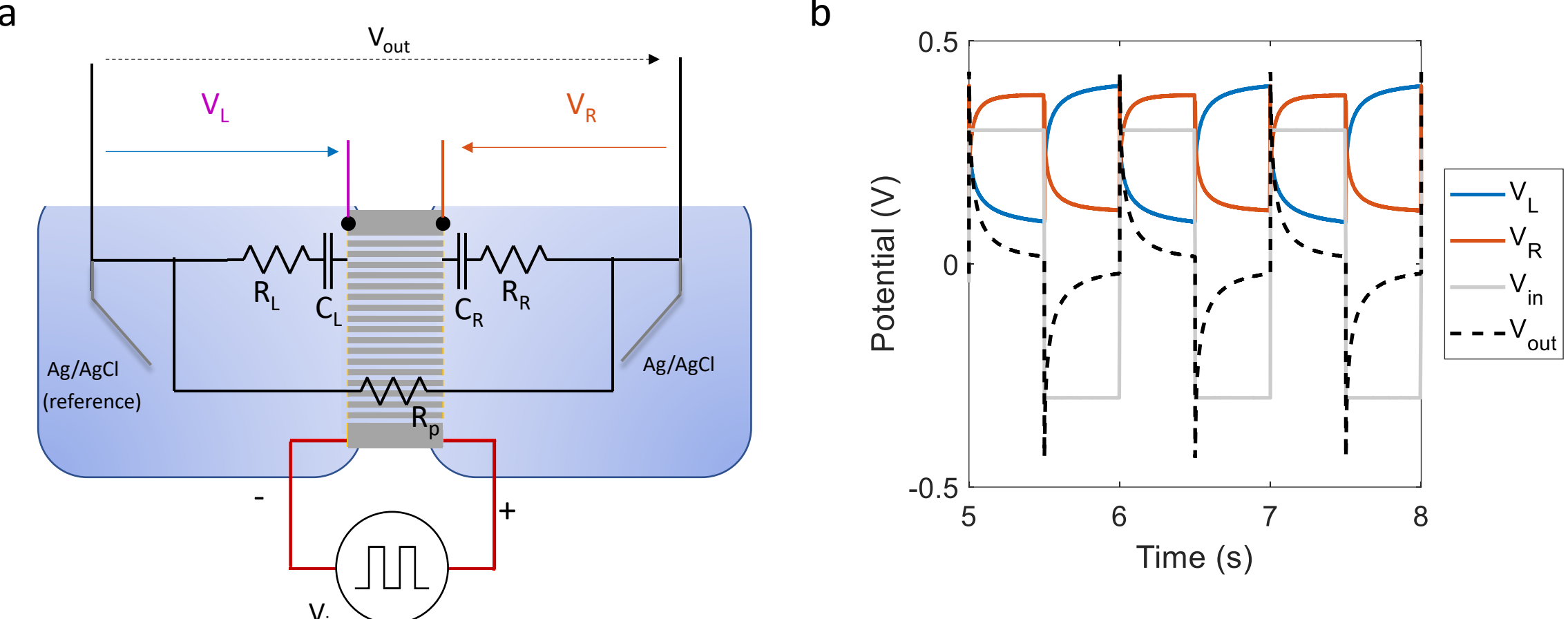

*Figure S31: (a) Illustration of the ratchet setup with all the measured signals and an equivalent circuit. (b) Measured voltages during ratchet operation. The frequency is 1 Hz, the duty cycle is 0.5 and the amplitude is $V_a$ =0.3 V. The RBIP was prepared as the one described in Figures 2a,b and the electrolyte is 1 mM KCl aqueous solution.*

Figure S31b shows an example for signals measured in response to an input signal with a 0.5 duty cycle, a frequency of 1 Hz, and an amplitude of $V_a$ = 0.3 V. The RBIP was prepared as the one discussed in Figures 2a,b, and the electrolyte is 1 mM KCl aqueous solution. The voltage signals $V_L$ and $V_R$ can be approximately described by a single exponent charging function defined by their initial voltage $V_i$, its time constant, $\tau$, and the steady state voltage, $V_f$:

$$V(V_i, V_f, \tau, t) = V_f + (V_i - V_f) \exp\left(-\frac{t}{\tau}\right). \tag{S11}$$

In non-linear systems such as electric double layers, the time constant depends on potential, frequency, and duty cycle. The acceptable exponential fit to our experimental data (Figure 1, Figure S2, Figure S4, Figure S6 ) suggest that that for our RBIPs, this non-linearity can be described through the different time constants for charging and discharging the contacts. Thus, for a given frequency and duty cycle the transients will be determined by two time constants, $\tau_1(d_c, f)$ and $\tau_2(d_c, f)$.

In the first part of every period, the input voltage is $V_a$. Thus, $V_L$ discharges towards $V_{Lf,1}$ with a time constant of $\tau_{L,1}$:

$$\begin{aligned} V_{L,1}(t) &= V_{Lf,1} + (V_{Li,1} - V_{Lf,1}) \exp\left(-\frac{t}{\tau_{L,1}}\right) \\ &= V_{L,eq} - aV_a + (V_{Li,1} - V_{L,eq} + aV_a) \exp\left(-\frac{t}{\tau_{L,1}}\right). \end{aligned} \tag{S12}$$

$V_R$ charges towards $V_{Rf,1}$:

$$\begin{aligned} V_{R,1}(t) &= V_{Rf,1} + (V_{Ri,1} - V_{Rf,1}) \exp\left(-\frac{t}{\tau_{R,1}}\right) \\ &= V_{R,eq} + (1-a)V_a \\ &\quad + (V_{Ri,1} - V_{R,eq} - (1-a)V_a) \exp\left(-\frac{t}{\tau_{R,1}}\right). \end{aligned} \tag{S13}$$

$V_{Li,1}$ and $V_{Ri,1}$ are calculated assuming that the $V_R$ and $V_L$ are continuous, and that the system is operating periodically in steady state.

For the second part of every period, we get:

$$\begin{aligned} V_{L,2}(t) &= V_{Lf,2} + \left(V_{Li,2} - V_{Lf,2}\right) \exp\left(-\frac{t - d_c T}{\tau_{L,2}}\right) \\ &= V_{L,eq} + aV_a + \left(V_{Li,2} - V_{leq} - aV_a\right) \exp\left(-\frac{t - d_c T}{\tau_{L,2}}\right), \end{aligned} \tag{S14}$$

and:

$$\begin{aligned} V_{R,2}(t) &= V_{Rf,2} + \left(V_{Ri,2} - V_{rf,2}\right) \exp\left(-\frac{t - d_c T}{\tau_{R,2}}\right) \\ &= V_{R,eq} - (1-a)V_a \\ &+ \left(V_{Ri,1} - V_{Req} + (1-a)V_a\right) \exp\left(-\frac{t - d_c T}{\tau_{R,2}}\right). \end{aligned} \tag{S15}$$

Periodicity implies that:

$$\begin{aligned} V_{L,1}(t=0) &= V_{L,2}(t=T), \\ V_{L,1}(\,t = d_c T) &= V_{L,2}(t = d_c T). \end{aligned} \tag{S16}$$

Inserting and noting $A_L = \exp\left(-\frac{d_c T}{\tau_{L,1}}\right)$, $\mathrm{B}_L = \exp\left(-\frac{(1-d_c)T}{\tau_{L,2}}\right)$:

$$\begin{aligned} V_{Li,1} &= V_{Lf,2} + \left(V_{Li,2} - V_{Lf,2}\right) \exp\left(-\frac{T(1-d_c)}{\tau_{L,2}}\right) = V_{Lf,2} + \left(V_{Li,2} - V_{Lf,2}\right)\mathrm{B}_L, \\ V_{Li,2} &= V_{Lf,1} + \left(V_{Li,1} - V_{Lf,1}\right) \exp\left(-\frac{T d_c}{\tau_{L,1}}\right) = V_{Lf,1} + \left(V_{Li,1} - V_{Lf,1}\right)A_L. \end{aligned} \tag{S17}$$

Inserting one into the other and extracting $V_{Li,2}$:

$$V_{Li,2} = \frac{V_{Lf,1} + (V_{Lf,2} - V_{Lf,1})A_L - \mathrm{B}_L A_L V_{Lf,2}}{1 - \mathrm{B}_L A_L}. \tag{S18}$$

Similarly for $V_R$:

$$\begin{aligned} V_{R,1}(t=0) &= V_{R,2}(t=T\,), \\ V_{R,1}(\,t = d_c T) &= V_{R,2}(t = d_c T). \end{aligned} \tag{S19}$$

Noting $A_R = \exp\left(-\frac{d_c T}{\tau_{r,1}}\right)$, $B_R = \exp\left(-\frac{(1-d_c)T}{\tau_{r,2}}\right)$, we obtain:

$$V_{Ri,2} = \frac{V_{Rf,1} + \left(V_{Rf,2} - V_{Rf,1}\right)A_R - B_R A_R V_{Rf,2}}{1 - B_R A_R}. \tag{S20}$$

We can now find the output voltage at each stage:

$$V_{out} = V_L + V_{in} - V_R. \tag{S21}$$

Inserting:

$$
\begin{aligned}
V_{out,1} = V_{L,eq} - aV_a + \left(V_{Li,1} - V_{L,eq} + aV_a\right)\exp\left(-\frac{t}{\tau_{L,1}}\right) + V_a \\
- \Bigg[ V_{Req} + (1-a)V_a \\
+ \left(V_{Ri,1} - V_{R,eq} - (1-a)V_a\right)\exp\left(-\frac{t}{\tau_{R,1}}\right)\Bigg].
\end{aligned}
\tag{S22}
$$

Rearranging:

$$
\begin{aligned}
V_{out,1} &= \left(V_{L,eq} - V_{R,eq}\right) + \left(V_{Li,1} - V_{Leq} + aV_a\right)\exp\left(-\frac{t}{\tau_{L,1}}\right) \\
&\quad - \left(V_{Ri,1} - V_{Req} - (1-a)V_a\right)\exp\left(-\frac{t}{\tau_{R,1}}\right) \\
&= \Delta V_{eq} + \left(V_{Li,1} - V_{Lf,1}\right)\exp\left(-\frac{t}{\tau_{L,1}}\right) \\
&\quad - \left(V_{Ri,1} - V_{Rf,1}\right)\exp\left(-\frac{t}{\tau_{R,1}}\right).
\end{aligned}
\tag{S23}
$$

And for the second part of the period:

$$
\begin{aligned}
V_{out,2} = V_{L,eq} + aV_a + \left(V_{Li,1} - V_{Leq} - aV_a\right)\exp\left(-\frac{t-d_cT}{\tau_{L,2}}\right) - V_a \\
- \Bigg[ V_{R,eq} - (1-a)V_a \\
+ \left(V_{Ri,1} - V_{R,eq} + (1-a)V_a\right)\exp\left(-\frac{t-d_cT}{\tau_{R,2}}\right)\Bigg].
\end{aligned}
\tag{S24}
$$

Rearranging:

$$
\begin{aligned}
V_{out,2} &= V_{L,eq} - V_{R,eq} + \left(V_{Li,2} - V_{L,eq} - aV_a\right)\exp\left(-\frac{t-d_cT}{\tau_{L,2}}\right) \\
&\quad - \left[\left(V_{R,i,2} - V_{R,eq} + aV_a\right)\exp\left(-\frac{t-d_cT}{\tau_{R,2}}\right)\right] \\
&= \Delta V_{eq} + \left(V_{Li,2} - V_{Lf,2}\right)\exp\left(-\frac{t-d_cT}{\tau_{L,2}}\right) \\
&\quad - \left(V_{Ri,2} - V_{Rf,2}\right)\exp\left(-\frac{t-d_cT}{\tau_{R,2}}\right).
\end{aligned}
\tag{S25}
$$

From equations (S23) and (S25) we can see that the net voltage is a result of the difference between the charging and the discharging of the two surfaces over the course of one period.

The time averaged voltage is found by integrating $V_{out}$:

$$\bar{V}_{out} = \frac{1}{T}\left[\int_0^{d_cT} V_{out,1}dt + \int_{d_cT}^{T} V_{out,2}dt\right] =$$

$$= \Delta V_{eq} + \frac{1}{T}\left\{\tau_{L,1}\left(V_{Li,1} - V_{Lf,1}\right)\left[1 - \exp\left(-\frac{d_cT}{\tau_{L,1}}\right)\right] - \tau_{R,1}\left(V_{Ri,1} - V_{Rf,1}\right)\left[1 - \exp\left(-\frac{d_cT}{\tau_{R,1}}\right)\right] + \tau_{L,2}\left(V_{Li,2} - V_{Lf,2}\right)\left[1 - \exp\left(-\frac{(1-d_c)T}{\tau_{L,2}}\right)\right] - \tau_{R,2}\left(V_{Ri,2} - V_{Rf,2}\right)\left[1 - \exp\left(-\frac{(1-d_c)T}{\tau_{R,2}}\right)\right]\right\}. \quad (S26)$$

It can be easily shown that in a linear system where $\tau_{L,1} = \tau_{L,2}$ and $\tau_{R,1} = \tau_{R,2}$ the $\bar{V}_{out}$ is 0. However, if the capacitances are nonlinear such that the time constants are different, non-zero values for $\bar{V}_{out}$ can be obtained. If the time constants are only determined by the duration of the two portions of the input signal (as in Figure 1), then $\tau_{L,2} = \tau_{R,2}$ and $\tau_{L,1} = \tau_{R,1}$. In all the results presented here we have assumed that $\Delta V_{eq}$ is 0 V, and unless stated otherwise $a$ is 0.5. Figure S32 shows the calculated ratchet output as a function of the duty cycle for several input signal frequencies using the time constants shown in Figure 1h for a duty cycle of 0.5: $\tau_{L,1} = 6.59$ ms, $\tau_{R,1} = 7.23$ ms, $\tau_{L,2} = 8.5$ ms, and $\tau_{R,2} = 8.21$ ms. Even for such a small difference in time constants a significant output is obtained. However, since the model does not account for changes in the saturation voltage $V_f$, and the effective amplitude $V_i$-$V_f$ with the duty cycle, it cannot fully reproduce the measured outputs.

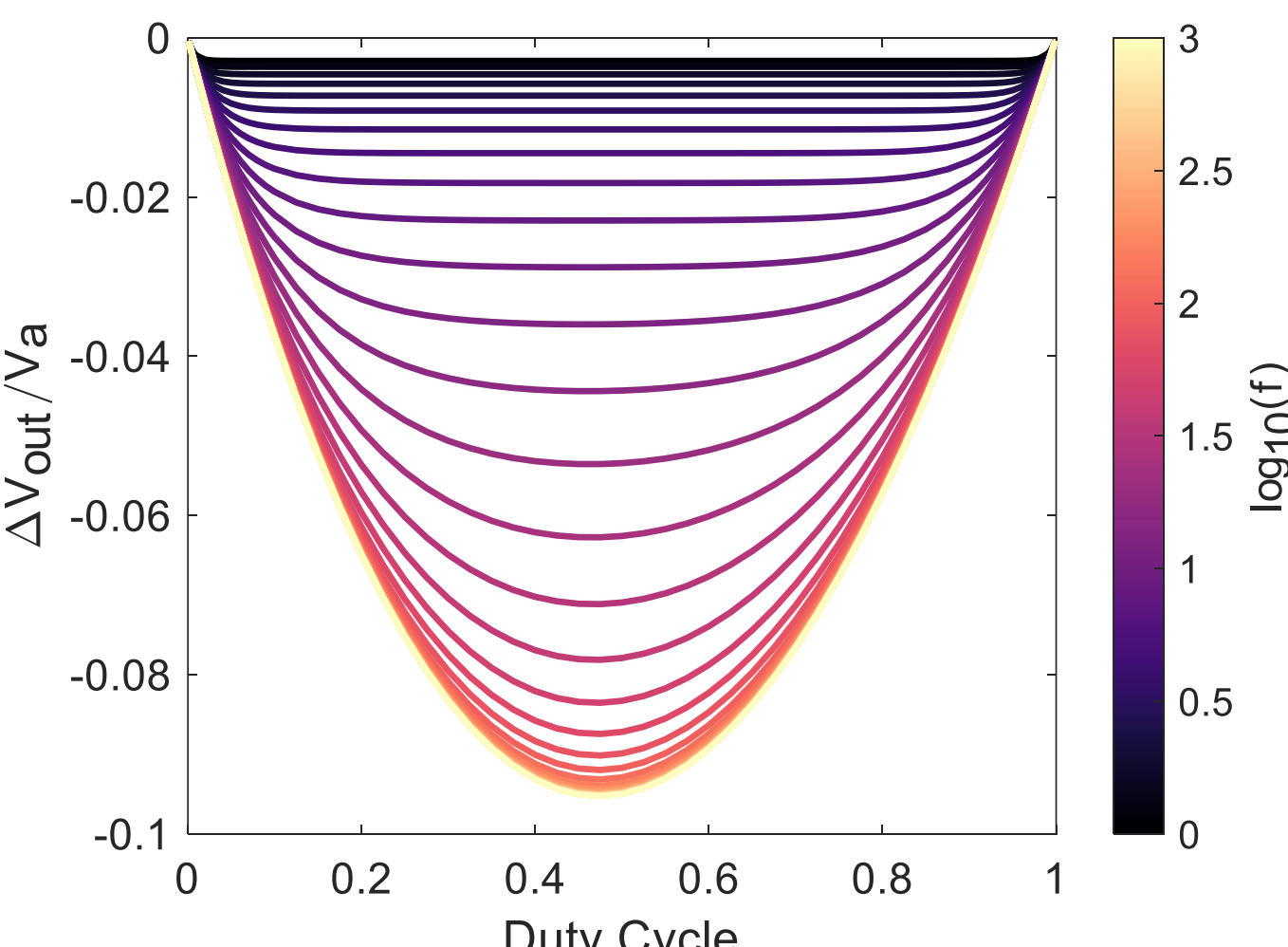

*Figure S32: The ratchet normalized output, $\Delta\bar{V}_{out}/V_a$, as a function of the input signal duty cycle for several input signal frequencies. The time constants are taken from Figure 1h for a duty cycle of 0.5: $\tau_{L,2} = 8.5$ ms, $\tau_{R,2} = 8.21$ ms, $\tau_{L,1} = 6.59$ ms, $\tau_{R,1} = 7.23$ ms.*

It is interesting to show the requirements for obtaining a non-zero output for a time-symmetric input signal ($d_c$ = 0.5). Inserting $d_c = 0.5$, $\tau_{L,1} = \tau_{R,1}$, $\tau_{L,2} = \tau_{R,2}$, and $\Delta V_{eq} = 0V$ into equation (S26) we obtain:

$$\frac{\Delta\bar{V}_{out}}{V_a} = \frac{2}{T}\frac{\left(1-\mathrm{e}^{-\frac{T}{2\tau_{L,1}}}\right)\left(1-\mathrm{e}^{-\frac{T}{2\tau_{R,2}}}\right)}{1-\mathrm{e}^{-\frac{T}{2}\left(\frac{1}{\tau_{L,1}}+\frac{1}{\tau_{R,2}}\right)}}\left(\tau_{L,1}-\tau_{R,2}\right) \quad \text{(S27)}$$

Thus, any difference between the charging (and discharging) time constants of the two surfaces results in a non-zero output and the longer time constant determines the direction of the driving force. Figure S33 shows the calculated normalized output $\Delta\bar{V}_{out}/V_a$ as a function of $\tau_{L,1} = \tau_{R,1} = \tau$ with $\tau_{L,2} = \tau_{R,2} = \tau_{ref} = 8.5\ ms$ and $d_c = 0.5$.

In a symmetric device both surfaces charge (and discharge) with the same time constants: $\tau_{R,1} = \tau_{L,2}$ and $\tau_{R,2} = \tau_{L,1}$. Under such conditions the output is zero for a $d_c = 0.5$, but can be non-zero for other duty cycle values. Figure S34 shows the ratchet normalized output as a function of the input signal duty cycle for several input signal frequencies. The charging time constants for both surfaces are $\tau_{R,1} = \tau_{L,2} = 8\ ms$, and the discharging time constants are $\tau_{R,2} = \tau_{L,1} = 6\ ms$. Similar to an anti-symmetric flashing ratchet,[10] the ratchet output is zero at duty cycles of 0,0.5 and 1 and is anti-symmetric about $d_c = 0.5$. Thus, in a symmetric system, a non-zero output can be obtained when the capacitances are nonlinear, and the input signal is not time symmetric ($d_c \neq 0.5$).

The suggested model shows that a non-linear capacitance is essential for an RBIP to operate. However, to fully predict and model the performance of real devices, the non-linear correlation between the input signal parameters and the charging and discharging time constants must be found. Since the frequency dispersion of the double layer capacitance is heavily affected by surface roughness, material crystallinity, and other micro-scale properties,[19,20] such analyses must be conducted for every RBIP material and fabrication process separately and is left for future work. The model assumes that the resistance for ion transport through the pores, $R_p$ is very high thus it accounts only for open-circuit conditions as in Figure 1i and Figure 2a,f. An equivalent circuit model that accounts for ion transport through the pores and the Ag/AgCl wires electrochemical reactions can be used to analyze other operating points. Nevertheless, when the Ag/AgCl auxiliary electrodes assure that charge neutrality is maintained, the RBIP operation can be modeled as a voltage source in parallel to a resistive element which accounts for charge transport through the pores (figure 2c). A more thorough analysis of equivalent circuit models for the RBIP is left for future work.

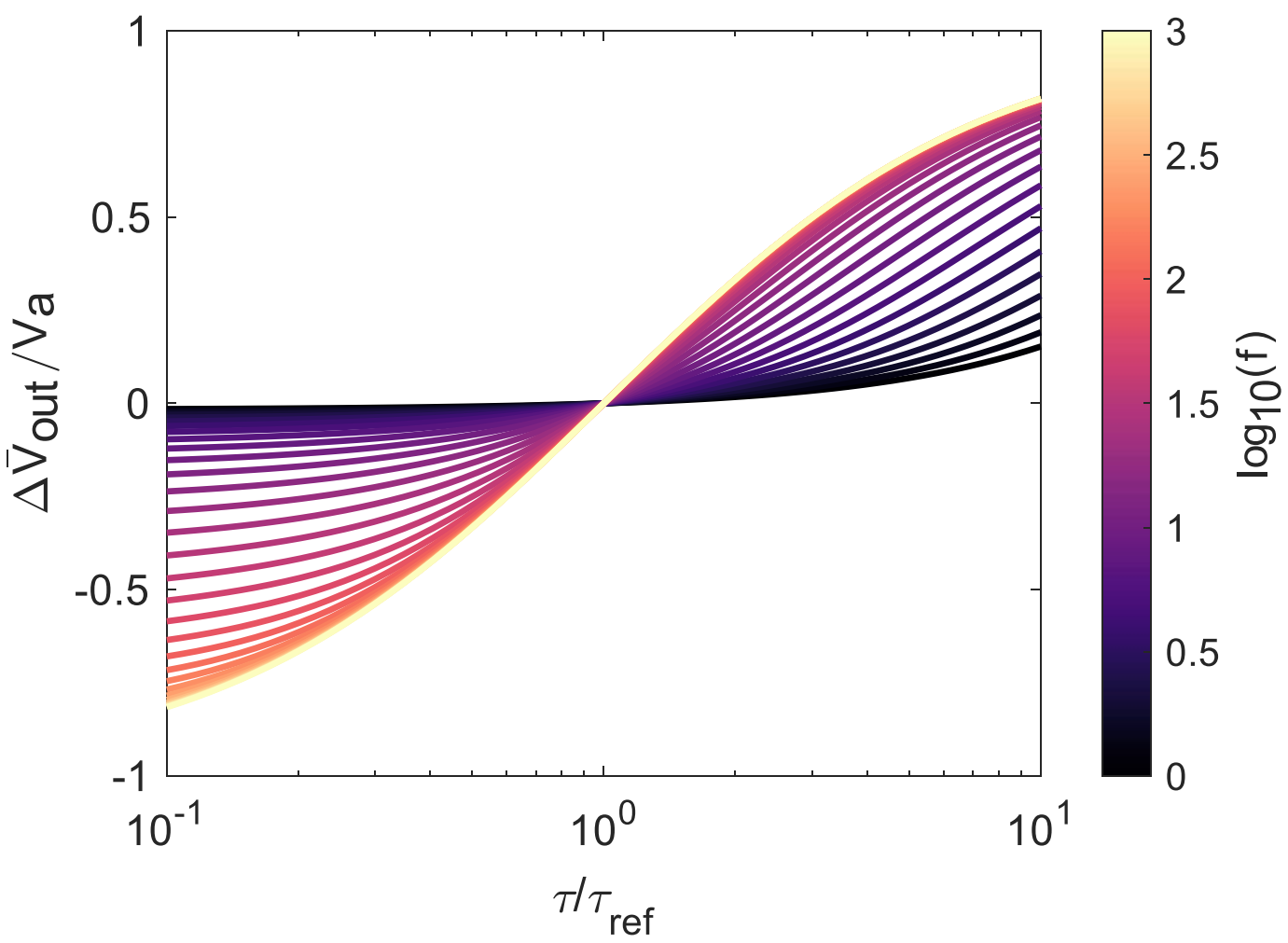

*Figure S33: the calculated normalized output $\Delta\bar{V}_{out}/V_a$ as a function of $\tau_{L,1} = \tau_{R,1} = \tau$ with $\tau_{L,2} = \tau_{R,2} = \tau_{ref} = 8.5\ ms$ and $d_c = 0.5$.*

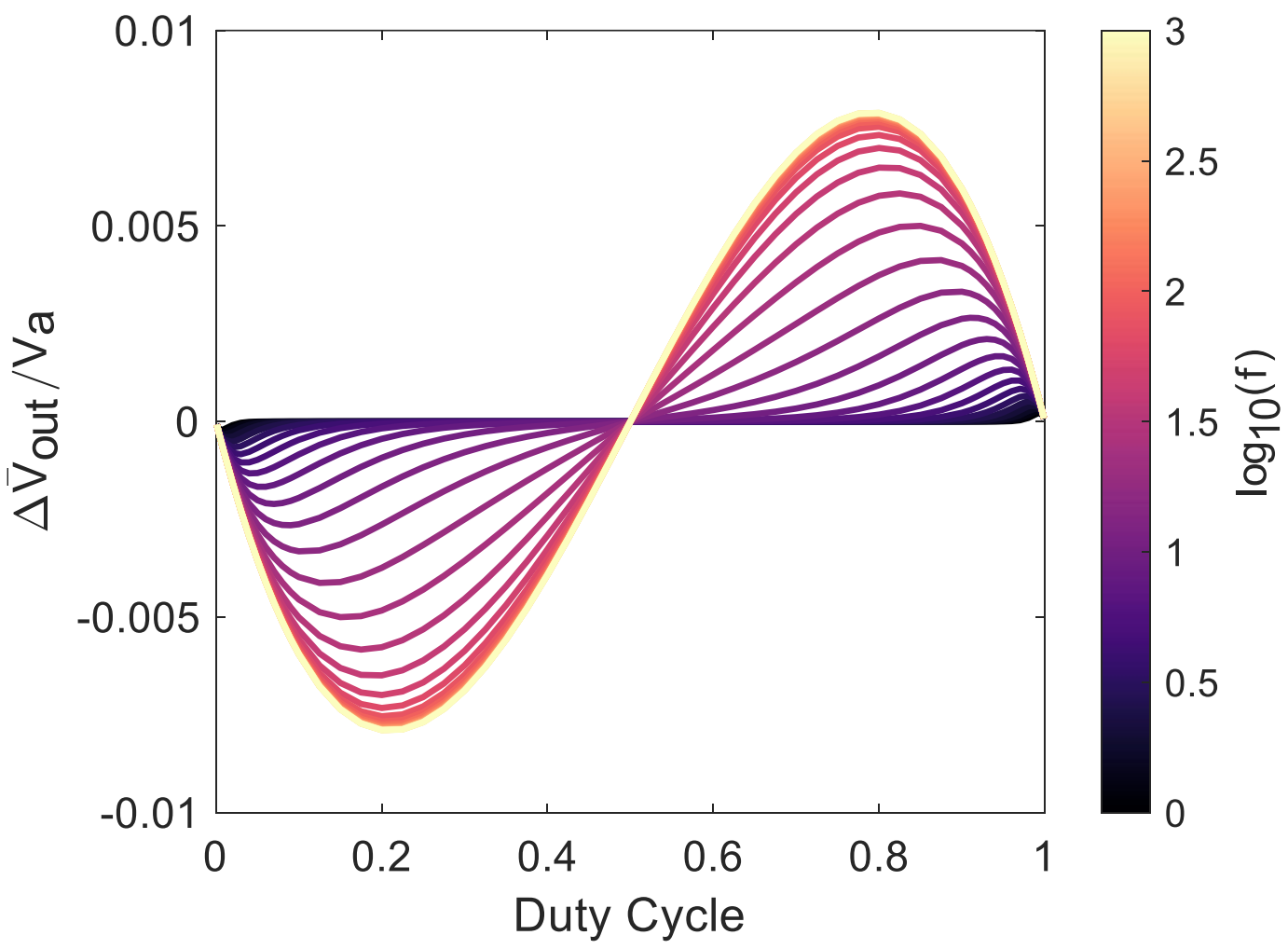

*Figure S34: the ratchet normalized output, $\Delta\bar{V}_{out}/V_a$, as a function of the input signal duty cycle for several input signal frequencies. The charging time constants for both surfaces are $\tau_{R,1} = \tau_{L,2} = 8\ ms$, and the discharging time constants are $\tau_{R,2} = \tau_{L,1} = 6\ ms$.*

Figure S35 shows the calculated ratchet output as a function of the time constants $\tau_{\mathrm{R},1}$ and $\tau_{\mathrm{R},2}$ for asymmetry factors of 0.05, 0.5 and 0.95 (a-c respectively). The other time constants are fixed at $\tau_{L,1} = 8$ ms, $\tau_{L,2} = 2$ ms. For all three asymmetry coefficients the highest output is obtained when the difference in time constants is the largest. The asymmetry coefficient acts as a weight that defines which of the two surfaces is more dominant in determining the output. For $a = 0.95$ (Figure S35c), the signal acts almost entirely on the left surface. As a result, the output is determined almost completely by it and changes in $\tau_{\mathrm{R},1}$ and $\tau_{\mathrm{R},2}$ have a negligible effect on the output. On the other hand, for $a = 0.05$ (Figure S35a), only the right surface is affected by the signal and the output is very sensitive to changes in $\tau_{\mathrm{R},1}$ and $\tau_{\mathrm{R},2}$. In all cases, the maximal output is obtained when there is a maximal difference in time constants of each surface i.e. $\tau_{L,1} = \tau_{R,1}$ and $\tau_{L,2} = \tau_{R,2}$.

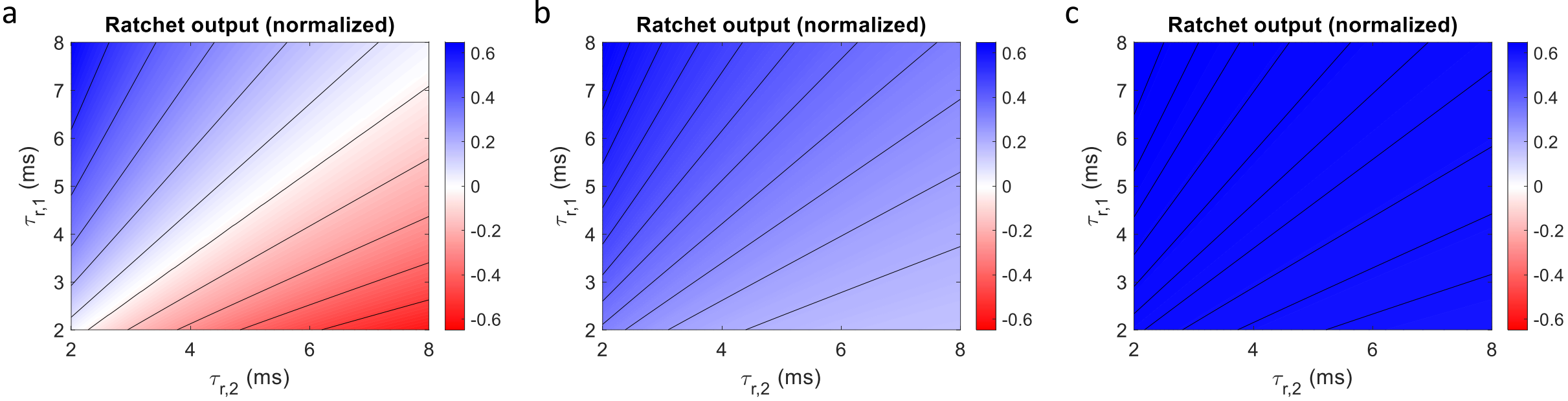

*Figure S35: The calculated ratchet normalized output, $\Delta\bar{V}_{out}/V_a$, as a function of the time constants $\tau_{R,1}$ and $\tau_{R,2}$ for asymmetry coefficients of 0.05, 0.5 and 0.95 (a-c respectively). $\tau_{L,1} = 8$ ms, $\tau_{L,2} = 2$ ms. The duty cycle is optimized for every pair of time constants and the frequency is 200 Hz.*